\documentclass[journal]{IEEEtran} 
\usepackage{times}
\usepackage{cite}
\usepackage{amsmath,amssymb,amsfonts}
\usepackage{textcomp}
\usepackage{xcolor}
\usepackage{xspace}
\usepackage{balance}
\usepackage{bm}
\usepackage{scalerel}
\usepackage{multicol}
\usepackage{graphicx}
\usepackage{url}
\usepackage{algorithm}
\usepackage[noend]{algpseudocode}
\usepackage{subcaption}
\usepackage{mwe}
\usepackage{breqn}

\usepackage{booktabs}
\usepackage{tabularx}
\usepackage{siunitx}
\usepackage{{makecell}}
\usepackage{mathtools}
\setcellgapes{3pt}\makegapedcells
\usepackage{array,collcell}
\usepackage{hyperref} 
\usepackage{tikz}
\usepackage[normalem]{ulem}

\newcommand{\sref}[1]{(s\ref{#1})}

\newcommand{\src}{\mathtt{S}}
\newcommand{\inter}{\mathtt{I}}
\newcommand{\sink}{\mathtt{T}}
\newcommand{\sys}{\text{sys}}
\newcommand{\tester}{\text{TA}}
\newcommand{\test}{\text{test}}

\newcommand{\hist}{\text{hist}}
\DeclareMathOperator{\Feventually}{\rotatebox[origin=c]{45}{$\Box$}}

\newcolumntype{M}{>{\hfil$\displaystyle}X<{$\hfil}} 
\newcolumntype{L}{>{\collectcell\AddLabel}r<{\endcollectcell}}

\usepackage{amssymb}
\usepackage{amsthm}

\usepackage{dsfont}
\usepackage{tikz-network}
\usepackage[shortlabels]{enumitem}
\usepackage{float}
\theoremstyle{definition}

\newtheorem{problem}{Problem}
\makeatletter
\newtheorem*{rep@theorem}{\rep@title}
\newcommand{\newreptheorem}[2]{%
\newenvironment{rep#1}[1]{%
 \def\rep@title{#2 \ref{##1}}%
 \begin{rep@theorem}}%
 {\end{rep@theorem}}}
\makeatother
\newtheorem{theorem}{Theorem}
\newreptheorem{theorem}{Theorem}

\newtheorem{lemma}{Lemma}
\newtheorem{proposition}{Proposition}
\newtheorem{corollary}{Corollary}
\newtheorem{remark}{Remark}
\newtheorem{example}{Example}
\newtheorem{definition}{Definition}
\newtheorem{assumption}{Assumption}
\newtheorem{construction}{Construction}

\newcommand{\mc}[1]{\mathcal{#1}}

\renewcommand{\algorithmicrequire}{\textbf{Input:}}
\renewcommand{\algorithmicensure}{\textbf{Output:}}

\newcommand{\fvec}[0]{\mathbf{f}}
\newcommand{\dvec}[0]{\mathbf{d}}

\def\BibTeX{{\rm B\kern-.05em{\sc i\kern-.025em b}\kern-.08em
    T\kern-.1667em\lower.7ex\hbox{E}\kern-.125emX}}

\makeatletter
\newcommand\fs@betterruled{%
  \def\@fs@cfont{\bfseries}\let\@fs@capt\floatc@ruled
  \def\@fs@pre{\vspace*{5pt}\hrule height.8pt depth0pt \kern2pt}%
  \def\@fs@post{\kern2pt\hrule\relax}%
  \def\@fs@mid{\kern2pt\hrule\kern2pt}%
  \let\@fs@iftopcapt\iftrue}
\floatstyle{betterruled}
\restylefloat{algorithm} 
\makeatother

\newcommand{\deni}[1]{{\color{blue}[\textsc{Deni}: \emph{#1}]}}

\usepackage{graphicx,scalerel}
\newcommand\bigbullet[1][1.5]{\mathbin{\ThisStyle{\vcenter{\hbox{%
  \scalebox{#1}{$\SavedStyle\bullet$}}}}}%
}
\definecolor{blue}{HTML}{648FFF}
\definecolor{orange}{HTML}{FE6100}
\definecolor{magenta}{HTML}{DC267F}
\definecolor{yellow}{HTML}{FFB000}
\newcommand\bluefilledcirc{\ensuremath{{\color{blue}\bigbullet}}}
\newcommand\pinkfilledcirc{\ensuremath{{\color{magenta}\bigbullet}}}
\newcommand\orangefilledcirc{\ensuremath{{\color{orange}\bigbullet}}}
\newcommand\yellowfilledcirc{\ensuremath{{\color{yellow}\bigbullet}}}
\def\BibTeX{{\rm B\kern-.05em{\sc i\kern-.025em b}\kern-.08em
    T\kern-.1667em\lower.7ex\hbox{E}\kern-.125emX}}
\begin{document}
\title{Flow-Based Synthesis of Reactive Tests for Discrete Decision-Making Systems with Temporal Logic Specifications}
\author{Josefine B. Graebener$^{*}$, \and Apurva S. Badithela$^{*}$, \and Denizalp Goktas, \and Wyatt Ubellacker, \and Eric V. Mazumdar, \and \newline Aaron D. Ames, \and Richard M. Murray
\thanks{This work was supported in by the U.S. Air Force Office of Scientific Research (AFOSR) under Grant FA9550-22-1-0333 and Grant FA9550-19-1-0302.}
\thanks{$*$ These authors contributed equally. Corresponding author: A.S. Badithela.}
\thanks{J.B. Graebener is with the Graduate Aerospace Laboratories of California Institute of Technology, Pasadena CA 91125 USA (e-mail: jgraeben@caltech.edu).}
\thanks{A.S. Badithela, W. Ubellacker, A.D. Ames, R.M. Murray are affiliated with Control and Dynamical Systems, California Institute of Technology, Pasadena CA 91125 USA. (e-mail: 
\{apurva, wubellac, ames, murray\}@caltech.edu).}
\thanks{D. Goktas is with the Department of Computer Science, Brown University, Providence RI 02912 USA (e-mail: denizalp\_goktas@brown.edu).}
\thanks{E.V. Mazumdar is with the Department of Computing and Mathematical Sciences, California Institute of Technology, Pasadena CA 91125 USA (e-mail: mazumdar@caltech.edu).}
}

\maketitle

\begin{abstract}
Designing tests to evaluate if a given autonomous system satisfies complex specifications is challenging due to the complexity of these systems. 
This work proposes a flow-based approach for reactive test synthesis from temporal logic specifications, enabling the synthesis of test environments consisting of static and reactive obstacles and dynamic test agents. The temporal logic specifications describe desired test behavior, including system requirements as well as a test objective that is not revealed to the system. The synthesized test strategy places restrictions on system actions in reaction to the system state. The tests are minimally restrictive and accomplish the test objective while ensuring realizability of the system's objective without aiding it (semi-cooperative setting). Automata theory and flow networks are leveraged to formulate a mixed-integer linear program (MILP) to synthesize the test strategy. For a dynamic test agent, the agent strategy is synthesized for a GR(1) specification constructed from the solution of the MILP. If the specification is unrealizable by the dynamics of the test agent, a counterexample-guided approach is used to resolve the MILP until a strategy is found. This flow-based, reactive test synthesis is conducted offline and is agnostic to the system controller. Finally, the resulting test strategy is demonstrated in simulation and experimentally on a pair of quadrupedal robots for a variety of specifications. 
\end{abstract}


\section{Introduction}
\label{sec:introduction}
\begin{figure*}
\centering
\begin{minipage}{\textwidth}
\centering
\includegraphics[width=\linewidth,trim={0.0cm 0.0cm 0cm 0.0cm}]{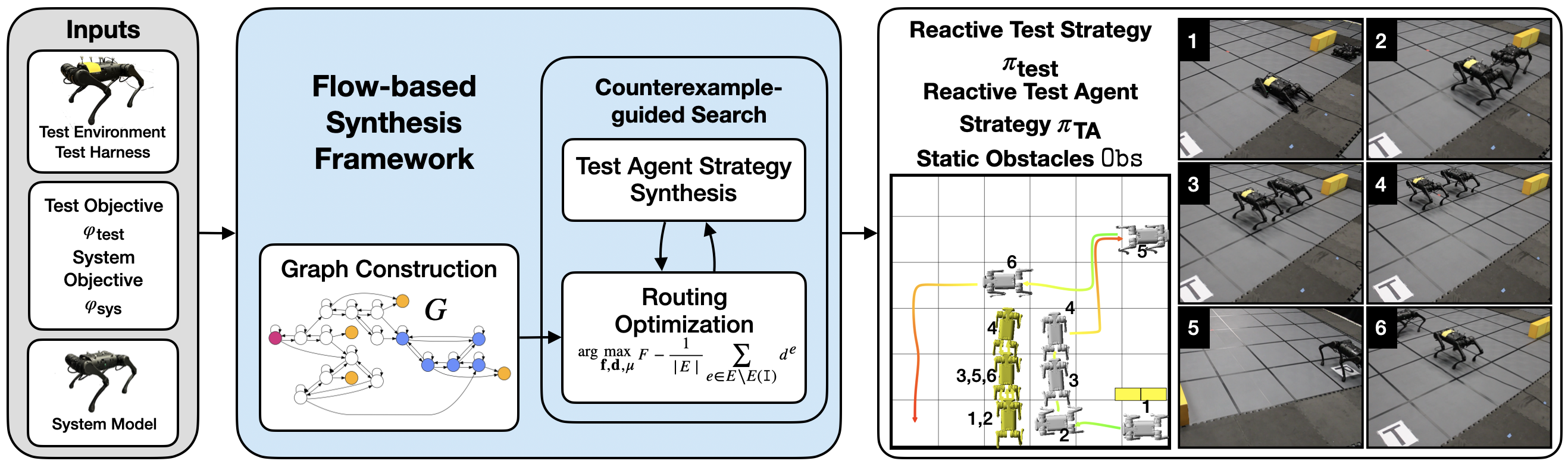}
\end{minipage}
\caption{Overview of the flow-based test synthesis framework which consists of three key parts: i) graph construction, ii) routing optimization, and iii) test environment synthesis (e.g., reactive test strategy / test agent strategy, static obstacles).}
\label{fig:overview}
\end{figure*}

Safety is imperative for a wide range of autonomous systems, from self-driving vehicles, to autonomous flight and space missions, to assistive robotics, and medical devices. To ensure safety, various challenges need to be addressed~\cite{sotif}. For example, these systems need to be aware of their own state and adapt their behavior in response to the environment, which requires reasoning over both discrete and continuous inputs and states. 
Deployment of these safety-critical autonomous systems requires thorough testing, both in simulation and in the operating environment, which is crucial to validating the system's performance. Typically, test cases are designed to uncover bugs and corner cases in the system design that lead to safety-critical errors. However, for these tests to be successful, executing them requires setting up a test environment that is consistent with the test case while also allowing for correct system implementations. To make this process efficient, it is equally important to automatically synthesize these test environments for the desired test case, i.e., automatically synthesize test environments that reveal corner cases.  

In this work, we focus on synthesizing test environments (e.g., placement of obstacles, agent strategies) to test the discrete decision-making logic in an autonomous system. Tests are synthesized from temporal logic descriptions of desired test behavior which encodes aspects of the test unknown to the system (test objective) in addition to the system requirements (system objective). The test objective is meant to capture the ``challenging" aspect of the test in terms of high-level decision-making, and is not revealed to the system. 
The purpose of testing is to check that the system can take correct decisions despite being given opportunities to fail, i.e., verify correct decision in the presence of ``hard tests'' or corner cases. Our framework routes the system to the test objective while also giving the system freedom to make decisions and ensuring that the test is fair (i.e., system can satisfy its requirements if it makes correct decisions). 
 Therefore, we synthesize tests that minimally restrict the system's decision-making to realize the desired test behavior. Fig.~\ref{fig:overview} provides an overview of the flow-based test synthesis framework.

\subsection{Background on Test and Evaluation}

Tests are typically manually designed by test engineers --- identifying challenging test cases and manually constructing the test environments either from expert experience or failure reports. Examples of this include the qualification tests in the DARPA Urban Challenge, track testing by self-driving car companies~\cite{zoox_test,webb2020waymos}, and constructing test scenarios in simulation using tools such as CARLA~\cite{dosovitskiy2017carla} or Scenic~\cite{fremont2019scenic}, for which test engineers either partially specify the scenarios or recreate them from crash reports~\cite{gambi2019generating,stark2020generation,webb2020waymos}. Due to the time-intensive nature of this endeavor, automatically finding challenging tests for safety-critical systems is an active area of research~\cite{lou2022testing, corso2021survey}. For self-driving vehicles, there is ongoing effort to standardize the testing process~\cite{ winner2019pegasus, li2016intelligence}.

Black-box optimization algorithms~\cite{mullins2018adaptive} and reinforcement learning~\cite{corso2019adaptive,feng2023dense} have been used to search over a specified input domain to find a falsifying input that leads to a trajectory that violates a metric of mission success. This metric can be derived from formal temporal logic specifications~\cite{annpureddy2011s,abbas2011linear,fainekos2009robustness,donze2010breach,fremont2020formal,tuncali2018simulation,innes_9811579} or from control barrier functions~\cite{pakella}. However, falsification algorithms typically require a well-defined test environment, and find a falsifying trace by fine-tuning the parameters in that scenario. The framework proposed in this paper is complementary to these approaches --- our focus is on synthesizing high-level strategies for the test environment, and continuous parameters of the synthesized test environment (e.g., continuous pose values of test agents, friction coefficients, exact timing of events) can be inputs to falsification algorithms for fine-tuning. 

Typically, high-level choices of autonomous robotic systems exhibit discrete decision-making~\cite{wongpiromsarn2023formal,fainekos_gazit}. The use of linear temporal logic (LTL) model checkers for testing has been explored in~\cite{tan2004specification,plaku2013falsification, fraser2007using, fraser2008reachability}. In these works, counterexamples from model-checking are used to construct test cases for deterministic systems and are inconclusive if the system behavior deviates from the expected test case. However, since robotic systems are often reactive, and because we want to generate tests without specific knowledge of the system controller, the generated tests must be able to \emph{adapt} or \emph{react} to system behavior at runtime. Our test synthesis procedure is gray-box in the sense that it requires knowledge of a nondeterministic model of the system but is agnostic to the high-level controller of the system and is completely black-box to models and controllers at lower levels of abstraction.

Adaptive specification-based testing using discrete logics has been explored in~\cite{bloem2019synthesizing,tretmans1996conformance,aichernig2015killing, hierons2006applying, petrenko2014adaptive}. Particularly in~\cite{bloem2019synthesizing}, an adaptive test strategy is synthesized using reactive synthesis~\cite{pnueli1989synthesis} from LTL specifications of the system and the fault model, both of which are specified by the test engineer. This adaptive test strategy ensures that the resulting test trace demonstrates a fault if the system implementation is faulty according to the fault model. However, these fault models must be carefully specified over the outputs of the system. While this is incredibly useful for specifying and catching sub-system level faults, it becomes intractable for specifying complex system-level faults resulting from multiple outputs. Our test synthesis framework is also specification-based and adaptive, but we specify desired test behavior in the form of test objectives instead of specifying system-level faults. Furthermore, in~\cite{bloem2019synthesizing}, the adaptive test strategies are synthesized from fault models that are designed for coverage goals corresponding to specification coverage, without accounting for the freedom of the system to satisfy its own requirements. We seek to synthesize reactive test strategies that demonstrate the test objective while placing minimal restrictions on the system. The automata-theoretic tools used in this paper build on concepts used in correct-by-construction synthesis and model checking~\cite{bloem2012synthesis,baier2008principles}. This background is covered in Section~\ref{sec:background}.

In~\cite{yannakakis2004testing}, testing of reactive systems was introduced as a game between two players, where the tester and the system try to reveal and hide faults, respectively. Similarly, in~\cite{nachmanson2004optimal} the test strategy is found by reasoning over a game graph to optimize reachability and coverage metrics. Testing in cooperative game settings has been explored in~\cite{david2008cooperative, bartocci2021adaptive}. However, the reactive test synthesis problem we consider is neither fully \emph{adversarial} nor fully \emph{cooperative} --- a well-designed system is cooperative with the test environment in realizing the system objective, but since the system is agnostic to test objective, it need not cooperate with the test environment in realizing it. 

We consider test environments that can consist of the following: static obstacles that restrict the system throughout the test, reactive obstacles and a dynamic test agent that is reactive to system behavior at runtime. In particular, we leverage flow networks to pose the test synthesis problem as a mixed-integer linear program (MILP). In recent years, network flow optimization frameworks with tight convex relaxations have led to massive computational speed-ups in solving robot motion planning problems~\cite{marcucci2024shortest,marcucci2023motion}. Network flow-based mixed integer programs have also been to synthesize playable game levels in video games~\cite{zhang2020video}, which was then applied to construct playable scenarios in robotics settings~\cite{fontaine2021importance}. 

In previous work~\cite{badithela2023synthesizing}, we formulated this problem in a semi-cooperative setting as a min-max Stackelberg game with coupled constraints. Despite being defined over continuous variables with an affine objective and affine constraints, the prior formulation resulted in slow runtimes and did not guarantee that the optimal solution would realize the test objective. Furthermore, it could only reactively restrict system actions, and did not characterize how to translate these restrictions to the choice of a test agent strategy. In this work, we present a simpler formulation of the routing optimization as an MILP, which led to an improvement in runtime. Test strategies from optimal solutions of the MILP are guaranteed to realize the test objective in a least restrictive fashion. Finally, we present a formal approach to restricting system actions in the form of static/reactive obstacles and dynamic test agent strategies, including a counterexample-guided approach to synthesize a test agent strategy from the solution of the routing optimization. 


\subsection{Contributions}

In this work, we study the problem of synthesizing a reactive test strategy for a test environment for discrete decision-making systems given a formal test objective, unknown to the system under test. In particular, we ask whether such a test strategy exists without making it impossible for the system to meet its specification.

To obtain the main results of this paper, we first characterize system and test objectives using a variety of specification patterns commonly used in robotic missions~\cite{menghi2019specification}. We formalize both the restrictiveness and the feasibility of a test strategy, i.e., a system should have freedom to make decisions and a correct system should be able to pass the test. Secondly, these conditions are translated into a routing optimization on a flow network to capture the requirement that all test executions that satisfy the system objective should demonstrate the test objective. For each test environment, we set up an MILP to find cuts, corresponding to restrictions on system actions, on the flow network. For static and reactive obstacles, the solution of the MILP is realized in the form of an obstacle placement test strategy. We prove that the optimal solutions to the MILPs solve the aforementioned routing requirement. Third, in the case of dynamic agents, we match the restrictions on system actions to a test agent strategy via GR(1) synthesis~\cite{bloem2012synthesis,wongpiromsarn2011tulip}. Furthermore, we use a counterexample-guided approach to exclude unrealizable solutions from the MILP until we find a realizable test agent strategy. We prove that test agent strategies synthesized in this manner exactly correspond to the test strategy found from the MILP. In the extended version, we prove that the routing problem is NP-hard via a reduction from 3-SAT. Despite this, our framework can reliably handle medium-sized problems with thousands of integer variables. Empirical runtimes for parametrized problems are also provided. 

Finally, the test synthesis framework is demonstrated on simulated grid world settings and on hardware with a pair of quadrupedal robots. For all experiments, our framework synthesizes test strategies that place the fewest possible restrictions on the system over the course of the test either by obstacle placement or a dynamic agent. In experiments with reactive obstacles and dynamic agents, the reactive test strategy results in a different test execution depending on system behavior. Despite this, the system is always routed through the test objective (e.g., being put in low-fuel state or having to walk over challenging terrain). 

\section{Preliminaries}
\label{sec:background}
This section introduces concepts from automata theory and network flows that are relevant to this work. 
\subsection{Automata Theory and Temporal Logic}
\label{sec:bkgnd} 

\begin{definition}[Finite Transition System]
A \emph{finite transition system} (FTS) is the tuple
\vspace{-1mm}
\begin{equation*}
TS \coloneqq (S,A,\delta,S_0,AP,L),\,
\vspace{-1mm}
\end{equation*}
where $S$ denotes a finite set of states, $A$ is a finite set of actions, $\delta: S \times A \rightarrow S$ the transition relation, $S_0$ the set of initial states, $AP$ the set of atomic propositions, and $L: S \rightarrow 2^{AP}$ denotes the labeling function. We denote the transitions in $TS$ as $TS.E:= \{(s,s') \in S \times S \: \vert \: \text{ if } \exists a \in A \text{ s.t. } \delta(s,a) = s'\}$. We refer to the states of $TS$ as $TS.S$, and similarly denote the other elements of the tuple.
An execution $\sigma$ is an infinite sequence $\sigma = s_0 s_1 \dots$, where $s_0 \in S_0$ and $s_k \in S$ is the state at time $k$. We denote the finite prefix of the trace $\sigma$ up to the current time $k$ as $\sigma_k$. A strategy $\pi$ is a function \(\pi:(TS.S)^*TS.S\rightarrow TS.A\).
\end{definition}

\begin{definition}[System]
    The \emph{system under test} is modeled as a finite transition system \(T_{\sys}\) with a single initial state, that is, \(\vert T_{\sys}.S_0\vert  = 1\). Furthermore, at least one of the system states is terminal (i.e., no outgoing edges).
\end{definition}

The system designers provide the states \(S\), actions \(A\), transitions \(\delta\), and a set of possible initial conditions \(S_0\), set of atomic propositions, \(AP_{\sys}\) and a corresponding label function \(L_{\sys}: S\rightarrow 2^{AP_{\sys}}\). We require a unique initial condition $s_0 \in S_0$ to synthesize the test. If the test designer wishes to select an initial condition, then they can synthesize the test for each \(s_0 \in S_0\) and choose accordingly.
In addition to \(AP_{\sys}\), the test designer can choose additional atomic propositions \(AP_{\test}\) and define a corresponding labeling function \(L:S\rightarrow 2^{AP}\), where \(AP:= AP_{\sys} \cup AP_{\test}\). For test synthesis, the system model is \(T_{\sys} = (S,A,\delta,\{s_0\},AP,L)\) is defined for the specific initial condition \(s_0\) chosen by the test designer. 
The terminal state is used for defining test termination when the system satisfies its objective.

\begin{assumption}
\label{asm:bidirectionality}
Except for sink states, transitions between states of the system are bidirectional: \(\forall  (s, s') \in T_{\sys}.E\) where \(s'\) is not a terminal state, we also have \((s', s) \in T_{\sys}.E\).
\end{assumption}

This assumption is for a simpler presentation, and the framework can be extended to transition systems without this assumption (see Remark~\ref{rem:bidirectional}). 

\begin{definition}[Test Harness]
A \emph{test harness} is used to constrain a state-action \((s,a)\) pair of the system in the sense that the system is prevented from taking action \(a\) from state \(s \in T_{\sys}.S\). Let the actions \(A_H \subseteq T_{\sys}.A\) denote the subset of system actions that can be restricted by the test harness. The test harness \(H: T_{\sys}.S \rightarrow 2^{A_H}\) maps states of the transition system to actions that can be restricted from that state. 
\end{definition}

In the examples in this paper, every state of the system has a self-loop transition corresponding to stay-in-place action, but the proposed framework does not require this. 
Note that in our examples, \(A_H\) does not contain self-loop actions.

\begin{definition}[Test Environment]
    The \emph{test environment} consists of one or more of the following: static obstacles, reactive obstacles, and dynamic test agents.
    A \emph{static obstacle} on \((s,s') \in T_{\sys}.E\) is a restriction on the system transition \((s,s')\) that remains in place for the entire duration of the test. 
    A \emph{reactive obstacle} on \((s,s') \in T_{\sys}.E\) is a temporary restriction on the system transition \((s,s')\) that can be enabled/disabled over the course of the test. 
    A \emph{dynamic test agent} can occupy states in \(T_{\sys}.S\), thus restricting the system from entering the occupied state.
\end{definition}

In this work, we synthesize tests for high-level decision-making components of the system under test and therefore model it as a discrete-state system. Linear temporal logic (LTL) has been effective in formally specifying safety and liveness requirements for discrete-decision making~\cite{wongpiromsarn2012receding,kress2009temporal,belta2019formal}. For our problem, we use LTL to capture the system and test objectives.

\begin{definition}[Linear Temporal Logic~\cite{baier2008principles}]
\emph{Linear temporal logic} (LTL) is a temporal logic specification language that allows reasoning over linear-time trace properties.
The syntax of LTL is given as:
\vspace{-2mm}
$$\varphi ::= \emph{True} \: | \:a \:| \:\varphi_1 \land \varphi_2 \:| \neg \varphi \:|\: \bigcirc \varphi \: |\: \varphi_1 \mathcal{U} \varphi_2,$$
with $a \in AP$, where $AP$ is the set of atomic propositions, $\land$ (conjunction) and $\neg$ (negation) are the Boolean connectors from which other Boolean connectives such as \(\rightarrow\) can be defined, and $\bigcirc$ (next) and $\mathcal{U}$ (until) are temporal operators. Let $\varphi$ be an LTL formula over $AP$. We can define the operators $\Feventually$ (eventually) and $\square$ (always) as $\Feventually \varphi = \emph{True}\: \mathcal{U} \varphi$ and $\square \varphi = \neg \Feventually \neg \varphi$. 
For an execution $\sigma = s_0 s_1 \ldots$ and an LTL formula $\varphi$, $s_i \vDash \varphi$ iff $\varphi$ holds at $i \geq 0$ of $\sigma$. More formally, the semantics of LTL formula $\varphi$ are inductively defined over an execution $\sigma = s_0 s_1 \ldots$ as follows,
\begin{itemize}
    \item[] for $a \in AP$, $s_i \vDash a$ iff $a$ evaluates to \emph{True} at $s_i$,
    \item[] $s_i \vDash \varphi_1 \land \varphi_2$ iff $s_i \vDash \varphi_1$ and $s_i \vDash \varphi_2$,
    \item[] $s_i \vDash \neg \varphi$ iff $\neg(s_i \vDash \varphi)$,
    \item[] $s_i \vDash \bigcirc \varphi$ iff $s_{i+1} \vDash \varphi$, and
    \item[] $s_i \vDash \varphi_1 \mathcal{U} \varphi_2$ iff $\exists k \geq i, s_k \vDash \varphi_2$ and $s_j \vDash \varphi_1$, for all $i \leq j < k$.
\end{itemize}
An execution/trace $\sigma = s_0 s_1 \ldots$ satisfies formula $\varphi$, denoted by $\sigma \models \varphi$, iff $s_0 \models \varphi$. A \emph{strategy} \(\pi\) is correct (satisfies formula \(\varphi\)), if the trace \(\sigma_{\pi}\) resulting from the strategy satisfies \(\varphi\).
\end{definition}
Every LTL formula can be transformed into an equivalent non-deterministic B\"uchi automaton, which can then be converted to a deterministic B\"uchi automaton~\cite{baier2008principles}. 

\begin{definition}[Deterministic Büchi Automaton]
A \emph{non-deterministic B\"uchi automaton} (NBA)~\cite{Buchi1990,baier2008principles} is a tuple 
$\mc{B} \coloneqq (Q, \Omega, \delta, Q_0, F),$
where \(Q\) denotes the states, \(\Omega \coloneqq 2^{AP}\) is the set of alphabet for the set of atomic propositions \(AP\), \(\delta : Q \times  \Omega \rightarrow  Q\) denotes the transition function, \(Q_0 \subseteq Q\) represents the initial states, and \(F\subseteq Q\) is the set of acceptance states. The automaton is a \emph{deterministic B\"uchi automaton} (DBA) iff \(\vert Q_0\vert \leq 1\) and \(\vert \delta(q, A) \vert \leq 1\) for all \(q \in Q\) and \(A \in \Omega\).
\end{definition}

\begin{remark}
We use \emph{deterministic} B\"uchi automata since each input word corresponding to an execution should have a unique run on the automaton. While there are several different automata representations, deterministic B\"uchi automata are a natural choice for LTL specifications. 
\end{remark}

A \emph{product} of two deterministic Büchi automata, $\mc{B}_1$ and $\mc{B}_2$ over the alphabet $\Omega$, is defined as $\mc{B}_1 \otimes \mc{B}_2 \coloneqq (Q, \Omega, \delta, Q_0, F)$, with states $Q \coloneqq \mc{B}_1.Q \times \mc{B}_2.Q$, initial state $Q_0 \coloneqq \mc{B}_1.Q_0 \times \mc{B}_2.Q_0$, acceptance states $F\coloneqq \mc{B}_1.F \times \mc{B}_2.F$. The transition relation $\delta$ is defined as follows, for all $(q_1,q_2) \in Q$,  for all $A \in \Omega$,  $\delta((q_1,q_2), A) = (q'_1, q'_2)$ where $\mc{B}_1.\delta(q_1,A)=q_1'$ and $\mc{B}_2.\delta(q_2,A)=q_2'$. 

\newcommand{\thetagequation}{\theequation}
\renewcommand{\thetagequation}{{s}\arabic{equation}}
\newcommand\AddLabel[1]{%
  \refstepcounter{equation}
  (\thetagequation)
  \label{#1}
}

The desired test behavior can be captured via sub-tasks that are defined over atomic propositions \(AP\). Table~\ref{tab:spec_pattern} lists the sub-task specification patterns that are considered. These specification patterns are commonly used to specify robotic missions~\cite{menghi2019specification}. The desired test behavior is characterized by the system and test objectives, defined over the set of atomic propositions \(AP\) that can be evaluated on system states \(T_{\sys}.S\).

\begin{table}[htbp]
\caption{Sub-task specification patterns defined on atomic propositions.}
\label{tab:spec_pattern}\centering
\begin{tabularx}\columnwidth{@{}lXL@{}}
\toprule
Name & Formula & \multicolumn{1}{l}{}\\
\midrule
Visit & $ \bigwedge\limits_{i=1}^{m} \Feventually p^{i}$ & eq:reach_spec\\
Sequenced Visit & $ \Feventually (p^{0} \land (\Feventually (p^1 \land \ldots \Feventually p^m)))$ & eq:sequence_reach_spec\\
Safety & $ \square \neg p$ & eq:safe_spec\\
Instantaneous Reaction & $ \square  (p \rightarrow q)$ & eq:instant_reaction_spec\\
Delayed Reaction & $ \square  (p \rightarrow \Feventually q)$ & eq:delayed_reaction_spec\\
\bottomrule
\end{tabularx}
\end{table}

\begin{definition}[Test Objective]
The \emph{test objective} \(\varphi_{\test}\) consists of at least one visit or sequenced visit sub-task or a conjunction of these sub-tasks.
The B\"uchi automaton \(\mc{B}_{\test}\) corresponds to the test objective \(\varphi_{\test}\). 
\end{definition}

\begin{definition}[System Objective]
The \emph{system objective} \(\varphi_{\sys}\) consists of at least one visit or sequenced visit sub-task. The final visit proposition should be a terminal state of the system. In addition, it can also contain some conjuction of safety, instantaneous and/or delayed reaction, and visit and/or sequenced visit sub-tasks.
The B\"uchi automaton \(\mc{B}_{\sys}\) corresponds to the system objective \(\varphi_{\sys}\). We say that the \emph{system reaches its goal} or that the \emph{test execution satisfies the system objective} if the system trace is accepted \(\mc{B}_{\sys}\). 
\end{definition}

Typically, some aspects of a test are not revealed to the system until test time such as testing the persistence of a robot or prompting it to exhibit a difficult maneuver by placing obstacles in its path. This is formalized as a test objective which is not known to the system. In contrast, the system is aware of the system objective, which captures its requirements.
For example, to test for \emph{safety}, the system should know to avoid unsafe areas \text{\sref{eq:safe_spec}}. To test a \emph{reaction}, $\square (p \rightarrow q)$, the system needs to be aware of the reaction requirement~\text{\sref{eq:instant_reaction_spec}}, and the test objective needs to contain the corresponding visit requirement $\Feventually p$ to trigger the reaction. Furthermore, the test objective can contain standalone reachability (visit and/or sequenced visit) sub-tasks that are not associated with a system reaction sub-task, but require the system to reach/visit certain states. The test objective is accomplished by restricting system actions in reaction to the system state.

In addition to the system objective, the system must interact safely with the test environment. The system must also obey the initial condition set by the test designer. For each obstacle/agent of the test environment, the system controller must respect the corresponding restrictions on its actions (i.e., cannot crash into obstacles/agents). Furthermore, for a valid system implementation, all lower-level planners and controllers of the system must simulate transitions on \(T_{\sys}\).
\begin{definition}[System Guarantees]
     The system guarantees are a conjunction of the system objective, initial condition, safe interaction with the test environment, and a system implementation respecting the model \(T_{\sys}\).
\end{definition}
\begin{definition}[System Assumptions]
\label{def:sys_assume}
    The system \emph{assumes} that the test environment satisfies the following conditions:\\
    \noindent
    \textbf{A1.} The test environment can consist of: i) static obstacles (e.g., wall), ii) reactive obstacles (e.g., door), and iii) test agents whose dynamics are provided to the system. \\
    \noindent
    \textbf{A2.} The test environment will not take any action that will inevitably lead to unsafe behavior (e.g., not restricting a system action after the system has committed to it, test agents not colliding into the system).\\
    \noindent
    \textbf{A3.} The test environment will not take any action that will inevitably block all paths for the system to reach its goal (e.g., restrictions will not completely the enclose the system or block it from progressing to its goal).\\
    \noindent
    \textbf{A4.} If the system and test environment are in a livelock, the system will have the option to break the livelock and take a different path toward its goal. 
\end{definition}
A \emph{correct system strategy} satisfies the system guarantees when the test environment satisfies the system assumptions. This full system specification cannot always be expressed as an LTL formula. This is because, in an LTL synthesis setting, the system can assume that the test environment can behave in a worst-case manner and will never synthesize a satisfying controller. However, the system can assume that the test environment will always ensure that a path to achieving the system specification remains. For many examples, expressing that a satisfying path exists is not possible in LTL.

\begin{definition}[Specification Product]
\label{def:spec_prod}
The \emph{specification product} is the product $\mc{B}_{\pi} := \mc{B}_{\text{sys}} \otimes \mc{B}_{\text{test}}$, where $\mc{B}_{\text{sys}}$ is the B\"uchi automaton corresponding to the system specification, and $\mathcal{B}_{\text{test}}$ is the B\"uchi automaton corresponding to the test objective. The states \((q_{\text{sys}}, q_{\text{test}}) \in \mc{B}_{\pi}.Q\), where \(q_{\sys} \in \mc{B}_{\text{sys}}.Q\) and \(q_{\test} \in \mc{B}_{\text{test}}.Q\), capture the event-based progression of the test and are referred to as history variables.
\end{definition}

The system reaching its goal would typically mark the end of a test execution. However, the test engineer can also decide to terminate the test if the system appears to be stuck or enters an unsafe state. Tests that are terminated prematurely might result in inconclusive results~\cite{bauer2011runtime}, so we rely on the test engineer to determine the termination condition. We assume that the test engineer gives the system a reasonable amount of time to complete the test. Upon test termination in state \(s_n\), we augment the trace \(\sigma\) with the infinite suffix \(s_n^{\omega}\) for evaluation purposes. 

\begin{remark}
As tests have a defined start and end point, we need to bridge the gap between the finiteness of test executions and the infinite traces that are needed to evaluate LTL formulae. Augmenting the trace with the infinite suffix allows us to leverage useful tools available for LTL. Other research on interpreting LTL over finite traces can be found in~\cite{bloem2019synthesizing, havelund2001monitoring,morgenstern2012asymptotically}.
\end{remark}

\begin{remark}
    The states of the specification product automaton track the states of the individual B\"uchi automata, \(\mc{B}_{\text{sys}}\) and \(\mc{B}_{\text{test}}\), in the form of the Cartesian product to remember accepting states of the individual automata, which will be necessary for our framework (see Definitions~\ref{def:spec_prod},~\ref{def:S_I_T}).
\end{remark}

\begin{figure}
    \centering
    \begin{minipage}{.25\textwidth}
    \centering
\includegraphics[width=\textwidth,trim={0.0cm 0.0cm 0cm 0.0cm}]{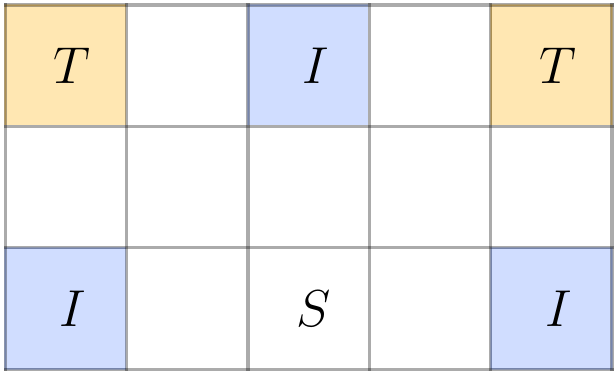}
    \subcaption{Example~\ref{ex:med_ex}.}
    \label{fig:med_empty}
    \end{minipage}
\hspace{1mm}
    \begin{minipage}{0.15\textwidth}
    \centering
\includegraphics[width=\linewidth,trim={0.0cm 0.0cm 0cm 0.0cm}]{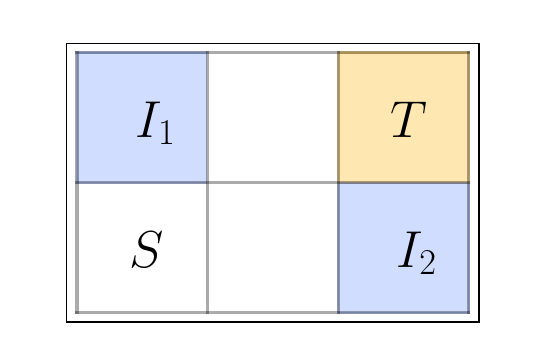}
\subcaption{Example~\ref{ex:small_reactive}.}
    \label{fig:small_reactive_empty}
\end{minipage}
\caption{Grid world layouts for examples.}
\end{figure}

\begin{example}
\label{ex:med_ex}
The system under test can transition (N-S-E-W) on the grid world as illustrated in Fig.~\ref{fig:med_empty}. The initial condition of the system is marked by \(S\), and the system is required to visit one of the terminal goal states marked by \(T\), $\varphi_{\sys}= \Feventually T$. The test objective is to observe the system visit at least one of the \(I\) states before the system reaches its goal, encoded as $\varphi_{\test}= \Feventually I$. 
\end{example}

\begin{figure}
    \centering
    \begin{minipage}{.18\textwidth}
    \centering
\includegraphics[width=0.65\linewidth]{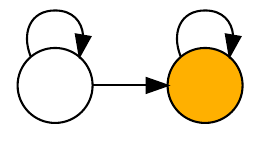}
    \subcaption{$\mc{B}_\sys$}
    \label{fig:small_reactive_Bsys}
    \includegraphics[width=\linewidth]{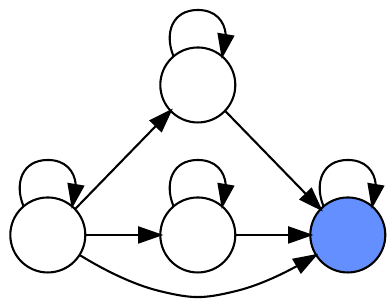}
    \subcaption{$\mc{B}_\test$}
    \label{fig:small_reactive_Btest}
    \end{minipage}
    \begin{minipage}{.28\textwidth}
    \centering
\includegraphics[width=\linewidth]{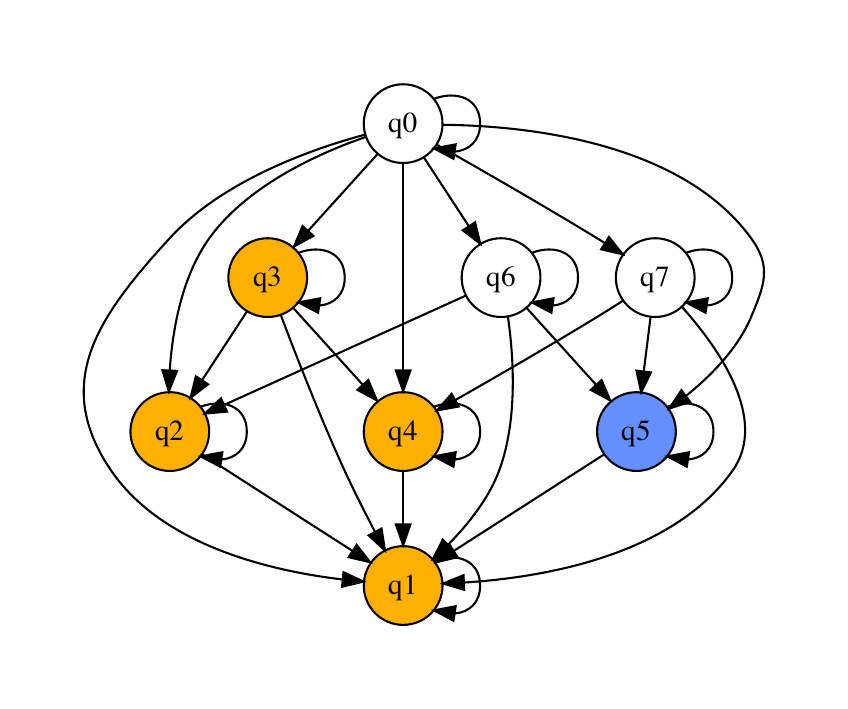}
    \subcaption{$\mc{B}_{\pi}$}
    \label{fig:small_reactive_Bpi}
    \end{minipage}
\caption{Automata for Example~\ref{ex:small_reactive}. Yellow \yellowfilledcirc~and blue \bluefilledcirc~nodes in \(\mc{B}_{\sys}\) and \(\mc{B}_{\test}\) are the respective accepting states. In the product \(\mc{B}_{\pi}\), we continue to track these states for the system and test objectives. States in the product \(\mc{B}_{\pi}\) that are accepting to both objectives (e.g., q$1$) are also shaded yellow.}
\label{fig:automata_small_reactive}
\end{figure}

\begin{example}
\label{ex:small_reactive}
In this example, the system under test can transition (N-S-E-W) on the grid world as illustrated in Fig.~\ref{fig:small_reactive_empty}. The initial condition of the system is marked by \(S\), and the system objective is to visit terminal state \(T\), $\varphi_{\sys}= \Feventually T$. The test objective is to observe the system visit states \(I_1\) and \(I_2\): $\varphi_{\test}= \Feventually I_1 \wedge \Feventually I_2$. The corresponding B\"uchi automata are illustrated in Fig.~\ref{fig:automata_small_reactive}.
\end{example}

The synchronous product operator is used to construct a product of a transition system and a B\"uchi automaton. In particular, we will use this operator to construct the virtual product graph and the system product graph (see Section~\ref{sec:problem}).
\begin{definition}[Synchronous Product]
The \emph{synchronous product} of a DBA $\mathcal{B}$ and a FTS $T_{\sys}$, where the alphabet of $\mathcal{B}$ is the labels of $T_{\sys}$, is the transition system $P \coloneqq T_{\sys}\otimes \mathcal{B}$, where:
\begin{equation*}
\begin{aligned}
& P.S  \coloneqq T_{\sys}.S\times \mc{B}.Q,\\
&P.\delta((s,q), a) \coloneqq (s',q') \text{ if }
\forall s,s' \in T_{\sys}.S, \forall q,q' \in B.Q, \\& \quad \exists a \in T_{\sys}.A, \text{ s.t. } T_{\sys}.\delta(s, a) = s' \text{ and } \mc{B}.\delta(q,T_{\sys}.L(s')) = q',\\
&P.S_0 \coloneqq \{(s_0,q)\, \vert \, s_0 \in T_{\sys}.S_0,\, \exists q_0\in \mc{B}.Q_0 \text{ s.t. } \\ & \quad  \mc{B}.\delta(q_0,T_{\sys}.L(s_0)) = q\},\\
&P.AP\coloneqq \mc{B}.Q,\\
& P.L((s,q)) \coloneqq \{q\}, \quad \forall (s,q) \in P.S.\\
\end{aligned}
\end{equation*}
We denote the transitions in $P$ as \(P.E \coloneqq \{(s,s') \,|\, s,s'\in P.S \text{ if } \exists a \in P.A \text{ s.t. } P.\delta(s,a) = s'\}\). An infinite sequence on $P$ corresponds to a state-history trace $\vartheta = (s, q)_0, (s, q)_1, \dots(s, q)_n^\omega$. We refer to \((s,q) \in P.S\) as the state-history pair and define the corresponding path to be the finite prefix: \(\vartheta_{n} = (s, q)_0, (s, q)_1,\ldots,(s, q)_n\).
\end{definition}

\subsection{Network Flows}

\begin{definition}[Flow Network~\cite{cormen2022introduction}]
 A \emph{flow network} is a tuple $\mathcal{N} = (V, E, (V_s, V_t))$, where $V$ denotes the set of nodes, $E \subseteq V \times V$ the set of edges excluding self-loops, $V_s \subseteq V$ the source nodes, and $V_t \subseteq V$ the sink nodes. We assume unit capacity for all edges. On the flow network $\mc{N}$, we can define the \emph{flow} vector $\textbf{f} \in \mathbb{R}^{\vert E \vert}_{\geq 0}$  to satisfy the following constraints: i) the capacity constraint 
 \begin{equation}
 \label{eq:flow_capacity}
0 \leq f^e \leq 1, \forall e \in E,
 \end{equation}
ii) the conservation constraint 
 \begin{equation}
 \label{eq:flow_conservation}
 \sum_{u \in V} f^{(u,v)}= \sum_{u \in V} f^{(v,u)}, \forall v \in V\setminus\{V_s, V_t\}, \text{ and}
 \end{equation}
iii) no flow into the source or out of the sink
 \begin{equation}
 \label{eq:flow_no_in_src_no_out_sink}
f^{(u,v)} = 0 \text{ if } u \in V_t \text{ or } v \in V_s.
 \end{equation}
 The flow value on the network $\mc{N}$ is defined as
 \begin{equation}
 \label{eq:total_flow}
F \coloneqq \sum_{\substack{(u,v) \in E,\\
u \in V_s}} f^{(u,v)}.
 \end{equation}
\end{definition}

\section{Problem Statement}
\label{sec:problem}
In this section, we will state the test environment synthesis problem.
The test engineer provides a system objective and a test objective, which describes the desired test behavior.
Then, we find a reactive test strategy for which every test execution that satisfies the system objective also satisfies the test objective.

\begin{definition}[Reactive Test Strategy]
A \emph{reactive test strategy} \(\pi_{\test}: (T_{\sys}.S)^* T_{\sys}.S \rightarrow 2^{A_H}\) defines the set of restricted system actions at each state during its execution \(\sigma\). For some finite prefix \(s_0\ldots s_i\) of execution \(\sigma\) starting from initial state \(s_0 \in T_{\sys}.S_0\), \(\pi_{\test}(s_0\ldots s_i) \subseteq H(s_i)\) is the set of actions that the system cannot take from state \(s_i\). A test environment is said to \emph{realize} a reactive test strategy $\pi_\test$ if it restricts system actions according to $\pi_\test$.
\end{definition}
Let \(\Sigma_{\text{fin}} := (T_{\sys}.S)^*T_{\sys}.S\) be the set of all finite prefixes of system traces. 
At each time step \(k\geq 0\), a correct system strategy \(\pi_{\sys}: \Sigma_{\text{fin}} \rightarrow T_{\sys}.A \setminus \pi_{\test}(\Sigma_{\text{fin}})\) must pick from available actions at state \(s_k\). The resulting execution is denoted as \(\sigma({\pi_{\sys} \times \pi_{\test}})\). 
\begin{remark}
\label{rem:system_resynt}
Note that the test environment externally blocks system transitions, and as a consequence, restricts corresponding actions that the system can safely take. 
When actions are restricted by the test environment, the system strategy \(\pi_{\sys}\) should select from the available actions at each state. Since these restrictions can be placed during the test execution, the system might have to re-plan and choose a different action than originally planned.
\end{remark}
\begin{definition}[Feasibility of a Test Strategy]
Given a test environment, system \(T_{\sys}\), system and test objectives, \(\varphi_{\sys}\) and \(\varphi_{\test}\), a reactive test strategy \(\pi_{\test}\) is said to be \emph{feasible} iff: i) the test environment can realize $\pi_\test$, ii) there exists a correct system strategy \(\pi_{\sys}\), and iii) any execution corresponding to a correct \(\pi_{\sys}\) satisfies the system and test objectives: \(\sigma({\pi_{\sys} \times \pi_{\test}}) \vDash \varphi_{\test} \wedge \varphi_{\sys}\).
\end{definition}

Note that the test strategy is not aiding the system in achieving the system objective; it only restricts system actions such that the test objective is realized. That is, the system is free to choose an incorrect strategy, in which case there are no guarantees. 
Furthermore, the test strategy should allow the system to make multiple decisions at each step of the execution, if possible, as opposed to leaving a single allowed action. For any system trace \(\sigma=s_0s_1\ldots\,\), every finite prefix of \(\sigma\) maps to a history variable \(q \in \mc{B}_{\pi}.Q\). For each \(\sigma\), we can define a corresponding state-history trace \(\vartheta = (s, q)_0, (s, q)_1, \ldots\), where history variable \(q\) at time step \(i\) corresponds to the prefix of \(s_0\ldots s_i\) of \(\sigma\). From now on, we will refer to \(\sigma\) and the associated \(\vartheta\) as the test execution, and clarify the context if necessary. 
\begin{definition}[Restrictiveness of a Test Strategy]
    State-history traces \(\vartheta_1\) and \(\vartheta_2\) are \emph{unique} if they do not share any consecutive state-history pairs. For a feasible \(\pi_{\test}\), let \(\Sigma\) be the set of all executions corresponding to correct system strategies, and let \(\Theta\) be the set of all state-history traces corresponding to \(\Sigma\). Let \(\Theta_{u} \subseteq \Theta\) be a set of unique state-history traces. A test strategy $\pi_{\test}$ is \emph{least restrictive} if the cardinality of \(\Theta_{u}\) is maximized. 
\end{definition}

\begin{remark}
Note that the set of all state history traces \(\Theta\) can be infinite. However, the set \(\Theta_{u}\) is finite because: i) the system has a finite number of states and the specification product has a finite number of history variables, and ii) every state-history trace in \(\Theta_{u}\) is \emph{unique} with respect to any other trace in \(\Theta_{u}\). 
\end{remark}

\begin{problem}[Finding a Reactive Test Strategy]
\label{prob:reactive_test_strategy}
Given a high-level abstraction of the system model \(T_{\sys}\), test harness \(H\), system objective \(\varphi_{\sys}\), test objective \(\varphi_{\test}\), find a feasible, reactive test strategy $\pi_{\text{test}}$ that is least restrictive.
\end{problem}
The restrictions on system actions placed by the test strategy can be realized in several ways in the test environment. For example, a dynamic test agent, together with any static obstacles, can be used to enforce the test strategy. This leads to the second problem of synthesizing a reactive strategy for a test agent to realize the test strategy. That is, at each time step of the test execution, the test environment consisting of an agent and static obstacles restricts the system actions according to $\pi_\test$.
\begin{problem}[Reactive Test Agent Strategy Synthesis]
\label{prob:match_agent}
Given a high-level abstraction of the system model \(T_{\sys}\), test harness \(H\), system objective \(\varphi_{\sys}\), test objective \(\varphi_{\test}\), and a test agent modeled by transition system \(T_{\tester}\).
Find the test agent strategy $\pi_\tester$ and the set of static obstacles \(\mathtt{Obs}\) that: i) satisfy the system's assumptions on its environment, and ii) realize a reactive test strategy $\pi_{\test}$ that is least-restrictive and feasible.
\end{problem} 

\section{Graph Construction}
\label{sec:graph_construction}
To reason about executions of the system in relation to the system and test objectives, we leverage automata theory to construct the following graphs. 

\begin{definition}[Virtual Product Graph and System Product Graph]
A \emph{virtual product graph} is the product transition system $G := T_{\sys} \otimes \mathcal{B}_{\pi}$. Similarly, the system product graph is defined as \(G_{\text{sys}} := T_{\sys}\otimes \mc{B}_{\text{sys}}\).
\end{definition}
The virtual product graph \(G\) tracks the test execution in relation to both the system and test objectives while the system product graph \(G_{\sys}\) tracks the system objective. We will find the restrictions on system actions on $G$, while $G_\sys$ represents the system's perspective concerning the system objective during the test execution. For each node \(u = (s,q) \in G.S\), we denote the corresponding state in $ s \in T_{\sys}.S$ as \(u.s := s\). Similarly, the state corresponding to $v \in G_\sys.S$ is denoted by $v.s\coloneqq s$. For practical implementation, we remove nodes on the product graphs that are not reachable from the corresponding initial states, \(G.S_0\) or \(G_{\sys}.S_0\).
\begin{definition}[Projection]
\label{def:projections}
We map states from \(G\) to \(G_{\sys}\) using the \emph{projection} $\mc{P}_{G\rightarrow G_{\text{sys}}}: G.S \rightarrow G_{\text{sys}}.S$ as 
\begin{equation}
\mc{P}_{G\rightarrow G_{\text{sys}}}(s, (q_{\sys}, q_{\test})) = (s, q_{\sys}).
\end{equation}
\end{definition}
These projections help us to reason about how restrictions found on \(G\) map to the system \(T_{\sys}\) and the system product graph \(G_{\sys}\). We can now define the edges on $G$ that we can restrict with the test harness as follows,
\begin{equation}
\begin{split}
    E_H =& \{((s,q),(s',q')) \in G.E \vert \; \forall s \in T_{\sys}.S, \\& \forall a \in H(s)  \text{ s.t. }s' = T_{\sys}.\delta(s,a)\}.
\end{split}
\end{equation}

\begin{lemma}
For every path \((s, q_{\sys})_0, (s, q_{\sys})_1,\ldots,(s, q_{\sys})_n\) on \(G_{\sys}\), there exists at least one corresponding path on \(G\). 
    \label{lem:path_existence}
\end{lemma}
\begin{proof}
Suppose there exists some \(q_{\test\:0}, \ldots, q_{\test\:n} \in \mc{B}_{\test}.Q\) such that \((s, (q_{\sys}, q_{\test}))_0,\ldots, (s,(q_{\sys}, q_{\test}))_n\) is a path on \(G\). Then, by construction, there exists a path on \(G_{\sys}\) where \((s,(q_{\sys}, q_{\test}))_k\) maps to \((s,q_{\sys})_k\) for all \(0\leq k \leq n\).
\end{proof}

Paths on the virtual product graph $G$ correspond to possible test executions. We identify the nodes on $G$ that capture the acceptance conditions for the system and test objectives.
\begin{definition}[Source, Intermediate, and Target Nodes] 
\label{def:S_I_T}
The \emph{source node} \(\mathtt{S}\) represents the initial condition of the system.
The \emph{intermediate nodes} \(\mathtt{I}\) correspond to system states in which the test objective acceptance conditions are met. Finally, the \emph{target nodes} \(\mathtt{T}\) represent the system states for which the acceptance condition for the system objective is satisfied. Formally, these nodes are defined as follows,
\begin{equation*}
    \begin{aligned}
        &\mathtt{S} \coloneqq \{(s_0, q_0) \in G.S \,\vert \, s_0 \in T_{\sys}.S_0, q_0 \in \mc{B}_{\pi}.Q_0\}, \\
        &\mathtt{I} \coloneqq  \{(s, (q_{\text{sys}}, q_{\text{test}})) \in G.S \,\vert \, q_{\text{test}}\in \mc{B}_{\text{test}}.F,\, q_{\text{sys}}\notin \mc{B}_{\text{sys}}.F\}, \\
       &\mathtt{T} \coloneqq \{(s, (q_{\text{sys}}, q_{\text{test}})) \in G.S \,\vert \, q_{\text{sys}}\in \mc{B}_{\text{sys}}.F\}.\\
    \end{aligned}
\end{equation*}
\end{definition}
In addition, we define the set of states corresponding to the system acceptance condition on $G_{\sys}$ as $\mathtt{T}_{\sys}:= \{(s,q) \in G_{\sys}.S \: \vert \: q \in \mathcal{B}_{\sys}.F \}$.

\begin{proposition}
\label{prop:correct_system_tester_path}
Every test execution corresponds to a path \(\vartheta_{n} = (s,q)_0, (s,q)_1,\ldots,(s, q)_n\) on \(G\) where \((s,q)_0 \in \src\). The corresponding system trace $\sigma_n$ satisfies the system objective, $\sigma \models \varphi_\sys$ iff \((s,q)_n \in \sink\). Furthermore, if \(\sigma \models \varphi_{\test}\), then the path \(\vartheta_{n}\) contains a state-history pair \((s, q)_i \in \inter\) for some \(0 \leq i \leq n\). 
\end{proposition}

Provided that there exists a path on \(G\) from \(\src\) to \(\sink\), identifying a feasible reactive test strategy corresponds to identifying edges to cut on \(G\). These edge cuts correspond to restricted system actions. In particular, these edge cuts are such that all paths on $G$ from source \(\src\) to target \(\sink\) visit the intermediate \(\inter\). 

\section{Network Flow Optimization for Identifying Restrictions on System Actions}
\label{sec:optimization}

\begin{figure*}
\centering
 \vspace{2mm}
\begin{minipage}{.5\linewidth}
    \centering
\includegraphics[width=\linewidth,trim={0.0cm 0.0cm 0cm 0.0cm}]{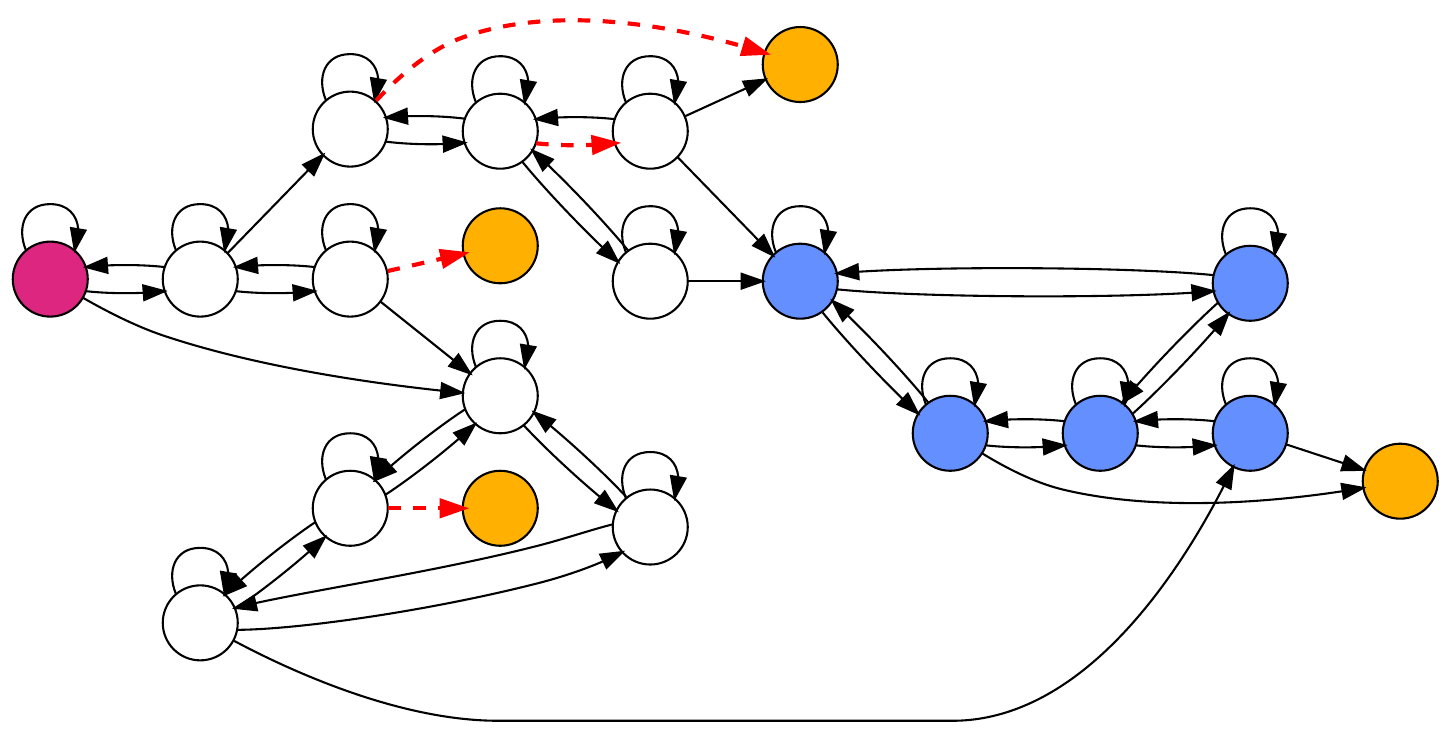}
\subcaption{\label{fig:small_reactive_virtual} Virtual product graph $G$.}
  \end{minipage}
  \hspace{10mm}
  \begin{minipage}{.1\textwidth}
    \centering
\includegraphics[width=\linewidth,trim={3.0cm 0.0cm 0cm 0.0cm}]{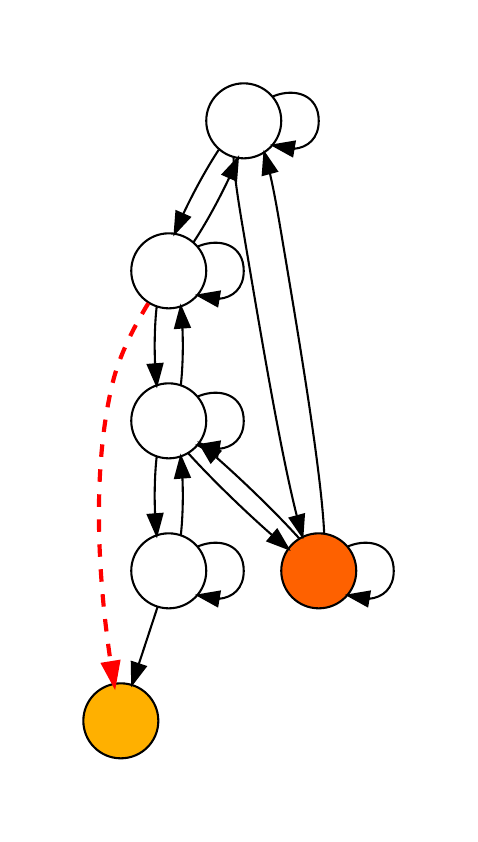}
    \subcaption{\label{fig:small_reactive_q0} $G_{\sys}^{(\text{q}0,s_3)}$}
  \end{minipage}
  \hspace{2mm}
  \begin{minipage}{.1\textwidth}
    \centering
\includegraphics[width=\linewidth,trim={3.0cm 0.0cm 0cm 0.0cm}]{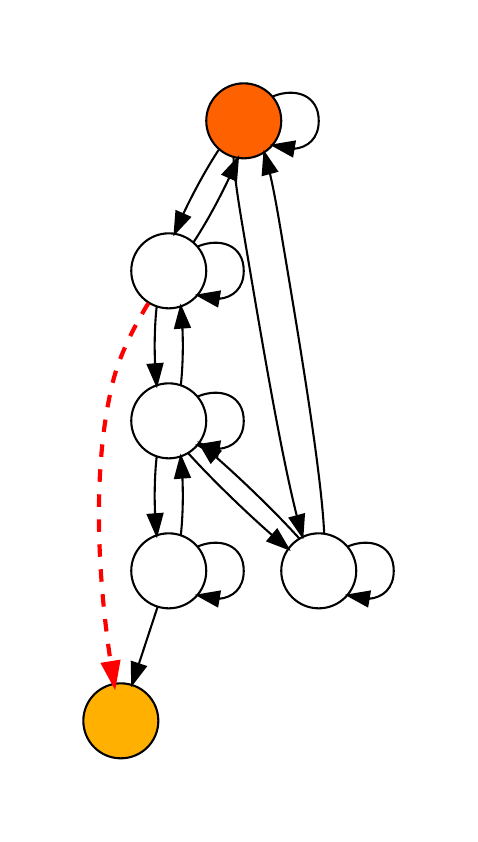}
    \subcaption{\label{fig:small_reactive_q6} $G_{\sys}^{(\text{q}6,s_1)}$}
  \end{minipage}
\hspace{2mm}
  \begin{minipage}{.1\textwidth}
    \centering
\includegraphics[width=\linewidth,trim={3.0cm 0.0cm 0cm 0.0cm}]{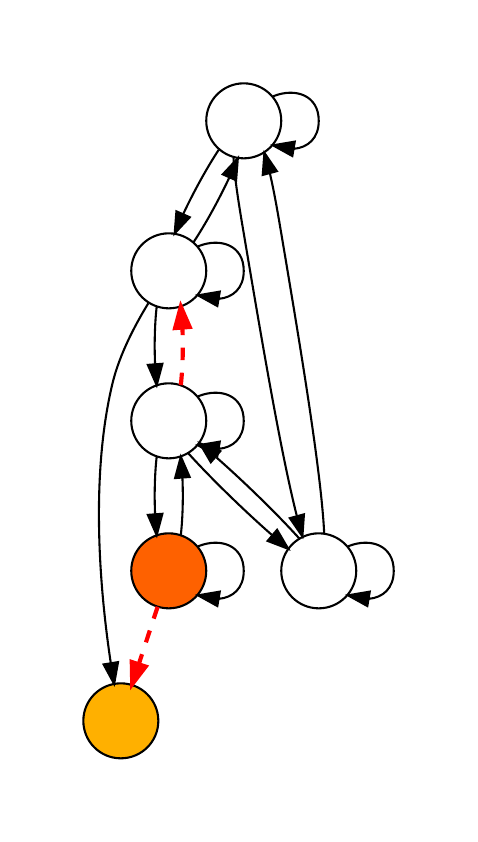}\subcaption{\label{fig:small_reactive_q7} $G_{\sys}^{(\text{q}7,s_{11})}$}
  \end{minipage}
\caption{Virtual product graph and system product graphs for Example~\ref{ex:small_reactive}. Fig.~\ref{fig:small_reactive_virtual} shows the virtual product graph $G$, with the source $\src$ (magenta \pinkfilledcirc), the intermediate nodes $\inter$ (blue \bluefilledcirc), and the target nodes (yellow \yellowfilledcirc). Edge cut values for each edge in $G$ are grouped by their history variable $q$ and projected to the corresponding copy of $G_\sys$. Red dashed lines indicate edge cuts. Figs.~\ref{fig:small_reactive_q0}-\ref{fig:small_reactive_q7} show the copies of $G_\sys$ with their source (\(s_3\), \(s_6\) or \(s_{11}\) in orange \orangefilledcirc) and target nodes (yellow \yellowfilledcirc). The graphs in Figs.~\ref{fig:small_reactive_q0}-\ref{fig:small_reactive_q7} correspond to the history variables q$0$, q$6$, and q$7$ from $\mc{B}_\pi$ shown in Fig.~\ref{fig:small_reactive_Bpi}. The constraints~\eqref{eq:flow_constraints_on_G_sys}-\eqref{eq:feasibility_flow_on_Gsys} ensure that the edge cuts are such that a path from each source to the target node exists for each history variable $q$.}
\vspace{-4mm}
\label{fig:small_reactive_example}
\end{figure*}

To identify which edges to cut on \(G\), we use network flow optimization, a commonly used paradigm for flow-cut problems on graphs. On $G$, which characterizes all possible test executions, all paths from the initial condition $\src$ to the system goal $\sink$ must be routed through the intermediate \(\inter\). Furthermore, the edge cuts should be least-restrictive and such that the system can satisfy the test objective and system objective. Maximum flow can be a proxy for freedom of the system under test to make decisions --- a higher network flow corresponds to more unique paths on \(G\). Since we use flow networks with unit edge capacities, a realization of maximum flow corresponds to a set of paths that do not share an edge. Furthermore, this flow should be achieved with the fewest possible cuts to not unnecessarily restrict system actions. A high network flow with the minimum possible edge cuts corresponds to a least restrictive test for the system.

\subsection{Optimization Setup}
We define the flow network \(\mc{G}\coloneqq (V,E, (\src, \sink))\), where $V \coloneqq G.S$, $E \coloneqq G.E$, source and target nodes correspond to \(\src\) and \(\sink\), with the corresponding flow $\fvec \in \mathbb{R}^{|E|}$. For simplicity, we use the same notation to refer to nodes and edges on the graph and the corresponding flow network. The Boolean edge cut vector \(\dvec \in \mathbb{B}^{\vert E \vert}\) represents whether edges are cut or not. That is, \(d^e = 1\) refers to edge \(e \in E\) being cut, and \(d^e = 0\) implies that edge \(e\) is not cut,
\begin{equation}
\label{eq:binary_cuts}
\tag{c1}
      d^e \in \{0,1\}, \quad \forall e\in E, \text{ and } d^e = 0, \quad \forall  e \notin E_H.
\end{equation}
The edges into and out of the intermediate \(\inter\) nodes are denoted as $E(\inter) := \{(u,v) \in E\: \vert\: u \in \inter \text{ or } v \in \inter\}$. To solve Problem~\ref{prob:reactive_test_strategy}, we formulate a mixed-integer linear program (MILP).
\smallbreak
\noindent
\textbf{Objective.} To find the least restrictive test, we want to maximize the system's freedom in satisfying the test objective. To capture this, we optimize for edge cuts that maximize the flow value on \(\mc{G}\).
However, a realization of maximum flow on a network is not unique. To ensure that we do not cut any edges unnecessarily, we subtract the sum of the edge cuts from the flow value:
\begin{equation}
\label{eq:objective}
    \sum_{\substack{(u,v) \in E,\\
    u \in \src}} f^{(u,v)} - \frac{1}{\vert E \vert} \sum_{e \in E} d^e.
\end{equation}
The regularizer \(\frac{1}{\vert E \vert}\) on the sum of edge cuts is chosen such that it will not compete with the maximum flow value on the network. The weighted sum \(\frac{1}{\vert E \vert} \sum_{e \in E} d^e\) is always between \(0\) and \(1\), and binary edge cuts and unit capacity will always result in maximum flow being integer-valued. Thus, the optimization will always favor increasing the maximum flow value rather than reducing edge cuts.
\smallbreak
\noindent
\textbf{Network flow constraints. } 
First, the network flow optimization is subject to the following standard constraints on flow \(\fvec\):
\begin{equation}
\label{eq:flow_constraints_on_G}
\tag{c2}
    \text{Flow constraints~\eqref{eq:flow_capacity}, ~\eqref{eq:flow_conservation}, and~\eqref{eq:flow_no_in_src_no_out_sink} on flow network } \mc{G}.
\end{equation}
An edge that is cut restricts flow completely, while an edge that is not cut may or may not have flow,
\begin{equation}
\label{eq:cut_const}
\tag{c3}
    \forall e\in E,   \quad  d^e + f^e \leq 1.
\end{equation}
\smallbreak
\noindent
\textbf{Partition constraints. } 
The following constraints ensure that all flow across the network will be routed through $\inter$. To accomplish this, we adapt the partitioning conditions given in~\cite{vazirani2001approximation} as follows. Except for the $\inter$ nodes, we divide the remaining nodes into two groups defined by the partition variable $\bm{\mu} \in \mathbb{R}^{\vert V \setminus \inter \vert}$, and ensure that the nodes $\src$ belong to one group, and $\sink$ belong to the other:
\begin{equation}
\label{eq:mu_partition}
\tag{c4}
     0 \leq \mu^v\leq1, \quad \mu^{\src} - \mu^{\sink} \geq 1, \forall v \in V \setminus \inter.
\end{equation}
The two groups are partitioned by the edge cut vector $\dvec$, where this constraint is only defined over the edges that do not go into or out of nodes in $\inter$,
\begin{equation}
\label{eq:cuts_partition}
\tag{c5}
    d^{(u,v)} - \mu^u + \mu^v \geq 0,\, \forall (u,v) \in E\setminus E(I).
\end{equation}

\smallbreak
\noindent
\textbf{Feasibility constraints. }To ensure that the test is not impossible from the system's perspective, we map restrictions found on \(G\) to \(G_{\sys}\) via the following feasibility constraints. For each history variable $q \in \mathcal{B}_{\pi}.Q$, we define the set of state-history pairs that captures the possible first observations of the history variable in a test execution via the function \(\mathtt{S}_G: \mathcal{B}_{\pi}.Q \rightarrow G.S\) defined as follows,
\begin{equation}
\begin{split}
\label{eq:first_occurence_of_q_on_G}
\mathtt{S}_G(q) :=& \{(s,q) \in G.S\: \vert \: \\&\forall ((\bar{s}, \bar{q}), (s,q)) \in G.E,\, \bar{q} \neq q\}.
\end{split}
\end{equation}
These sets of states are mapped to \(G_{\sys}\) as follows: 
\begin{equation}
\begin{split}
\mathtt{S}_{G_{\sys}}(q) := & \{u \in G_\sys.S \: \vert \:u= \mc{P}_{G \rightarrow G_{\text{sys}}}(v),\\& \: v \in \src_G(q), \text{ and } \exists \:\text{path}(u, \sink_{\sys})\},
\end{split}
\end{equation}
where this set is empty if no path from the node $u$ to $\sink_{\sys}$ exists on $G_{\sys}$.
For each \(q \in \mathcal{B}_{\pi}.Q\), for each source in $\mathtt{s} \in \src_{G_\sys}(q)$, we define a flow network $\mc{G}_{\text{sys}}^{(q,\mathtt{s})} \coloneqq (V_{\text{sys}}, E_{\text{sys}}, c, (\mathtt{s}, \mathtt{T}_{\sys}))$, where $V_{\text{sys}} \coloneqq G_{\sys}.S$, and $E_{\text{sys}} \coloneqq G_{\sys}.E$, with the corresponding flow variable $\fvec_\sys^{(q,\mathtt{s})}$. 
For each of these flow networks, we define a flow subject to the standard flow constraints:
\begin{equation}
\label{eq:flow_constraints_on_G_sys}
\tag{c6}
\begin{split}
&\forall q \in \mc{B}_{\pi}.Q, \forall \mathtt{s} \in \src_{G_{\sys}}(q), \\
&\text{Flow constraints~\eqref{eq:flow_capacity},~\eqref{eq:flow_conservation}, and~\eqref{eq:flow_no_in_src_no_out_sink} on network } \mc{G}_{\sys}^{(q,\mathtt{s})}.
\end{split}
\end{equation}
For each $\mc{G}_{\text{sys}}^{(q,\mathtt{s})}$, we map the edge cuts \(\dvec\) and check that there is still a path from \(\mathtt{s}\) to some node in \(\mathtt{T}_{\sys}\). This ensures that reactively placing restrictions on system actions does not make it impossible for a correct system strategy to make progress toward its goal. Intuitively, the edge cuts are grouped by the history variable $q$ and checked to ensure that the system has a feasible path when these restrictions are placed on system actions. 
The edges are grouped by their history variable using the mapping $\mathtt{Gr}: \mc{B}_{\pi}.Q \rightarrow 2^{G.E}$, defined as follows:
\begin{equation}
\label{eq:grouping_by_q}
\mathtt{Gr}(q) \coloneqq \{((s,q), (s',q')) \in G.E\}.
\end{equation}
The edge cuts are mapped onto the corresponding $\mc{G}_{\text{sys}}^{(q,\mathtt{s})}$ to cut the corresponding flow $\fvec_{\text{sys}}^{(q,\mathtt{s})}$ as follows:

\begin{equation}
\label{eq:match_cuts_to_Gsys}
\tag{c7} 
\begin{split}
\forall q \in \mc{B}_{\pi}.Q, \forall \mathtt{s} \in \src_{G_{\sys}}(q), \forall (u,v) \in \mathtt{Gr}(q), \forall (u',v') \in E_{\sys},\\
d^{(u,v)} + {f^{(q,\mathtt{s})}_{\text{sys}}}^{(u',v')} \leq 1,\, \text{ if } u'.s = u.s\text{ and } v'.s = v.s.
\end{split}
\end{equation}
Since we are agnostic to the system controller, we need to ensure that a path to the system's goal exists at all times during the test execution. To enforce this, we require a flow of at least 1 on each system flow network $\mc{G}_{\text{sys}}^{(q,\mathtt{s})}$,
\begin{equation}
\label{eq:feasibility_flow_on_Gsys}
\tag{c8}
    \sum_{(\mathtt{s},v) \in E_{\text{sys}}} {f^{(q, \mathtt{s})}_{\text{sys}}}^{(\mathtt{s},v)} \geq 1, \: \forall q \in \mathcal{B}_{\pi}.Q, \, \forall \mathtt{s} \in \src_{G_{\sys}}(q).
\end{equation}
These feasibility cuts correspond to the reactive constraint setting since edge cuts are placed on \(\mc{G}\) and depend on the history variable \(q\). For an illustrated explanation for Example~\ref{ex:small_reactive}, refer to Fig.~\ref{fig:small_reactive_example}. Finally, the optimization to identify edge cuts for the reactive test strategy is characterized by the following mixed-integer linear program (MILP) with the cuts \(\dvec\) as the integer variables, and the flow and partition variables taking continuous values. 

\smallbreak
\noindent
\rule{\linewidth}{0.6pt}
\textbf{\textproc{MILP-reactive}:}
\begin{equation}
\label{milp:reactive}
\begin{aligned}
    \max_{\substack{\fvec, \dvec, \bm{\mu}, \\\fvec_{\sys}^{(q,\mathtt{s})} \,\forall q \in \mc{B}_{\pi}.Q \, \forall \mathtt{s} \in \src_{\mc{G}_{\sys}}(q)}} F - \frac{1}{\vert E \vert} \sum_{e \in E}d^e \\
   \text{s.t.} \quad \text{\eqref{eq:binary_cuts}-\eqref{eq:cut_const}}, 
   \text{\eqref{eq:mu_partition}-\eqref{eq:cuts_partition}}, \text{\eqref{eq:flow_constraints_on_G_sys}-\eqref{eq:feasibility_flow_on_Gsys}}.\\
   \end{aligned}
\end{equation}
\rule{\linewidth}{0.6pt}
\smallbreak
\noindent
\textbf{Static Constraints. }We can simplify the feasibility constraints in the case of static obstacles. This corresponds to the requirement that any transition that is restricted will remain restricted for the entire duration of the test. From the system's perspective, the restrictions will not change depending on the history variable $q$. That is, edges in $G$ corresponding to the same transition in $T_{\sys}.E$ are grouped and share the same cut value:
\begin{equation}
\label{eq:static_map_cuts_in_G}
    \tag{c9}
\begin{split}
    d^{(u,v)} = d^{(u',v')},\: \forall (u,v), (u',v') \in E, \\
    \text{if } u.s = u'.s \text{ and }  v.s = v'.s.
\end{split}
\end{equation}
\smallbreak
\noindent
Similarly, the optimization to find edge cuts in a static setting is as follows.\\
\rule{\linewidth}{0.6pt}
\textbf{\textproc{MILP-static}:}
\begin{equation}
\label{milp:static}
\begin{aligned}
    \max_{\fvec, \dvec, \bm{\mu}} F - \frac{1}{\vert E \vert} \sum_{e \in E}d^e \\
   \text{s.t.} \quad \text{\eqref{eq:binary_cuts}-\eqref{eq:cut_const}}, 
   \text{\eqref{eq:mu_partition}-\eqref{eq:cuts_partition}}, \eqref{eq:static_map_cuts_in_G}.\\
   \end{aligned}
\end{equation}
\rule{\linewidth}{0.6pt}
\begin{lemma}
\label{lem:equivalence_static}
For the case of static constraints, due to~\eqref{eq:static_map_cuts_in_G}, ensuring feasibility from the system's perspective is guaranteed by checking $F > 0$ on $G$.
That is, $F> 0$ on $G$ is equivalent to checking~\eqref{eq:flow_constraints_on_G_sys}-\eqref{eq:feasibility_flow_on_Gsys}.
\end{lemma}
\begin{proof}
Under~\eqref{eq:static_map_cuts_in_G}, the edge groupings $\mathtt{Gr}(q)$ become the same for all $q \in \mc{B}_\pi.Q$. Thus, the constraints~\eqref{eq:flow_constraints_on_G_sys}-\eqref{eq:feasibility_flow_on_Gsys} can be reduced onto a single flow network $\mc{G}_{\text{sys}} = (V_{\text{sys}}, E_{\text{sys}}, (\mathtt{S}_{\sys}, \mathtt{T}_{\sys}))$, where $\mathtt{S}_{\sys} := G_{\text{sys}}.I$.
Equation~\eqref{eq:feasibility_flow_on_Gsys} being satisfied on  \(\mc{G}_{\sys}\) implies that there is a path on \(G\) from \(\src\) to \(\sink\) via Lemma~\ref{lem:path_existence}. 
Additionally, if there is a path on \(G\) from \(\src\) to \(\sink\) with the static constraints~\eqref{eq:static_map_cuts_in_G}, then it must be that there exists a path from \(\src_{\sys}\) to \(\sink_{\sys}\) on \(G_{\sys}\).
\end{proof}

\begin{remark}
\label{remark:reactive_grouping_on_G}
For the reactive constraint setting, we can replace the feasibility constraints~\eqref{eq:flow_constraints_on_G_sys}-\eqref{eq:feasibility_flow_on_Gsys} by several static constraints. That is, we  
introduce a copy of $\mc{G}$ for each history variable $q \in \mc{B}_\pi.Q$ and each source $\mathtt{s} \in \src_G(q)$, denoted $\mc{G}^{(q, \mathtt{s})} = (V,E,\mathtt{s}, \sink)$, and require a path from $\mathtt{s}$ to $T$ to exist under a static mapping of the edges in the group $\mathtt{Gr}(q)$ by constraint~\eqref{eq:static_map_cuts_in_G}. We choose the former since it reduces the number of variables and constraints in the optimization.
\end{remark}
\smallbreak
\noindent
\textbf{Mixed Constraints. } In some cases, it might be desirable to define specific transitions $T_{\sys}.E_{\text{static}} \subseteq T_{\sys}.E$ which require static constraints. The mixed setting of reactive and static transition restrictions can be implemented by enforcing the feasibility constraints~\eqref{eq:flow_constraints_on_G_sys}-\eqref{eq:feasibility_flow_on_Gsys}, and the static constraints~\eqref{eq:static_map_cuts_in_G} on edges $(u,v) \in E$, where the corresponding transition $(u.s, v.s) \in T_{\sys}.E_{\text{static}}$. Finally, the optimization for the mixed constraint setting is as follows.
\smallbreak
\noindent
\rule{\linewidth}{0.6pt}
\textbf{\textproc{MILP-mixed}:}
\begin{equation}
\label{milp:mixed}
\begin{aligned}
    \max_{\substack{\fvec, \dvec, \bm{\mu}, \\\fvec_{\sys}^{(q,\mathtt{s})} \,\forall q \in \mc{B}_{\pi}.Q \, \forall \mathtt{s} \in \src_{\mc{G}_{\sys}}(q)}} F - \frac{1}{\vert E \vert} \sum_{e \in E}d^e \\
   \text{s.t.} \quad \text{\eqref{eq:binary_cuts}-\eqref{eq:cut_const}}, 
   \text{\eqref{eq:mu_partition}-\eqref{eq:cuts_partition}}, 
   \text{\eqref{eq:flow_constraints_on_G_sys}-\eqref{eq:feasibility_flow_on_Gsys}},
   \text{\eqref{eq:static_map_cuts_in_G}}.\\
   \end{aligned}
\end{equation}
\rule{\linewidth}{0.6pt}
\smallbreak
\noindent
\textbf{Auxiliary Constraints. } Additional constraints can be added to the optimization depending on the test harness or the desired test setup. For example, it might be required to enforce that if an edge is cut, the transition will be blocked in both directions. This can be enforced as follows,
\begin{equation}
\label{eq:bidirectional_cuts}
    \tag{c14}
\begin{split}
    d^{(u,v)} = d^{(u',v')}, \: \forall (u,v),\, (u',v') \in E,\\
    \text{if }u.s =v'.s \text{ and }  v.s = u'.s.
\end{split}
\end{equation}

\floatname{algorithm}{Algorithm}
\renewcommand{\algorithmicrequire}{\hspace*{\algorithmicindent}\quad\textbf{Input:}}
\renewcommand{\algorithmicensure}{\hspace*{\algorithmicindent}\quad\textbf{Output:}}
\begin{algorithm}
\caption{Finding the test strategy $\pi_\test$}\label{alg:find_test_strategy} 
\begin{algorithmic}[1] 
\Procedure{FindTestStrategy}{$T_{\sys},H, \varphi_{\text{sys}}, \varphi_{\text{test}}$}
\Require transition system $T_{\sys}$, test harness $H$, system objective $\varphi_{\text{sys}}$, test objective $\varphi_{\text{test}}$
\Ensure test strategy $\pi_\test$ 
\State $\mc{B}_{\text{sys}} \leftarrow \text{BA}(\varphi_{\text{sys}})$ \Comment{\small{System B\"uchi automaton}}\normalsize
\State $\mc{B}_{\text{test}} \leftarrow \text{BA}(\varphi_{\text{test}})$  \Comment{\small Tester B\"uchi automaton\normalsize}
\State \(\mc{B}_{\pi} \leftarrow \mc{B}_{\text{sys}} \otimes \mc{B}_{\text{test}}\) \Comment{\small Specification product\normalsize}
\State \(G_{\text{sys}} \leftarrow  T_{\sys} \otimes \mc{B}_{\text{sys}} \) \Comment{\small System product\normalsize}
\State \(G \leftarrow T_{\sys} \otimes \mc{B}_{\pi}\) \Comment{\small Virtual Product Graph\normalsize}
\State \(\mathtt{S}\), \(\mathtt{I}\), \(\mathtt{T} \leftarrow \) \textproc{IdentifyNodes}\((G, \mc{B}_\sys, \mc{B}_\test)\)
\State $\mc{G} \leftarrow$ \textproc{DefineNetwork} $(G,\src, \sink )$
\State $\mathfrak{G} \leftarrow \mathtt{set()}$ \Comment{\small System Perspective Graphs \normalsize}
\For{$q \in \mc{B}_\pi.Q$}
\For{$\mathtt{s} \in \src_{G_\sys}(q)$}
\State $\mc{G}_\sys^{(\mathtt{s},q)} \leftarrow $ \textproc{DefineNetwork}$(G_{\sys},\mathtt{s}, \sink_{\sys})$
\State $\mathfrak{G} \leftarrow \mathfrak{G} \cup \mc{G}_\sys^{(\mathtt{s},q)}$
\EndFor
\EndFor
\State \(\dvec^*\leftarrow\) \textproc{MILP}$(\mc{G}, T_{\sys}, \mathfrak{G}, \inter, H)$
\Comment{\small{Reactive, static, or mixed.}\normalsize}
\State \(C \leftarrow \{(u,v) \in G.E\, \vert\, {\dvec^*}^{(u,v)} = 1\}\) \Comment{\small Cuts on $G$ \normalsize}
\State $\pi_{\test} \leftarrow $ \small{ Define test strategy according to equation~\eqref{eq:reactive_test_strategy}}\normalsize
\State \textbf{return} \(\pi_{\test}\)
\EndProcedure
\end{algorithmic}
\label{alg:find_virtual_game_graph}
\end{algorithm}
\subsection{Characterizing Optimization Results}
\label{sec:characterizing_opt_results}
The flow value (Eq.~\eqref{eq:total_flow}) of the network is always integer-valued since the edge cuts are binary and edges have unit capacities, and therefore, any strictly positive flow value corresponds to at least one valid test execution. In the following cases, the problem data are \emph{inconsistent} and a flow value $\geq 1$ cannot be found.\\
\textbf{Case 1:} There is no path from \(\src\) to \(\sink\) on \(G\) (and equivalently, no path from \(\src_{\sys}\) to \(\sink_{\sys}\) on \(G_{\sys}\)). In this case, the optimization will not have to place any cuts because the only possible maximum flow value is \(0\).\\
\textbf{Case 2:} There is a path from \(\src\) to \(\sink\) on \(G\), but there is no path \(\src\) to \(\sink\) in \(G\) visiting an intermediate node in \(\inter\). In this case, the partition constraints will cut all paths from \(\src\) to \(\sink\), while by Lemma~\ref{lem:path_existence} the feasibility constraints require a path to exist from  \(\src\) to \(\sink\)---a contradiction. The routing optimization is infeasible in this instance.

For each MILP, the set of edges that are cut are found from the optimal \(\dvec^*\) as follows, \(C \coloneqq \{(u,v) \in E\setminus E(\inter)\, \vert\, {d^*}^{(u,v)} = 1\}\), resulting in the cut network \(\mc{G}_{\text{cut}} = (V, E\setminus C, \src, \sink)\). The bypass flow value is computed on the network \(\mc{G}_{\text{byp}} \coloneqq (V_{\text{byp}}, E_{\text{byp}}, \src, \sink)\), where $V_{\text{byp}}\coloneqq V\setminus \inter$, and $E_{\text{byp}}\coloneqq E\setminus \big(E(\inter) \cup C\big)$. A strictly positive bypass flow value indicates the existence of a \(Path(\src, \sink)\) on \(\mc{G}_{\text{cut}}\) that does not visit an intermediate node in \(\inter\). 

\begin{theorem}
\label{thm:partition}
For each MILP, the optimal cuts \(C\) result in a bypass flow value of \(0\).
\end{theorem}
\begin{proof}
 The partition constraints~\eqref{eq:mu_partition} and~\eqref{eq:cuts_partition} partition the set of vertices \(V\setminus \inter\) into two groups: nodes with potential \(\mu = 0\) (e.g., \(\sink\)) and nodes with potential \(\mu = 1\) (e.g., \(\src\)). On any path \(v_0\ldots v_k\) on \(\mc{G}_{\text{byp}}\), where $v_0 = \src$ and $v_k = \sink$, the difference in potential values can be expressed as a telescoping sum: \(\sum_{i=0}^{k-1} (\mu^i - \mu^{i+1}) = \mu^{\src} - \mu^{\sink}\). Then, by partition constraints~\eqref{eq:mu_partition} and~\eqref{eq:cuts_partition},
\begin{equation*}
    \sum_{i=0}^{k-1} d^{(v_i, v_{i+1})} \geq \sum_{i=0}^{k-1} (\mu^i - \mu^{i+1}) = \mu^{\src} - \mu^{\sink} \geq 1.
 \end{equation*}
Therefore, for at least one edge $(v_i,v_{i+1})$ on the path, where $0 \leq i \leq k-1$, the corresponding cut value is $d^{(v_i,v_{i+1})}=1$. These edges belong to the set of cut edges \(C\). Thus, the flow value on \(\mc{G}_{\text{byp}}\) is zero.
\end{proof}

\begin{theorem}
\label{thm:feasibility}
For each MILP, the optimal cuts \(C\) are such that there always exists a path to the goal from the system's perspective.
\end{theorem}
\begin{proof}
First, consider the MILP in the reactive setting. The optimal cuts \(C\) satisfy the feasibility constraints~\eqref{eq:flow_constraints_on_G_sys},~\eqref{eq:match_cuts_to_Gsys}, and~\eqref{eq:feasibility_flow_on_Gsys}. These constraints ensure that for each history variable \(q \in \mc{B}_\pi.Q\), there exists a path for the system from each state \(\mathtt{s} \in \src_{G_{\sys}}(q)\) to \(\sink_{\sys}\) on $G_{\sys}$. 
The edge cuts \(C\) are grouped by their history variable (see equation~\eqref{eq:grouping_by_q}) and mapped to the corresponding $\mc{G}_{\sys}^{(q,\mathtt{s})}$ (see equation~\eqref{eq:match_cuts_to_Gsys}). Then, each copy $\mc{G}_{\sys}^{(q,\mathtt{s})}$ represents all the cuts that can be simultaneously applied when the state of the test execution is at history variable $q$. Thus, all restrictions on system actions at history $q$ are captured by the cuts on $\mc{G}_{\sys}^(q,\mathtt{s})$. Since this is true for every \(q\) and every source state \(\mathtt{s}\) at which the test execution enters into \(q\), there always exists a path to the goal by equation~\eqref{eq:feasibility_flow_on_Gsys}.
The proof for the static and mixed settings follows similarly. 
\end{proof}

\begin{lemma}
\label{lem:least_restrictive}
For each MILP, the optimal cuts \(C\) correspond to maximizing the cardinality of \(\Theta_u\).
\end{lemma}
\begin{proof}
By construction, a realization of the flow \(\fvec\) on \(\mc{G}\) corresponds to a set of unique state-history traces \(\Theta_u\). The MILP objective maximizes the flow, and therefore the cardinality of \(\Theta_u\) is maximized.
\end{proof}

\begin{figure*}
\centering
 \vspace{2mm}
\begin{minipage}{.32\textwidth}
  \begin{minipage}{\textwidth}
    \centering
\includegraphics[width=\linewidth,trim={0.0cm 0.0cm 0cm 0.0cm}]{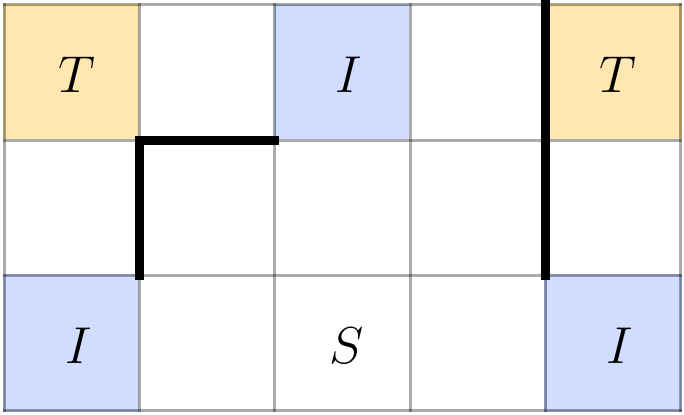}
    \vspace{0.05mm}
    \subcaption{\label{fig:med_result} Static obstacles in black.}
  \end{minipage}
  \end{minipage}
  \hspace{3mm}
  \begin{minipage}{.6\textwidth}
    \centering
\includegraphics[width=\linewidth,trim={0.0cm 0.0cm 0cm 0.0cm}]{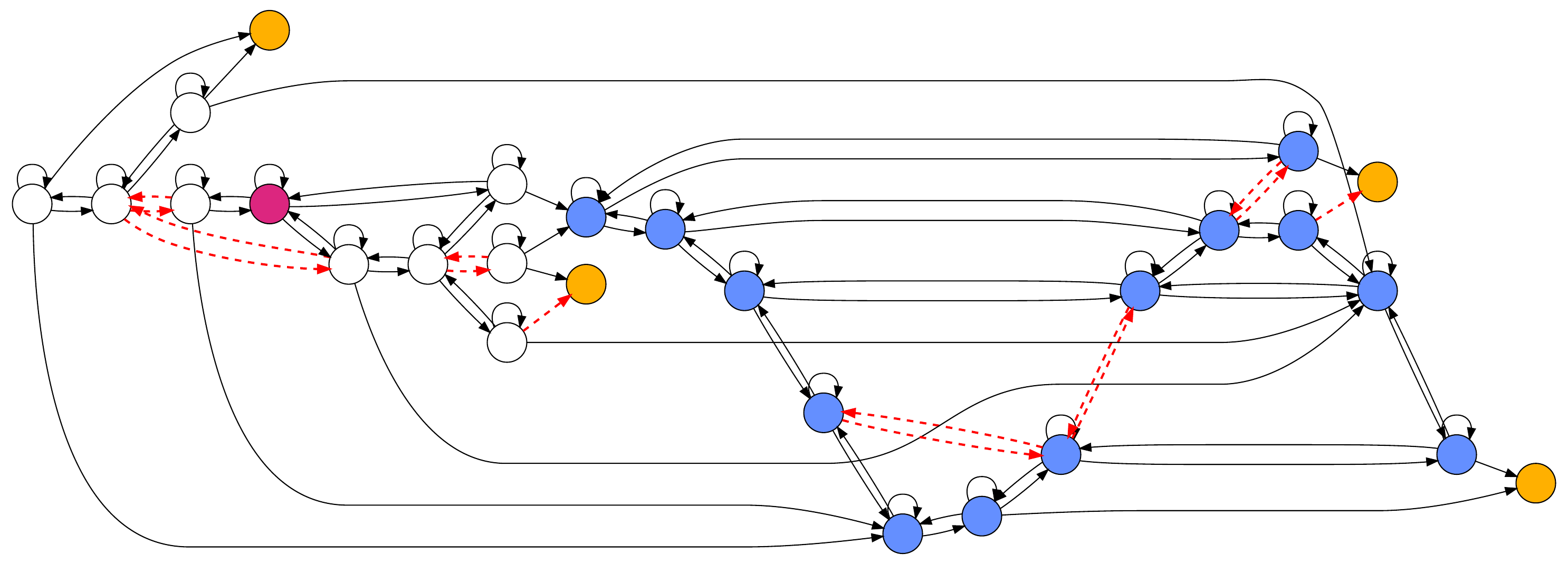}
    \subcaption{\label{fig:med_virtual} Virtual product graph with edge cuts in dashed red.}
  \end{minipage}
\caption{Static obstacles in (a) corresponding to edge cuts found on the virtual product graph (b) for Example~\ref{ex:med_ex}. States marked \(S\), \(I\), and \(T\) illustrated in (a) correspond to states \(\src\) (magenta \pinkfilledcirc), \(\inter\) (blue \bluefilledcirc), and \(\sink\) (yellow \yellowfilledcirc) on G as shown in (b).}
\vspace{-4mm}
\label{fig:med_ex_results}
\end{figure*}

\begin{figure}
\centering
  \begin{minipage}{.15\textwidth}
    \centering
\includegraphics[width=\linewidth]{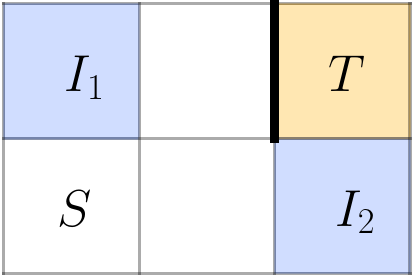}
    \subcaption{\label{fig:small_reactive_cuts_q0} q$0$}
  \end{minipage}
  \hspace{1mm}
  \begin{minipage}{.15\textwidth}
    \centering
\includegraphics[width=\linewidth]{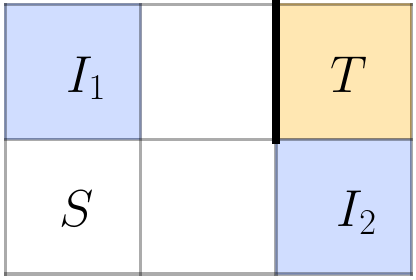}
    \subcaption{\label{fig:small_reactive_cuts_q6} q$6$}
  \end{minipage}
\hspace{1mm}
  \begin{minipage}{.15\textwidth}
    \centering
\includegraphics[width=\linewidth]{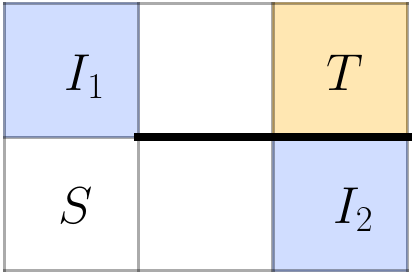}\subcaption{\label{fig:small_reactive_cuts_q7} q$7$}
  \end{minipage}
\caption{\label{fig:small_reactive_w_cuts} Test environment implementation of a reactive test strategy for Example~\ref{ex:small_reactive}.}
\end{figure}

\section{Test Strategy Synthesis}
In this section, we will outline how to find the reactive test strategy from the optimization result in the different settings.

\subsection{Test Environments with Static and/or Reactive Obstacles}
For each setting (static, reactive, and mixed), the optimal cuts from solving the corresponding MILP are used to realize a test strategy with static and/or reactive obstacles.
The optimal cuts \(C\) for each MILP are parsed into a reactive map \(\mc{C}: \mc{B}_{\pi}.Q \rightarrow T_{\sys}.E\), where
\begin{equation}
\label{eq:c_map}
\mc{C}(q) \coloneqq \{(s,s') \in T_{\sys}.E \: \vert \:((s,q), (s',q')) \in C\}.
\end{equation}
The set \(\mc{C}(q)\) captures cuts that will be used to restrict the system when the state of the test execution is at the history variable \(q\). When the test execution $\vartheta$ reaches a state-history pair $(s,q)$ at time step $k\geq 0$, and $\mc{C}(q)$ contains a system transition $(s,s') \in T_{\sys}.E$, then the reactive test strategy $\pi_{\test}$ will restrict the system action corresponding to this transition. That is, the set of restrictions on the system is given by  
\begin{equation}
\begin{split}
\label{eq:reactive_test_strategy}
    \pi_\test(\sigma_k) &\coloneqq \{a \in T_{\sys}.A \,|\,\\& s' \in T_{\sys}.\delta(s,a) \text{ and } (s,s') \in \mc{C}(q)\}.
\end{split}
\end{equation}
In practice, the reactive test strategy can be realized by the test environment by placing obstacles during the test execution. The set of \emph{active obstacles} \(\mathtt{Obs}(\sigma_k)\) at time step \(k\geq 0\) is defined as the set of all state-action restrictions at time \(k\). The test environment uses the test strategy \(\pi_{\test}\) to determine \(\mathtt{Obs}\) in the following settings.\\
\textbf{Instantaneous:} In this setting, the test environment \emph{instantaneously} places obstacles for the current history variable $q$. For any \(k\geq 0\), let \((s,q)\) be the state-history pair at time step \(k\) of the test execution. Therefore, the set of active obstacles at $\sigma_k$ is given as, $\mathtt{Obs}(\sigma_k) = \{(s',a) \:\vert\: \forall s'' \in T_{\sys}.\delta(s',a) \text{ and } (s',s'') \in \mc{C}(q)\}$.\\
\textbf{Accumulative:} 
In this setting, the test environment \emph{accumulates} obstacles according to the system state during the test execution. For any \(k\geq 0\), let \((\bar{s}, \bar{q})\) and \((s,q)\) be the state-history pairs at time steps \(k-1\) and \(k\) of the test execution, respectively. If $\bar{q} \neq q$, we set active obstacles to be \(\mathtt{Obs}(\sigma_k)= \{(s,a)\:\vert\: \forall a \in \pi_{\test}(\sigma_k)\}\). As the test execution progresses to state-history pair $(s',q)$ at time step \(l > k\), any transition restricted by the test strategy is added to the set of active obstacles $\mathtt{Obs}(\sigma_l) =\bigcup_{i=k}^l \mathtt{Obs}(\sigma_i)$ and is restricted by the test environment.
These obstacles remain in place until the test execution reaches a state history pair $(s'',q')$ at time step \(m > k\), where $q \neq q'$, at which point the test environment resets the set of active obstacles to be \(\mathtt{Obs}(\sigma_m)=  \{(s'',a) \: \vert \: \forall a \in \pi_\test(\sigma_m) \}\) and restrictions are accumulated until a different history variable is reached.\\

\begin{remark}
\label{rem:bidirectional}
Assumption~\ref{asm:bidirectionality} can be relaxed if we can ensure that cuts \(C\) do not introduce any livelocks, in which the system has no path the goal. The feasibility constraints ensure that there always exists a path to the goal from every source \(\mathtt{s} \in \src_{G_{\sys}}(q)\) for every history variable \(q\), and under the bidirectional setting of Assumption~\ref{asm:bidirectionality}, the system can navigate back to the corresponding source. Without Assumption~\ref{asm:bidirectionality} we need to check for every cut that the MILP returns, that a path to the goal still exists. If that is not the case, we can exclude the solution and re-solve the MILP in a counterexample-guided search similar to the approach presented in Section~\ref{sec:synth_test_agent_strategy}.
\end{remark}

\begin{proposition}
    \label{prop:no_new_cuts_same_F}
In both the instantaneous and accumulative settings, as long as no new restrictions that are not in $\mc{C}(q)$ are introduced, the flow value $F$ remains the same.
\end{proposition}

\addtocounter{example}{-1}
\begin{example}[Small Reactive (continued)]
Fig.~\ref{fig:small_reactive_w_cuts} illustrates the test environment implementing a reactive test strategy. The reactive test strategy is constructed from the optimal cuts (as depicted in Fig.~\ref{fig:small_reactive_virtual}) on \(\mc{G}\) found by solving \textbf{\textproc{MILP (reactive)}}. The test starts in history variable q$0$ and the system transitions are restricted according to Fig.~\ref{fig:small_reactive_cuts_q0}. If the system decides to visit \(\inter_1\) first, the test execution moves to history variable q$6$ shown in Fig.~\ref{fig:small_reactive_cuts_q6}, whereas if the system decides to visit $\inter_2$ first, the test execution moves to q$7$, as depicted in Fig.~\ref{fig:small_reactive_cuts_q7}. This test environment can be implemented in either the instantaneous or the accumulative setting.
\end{example}

\textit{Static and Mixed Test Environments. }
The cuts found from \textbf{\textproc{MILP-static}} result in a reactive map \(\mc{C}\) in which \(\mc{C}(q) = \mc{C}(q')\) for all \(q,q' \in \mc{B}_{\pi}.Q\). That is, restrictions on system actions remain in place for the entire duration of the test, and do not change depending on the history variable \(q\). In this fully static setting, every edge is in the static area, that is $T_{\sys}.E_{\text{static}} = T_{\sys}.E$. Therefore, the test environment realizes the test strategy by restricting all system actions corresponding to any cut in $\mc{C}(q)$ for all $q \in \mc{B}_\pi.Q$ with static obstacles simultaneously,
\begin{equation}
    \label{eq:static_obs}
    \mathtt{Obs} \coloneqq \{(u.s,v.s) \in T_{\sys}.E_{\text{static}} \: \vert \: (u,v) \in C \}.
\end{equation}
In the mixed setting of static and reactive obstacles, the test strategy resulting from \textbf{\textproc{MILP-mixed}} is implemented similarly to the reactive setting, except for system transitions in $T_{\sys}.E_{\text{static}}$ that are blocked by static obstacles. 

\addtocounter{example}{-2}
\begin{example}[continued]
For the grid world example, Fig.~\ref{fig:med_result} illustrates the static test on the grid world, and Fig.~\ref{fig:med_virtual} shows the corresponding cuts \(C^*\) on the virtual product network \(\mc{G}\). Here, the 14 cuts on \(\mc{G}\) map to 4 static obstacles since multiple edges on \(\mc{G}\) correspond to the same transition in \(T_{\sys}\). The optimal flow value is \(F^*=3\) and there is no bypass flow. Thus, as the system navigates from source \(\src\) to target \(\sink\), it must visit at least one of the intermediate nodes $\inter$. 
\end{example}

\begin{remark}
The instantaneous and accumulative implementations of the test environment guide when the obstacles are placed by the test environment. However, this does not have to be the same as when the system senses or observes these restrictions on its actions. We assume that the system can observe all restricted actions on its current state before it commits to an action. 
\end{remark}

The graph construction, network flow optimization, and finding the reactive test strategy are summarized in Algorithm~\ref{alg:find_test_strategy}.   
\begin{theorem}
If the problem data are not inconsistent (see Section~\ref{sec:characterizing_opt_results}), the reactive test strategy $\pi_\test$ found by Algorithm~\ref{alg:find_test_strategy} solves Problem~\ref{prob:reactive_test_strategy}.
\end{theorem}
\begin{proof}
The test environment informs the choice of the MILP (static, reactive, or mixed). Therefore, the resulting \(\pi_{\test}\) will be realizable by the test environment.
By construction of $G_\sys$, any correct system strategy corresponds to a $\mathtt{Path}(\src_\sys, \sink_\sys)$.
By Theorem~\ref{thm:feasibility}, at any point during the test execution, if the system has not violated its guarantees, there exists a path on $G_\sys$ to $\sink_\sys$. Therefore, there exists a correct system strategy $\pi_\sys$, and resulting trace $\sigma(\pi_\sys \times \pi_\test)$, which corresponds to the path $\vartheta_{\sys,n} = (s,q)_0\ldots(s,q)_n$ on $G_\sys$, where $(s,q)_0 \in \src_\sys$ to $(s,q)_n \in \sink_\sys$. 
By Lemma~\ref{lem:path_existence} any $\mathtt{Path}(\src_\sys, \sink_\sys)$ on $G_\sys$ has a corresponding $\mathtt{Path}(\src, \sink)$ on $G$ and by Theorem~\ref{thm:partition}, the cuts ensure that all such paths on $G$ are routed through the intermediate $\inter$. Therefore, for a correct system strategy $\pi_\sys$, the trace \(\sigma(\pi_{\sys} \times \pi_{\test}) \models \varphi_{\sys} \wedge \varphi_{\test}\). Thus, \(\pi_{\test}\) is feasible and by Proposition~\ref{prop:no_new_cuts_same_F} and Lemma~\ref{lem:least_restrictive}, \(\pi_{\test}\) is least-restrictive. Thus, Problem~\ref{prob:reactive_test_strategy} is solved. 
\end{proof}

\begin{algorithm}
\caption{Reactive Test Synthesis}\label{alg:synth_test_agent_strategy} 
\begin{algorithmic}[1] 
\Procedure{Test Synthesis}{$T_{\sys}, T_\tester, H, \varphi_{\text{sys}}, \varphi_{\text{test}}$}
\Require system $T_{\sys}$, test agent $T_\tester$, test harness $H$, system objective $\varphi_{\text{sys}}$, test objective $\varphi_{\text{test}}$
\Ensure test agent strategy $\pi_\tester$ 
\State $T_{\sys}.E_{\text{static}} \leftarrow $ Define from $T_{\sys}$, $T_\tester$ \Comment{\small Static area (Eq.~\eqref{eq:te_static} \normalsize}
\State $\mc{G}, \mathfrak{G}, \mathtt{I}, G \leftarrow $ Setup arguments \Comment{\small Lines 2-13 in Alg.~\ref{alg:find_test_strategy} \normalsize}
\State $\mathtt{C}_{\text{ex}} \leftarrow \emptyset$ \Comment{\small Initialize empty set of excluded solutions \normalsize}
\While{$\mathtt{True}$}
\State $\dvec^* \leftarrow $Solve \textbf{\textproc{MILP-agent}}($\mc{G},\mathfrak{G},\inter, T_{\sys},H,\mathtt{C}_{\text{ex}}$) 
\If{$\textproc{Status(MILP)} = \mathtt{infeasible}$}
\State \textbf{return } \(\mathtt{infeasible}\)
\EndIf
\State \(C \leftarrow \{(u,v) \in G.E\, \vert\, {\dvec^*}^{(u,v)} = 1\}\) \Comment{\small Cuts on $G$ \normalsize}
\State \(\mathtt{Obs} \leftarrow \) Define from $C$ \Comment{\small Static Obstacles (Eq.~\eqref{eq:static_obs})\normalsize}
\State \(\mc{R} \leftarrow\)Define from \(C\) \Comment{\small  Reactive map (Eq.~\eqref{eq:reactive_obs_map})\normalsize}
\State \textbf{A} \(\leftarrow\) Assumptions~\eqref{eq:sys_init}--\eqref{eq:no_collision_asm} from \(T_{\sys}\), \(T_\tester\), \(G\), \(\varphi_\sys\)
\State \textbf{G} \(\leftarrow\) Guarantees~\eqref{eq:testagent_init}--\eqref{eq:do_not_overconstrain} from \(T_{\sys}\), \(T_\tester\), \(\mc{R}\)
\State $\varphi \leftarrow (\textbf{A} \rightarrow \textbf{G})$ \Comment{\small Construct GR(1) formula\normalsize}
\If{$\textproc{Realizable}(\varphi)$}
\State \(\pi_{\tester} \leftarrow\) GR1Solve(\(\varphi\))
\State \textbf{return} $\pi_\tester$, $\mathtt{Obs}$
\EndIf
\State $\mathtt{C}_{\text{ex}} \leftarrow \mathtt{C}_{\text{ex}} \cup C$
\EndWhile
\EndProcedure
\end{algorithmic}
\label{alg:reactive_test}
\end{algorithm}

This framework results in a test that is not impossible (with respect to the system objective) for a correctly implemented system. On the other hand, a poorly designed system can still fail since the system is not aided in satisfying its guarantees.

\subsection{Synthesizing a Dynamic Test Strategy}
\label{sec:synth_test_agent_strategy}
In some test scenarios, it might be beneficial to make use of an available dynamic test agent. Thus, the challenge is to find a test agent strategy that corresponds to \(\mc{C}\) while ensuring that the system's operational environment assumptions are satisfied. To accomplish this, we adapt the \textbf{\textproc{MILP-mixed}} using information about the dynamic test agent. Then, we find the test agent strategy using reactive synthesis and counter-example guided search. From the optimal cuts of \textbf{\textproc{MILP-mixed}} and the resulting reactive map $\mc{C}$, we can find states that the test agent must occupy in reaction to the system state. Then, we synthesize a strategy for the dynamic test agent using the Temporal Logic and Planning Toolbox (TuLiP)~\cite{filippidis2016control}. If we cannot synthesize a strategy, we use a counterexample-guided approach to exclude the current solution and resolve the MILP to return a different set of optimal cuts until a strategy can be synthesized. Suppose we are given a test agent whose dynamics are given by the finite transition system \(T_{\tester}\), where \(T_{\tester}.S\) contains at least one state that is not in \(T_{\sys}.S\), denoted as \(\mathtt{park}\). During the test execution, the test agent can navigate to these \(\mathtt{park}\) states, if necessary. These states are required to synthesize a test agent strategy. 
From the test agent's transition system \(T_{\tester}\), we determine which states in $T_{\sys}$ the test agent can occupy. From these states, we can define the static area as, 
\begin{equation}
\label{eq:te_static}
    T_{\sys}.E_{\text{static}}\coloneqq \{(u,v) \in T_{\sys}.E \: \vert \: v \notin T_{\tester}.S\}.
\end{equation}

\smallbreak
\noindent
\textbf{Adapting the MILP: }
Since an agent can only occupy a single state at a time, we incentivize solutions in which multiple edge cuts can be realized by occupying the same state. For this, we introduce the variable $\dvec_{\text{state}} \in \mathbb{R}^{|V|}_+$, which represents whether an incoming edge into a state is cut. This is captured by the constraint
\begin{equation}
\label{eq:d_state_constraint}
\tag{c10}
\begin{split}
\forall (u,v) \in E,\quad d^{(u,v)} \leq d_{\text{state}}^v,
\end{split}
\end{equation}
where $d_{\text{state}}^v\geq 1$ corresponds to at least one incoming edge being cut. The adapted objective is then defined as
\begin{equation*}
F - \frac{1}{\vert E \vert} \sum_{v \in V} d^v_{\text{state}} - \frac{1}{\vert E \vert^2} \sum_{e \in E} d^e.
\end{equation*}
\smallbreak
\noindent
The objective is chosen such that the number of states that need to be blocked is minimized with the fewest possible edge cuts. The regularizers are chosen to reflect this order of priority. The optimal cuts from the resulting MILP are used to synthesize a reactive test agent strategy as follows. From the optimal cuts \(C\), we find the set of static obstacles $\mathtt{Obs} \subseteq T_{\sys}.E_{\text{static}}$ according to Eq.~\eqref{eq:static_obs} and the reactive map \(\mc{R}:\mc{B}_{\pi}.Q \rightarrow T_{\sys}.E\) as follows:
\begin{equation}
    \label{eq:reactive_obs_map}
    \begin{split}
    \mc{R}(q) \coloneqq \{(s,s') \in T_{\sys}.E \: \vert \: (s,s') \notin T_{\sys}.E_{\text{static}} \text{ and } \\((s,q), (s',q')) \in C\}.
    \end{split}
\end{equation}
The reactive map \(\mc{R}\) is used to synthesize a strategy for the test agent. If no strategy can be found, a counter-example guided approach is used to resolve the MILP. 
\smallbreak
\noindent
\textbf{Reactive Synthesis: }
From the solution of the MILP, we now construct the specification to synthesize the test agent strategy using TuLiP. In particular, we construct a GR(1) formula with assumptions being our model of the system and the guarantees capturing requirements on the test agent. Note that we are synthesizing a strategy for the test agent, where the environment is the system under test. The variables needed to define the GR(1) formula consist of variables capturing the system's state $\mathtt{x}_\sys \in T_{\sys}.S$ and $\mathtt{q}_\hist \in \mc{B}_\pi.Q$, which track how system transitions affect the history variable \(q\). The test agent state is represented in the variable $\mathtt{x}_\tester \in T_\tester.S$. 

First, we set up the subformulae constituting the assumptions on the system model. 
The initial conditions of the system are defined as
\begin{equation}
\label{eq:sys_init}
\tag{a1}
(\mathtt{x}_{\sys} = s_0 \land \mathtt{q}_{\hist} = q_0),
\end{equation}
where $s_0 \in T_{\sys}.S_0$ and $\mc{B}_\pi.Q_0$.
We define the dynamics of the system and the history variable for each state $(s,q) \in G.S$ as follows:
\begin{equation}
\label{eq:history_var_change}
\tag{a2}
    \square \Big( (\mathtt{x}_{\sys} = s \land \mathtt{q}_{\hist} = q )\rightarrow  \bigvee_{\substack{(s', q') \in \\ \mathtt{succ}(s, q)}} \bigcirc\big(\mathtt{x}_{\sys} = s' \land \mathtt{q}_{\hist} = q') \Big),
\end{equation}
where \(\mathtt{succ}(s,q)\) denotes the successors of state  \((s, q) \in G.S\). 
For simplicity, we choose a turn-based setting, in which each player will only take their action if it is their turn. To track this, we introduce the variable $\mathtt{turn} \in \mathbb{B}$ as a test agent variable. For the system, this is encoded as remaining in place when \(\mathtt{turn}=1\):
\begin{equation}
\label{eq:turn_based_asm}
\tag{a3}
\bigwedge_{s \in T_{\sys}.S} \square \Big((\mathtt{x}_{\sys} = s \land \mathtt{turn} = 1) \rightarrow \bigcirc (\mathtt{x}_{\sys} = s)\Big).
\end{equation}
If a turn-based setup is not used, we need to synthesize a Moore strategy for the test agent since it should account for all possible system actions. 
The system objective \(\varphi_{\sys}\) can be encoded as the formula
\begin{equation}
    \label{eq:sys_obj}
    \tag{a4}
    \square \Feventually (\mathtt{x}_{\sys} = x_{\text{goal}}) \land \varphi_{\text{aux}},
\end{equation}
where \(x_{\text{goal}}\) is the terminal state of the system and a reachability objective specified in 
\(\varphi_{\sys}\). The other objectives specified in \(\varphi_{\sys}\) are transformed to their respective GR(1) forms in \(\varphi_{\text{aux}}\). This transformation of LTL formulae into GR(1) form is detailed in~\cite{maoz2015gr}. In addition, the system is expected to safely operate in the test agent's presence. The set of states where collision is possible is denoted by \(S_{\cap} := T_{\sys}.S \cap T_{\tester}.S\). Thus, the safety formula encoding that the system will not collide into the tester is given as:
\begin{equation}
\label{eq:no_collision_asm}
\tag{a5}
\bigwedge_{s \in S_{\cap}} \square \Big(\mathtt{x}_{\tester} = s \rightarrow \bigcirc\neg (\mathtt{x}_{\sys} = s)\Big).
\end{equation}
Equations~\eqref{eq:sys_init}--~\eqref{eq:no_collision_asm} represent the test agent's assumptions on the system model. Next, we describe the subformulae for the guarantees of the GR(1) specification. The initial conditions for the test agent are
\begin{equation}
\label{eq:testagent_init}
\tag{g1}
   \bigvee_{s \in T_{\tester}.S_0} \mathtt{x}_{\tester} = s.
\end{equation}
The test agent dynamics are represented by
\begin{equation}
\label{eq:testagent_dynamics_guar}
\tag{g2}
\square \Big( (\mathtt{x}_{\tester} = s)\rightarrow  \bigvee_{(s,s') \in T_\tester.E}\bigcirc\big(\mathtt{x}_{\tester} = s') \Big).
\end{equation}
The test agent can also move only in its turn and will remain stationary when \(\mathtt{turn}=0\):
\begin{equation}
\label{eq:turn_based_guar}
\tag{g3}
\bigwedge_{s \in T_{\tester}.S} \square \Big((\mathtt{x}_{\tester} = s \land \mathtt{turn} = 0) \rightarrow \bigcirc (\mathtt{x}_{\tester} = s)\Big).
\end{equation}
The \(\mathtt{turn}\) variable alternates at each step:
\begin{equation}
\tag{g4}
\label{eq:turn_change_asm}
\begin{split}
    (\mathtt{turn} = 1) \rightarrow \bigcirc (\mathtt{turn} = 0) \: \land \\ (\mathtt{turn} = 0)\rightarrow \bigcirc (\mathtt{turn} = 1).
\end{split}
\end{equation}
To satisfy the system assumptions (Def.~\ref{def:sys_assume}), the test agent should not adversarially collide into the system. This is captured via the following safety formula,
\begin{equation}
\label{eq:no_collision_guar}
\tag{g5}
\bigwedge_{s \in S_{\cap}} \square\Big(\mathtt{x}_{\sys} = s \rightarrow \bigcirc\neg (\mathtt{x}_{\tester} = s)\Big).
\end{equation}

Now, we enforce the optimal cuts found from the MILP. To enforce cuts reactively during the test execution, the states occupied by the system are defined as follows,
\begin{equation}
\label{eq:realize_cuts}
\tag{g6}
\begin{split}
\bigwedge_{q \in \mc{B}_{\pi}.Q}\bigwedge_{(s,s') \in \mc{R}(q)} \square\Big((\mathtt{x}_{\sys} = s &\land \mathtt{q}_{\hist} = q \land \mathtt{turn}=0)\\ & \rightarrow (\mathtt{x}_{\tester} = s')\Big).
\end{split}
\end{equation}
Essentially, for some history variable \(q\), if \((s,s') \in \mc{R}(q)\) is an edge cut, then the test agent must occupy the state \(s'\) when the system is in the state \(s\) when the test execution is at history variable \(q\). However, the test agent should not introduce any additional restrictions on the system, which is formulated as
\begin{equation}
\label{eq:do_not_overconstrain}
\tag{g7}
\begin{split}
\bigwedge_{q \in \mc{B}_{\pi}.Q}\bigwedge_{\substack{(s,s') \in T_{\sys}.E \\(s,s') \not\in \mc{R}(q)}}\square\Big((\mathtt{x}_{\sys} = s &\land \mathtt{q}_{\hist} = q \land \mathtt{turn}=0)\\& \rightarrow \neg (\mathtt{x}_{\tester} = s')\Big).
\end{split}
\end{equation}
Intuitively, this corresponds to the requirement that the tester agent shall not restrict system transitions that are not part of the reactive map $\mc{R}$. 
A test agent strategy that satisfies the above specifications is guaranteed to not restrict any system action unnecessarily. However, the test agent can occupy a state that is not adjacent to the system and block all paths to the goal from the system's perspective. This could lead the system to not making any progress towards the goal at all, resulting in a livelock. To avoid this, we characterize the livelock condition as a safety constraint that the test agent must satisfy (e.g., if it occupies a livelock state, it must not occupy it in the next step). The specific safety formula that captures the livelock depends on the example. We find the states where the tester would block the system from reaching its goal $T_{\sys}.S_{\text{block}} \subseteq T_{\tester}.S$. The following condition  ensures that it will only transiently occupy blocking states:
\begin{equation}
\label{eq:only_transiently_block}
\tag{g8}
   \bigwedge_{s \in T_{\sys}.S_{\text{block}}} \square \Big(\mathtt{x}_{\tester} = s \rightarrow \bigcirc \neg(\mathtt{x}_{\tester} = s)\Big).
\end{equation}
Therefore, we synthesize a test agent strategy \(\pi_{\tester}\) for the GR(1) formula with assumptions~\eqref{eq:sys_init}--\eqref{eq:no_collision_asm} and guarantees ~\eqref{eq:testagent_init}--\eqref{eq:only_transiently_block}.

\smallbreak
\noindent
\textbf{Counterexample-guided Approach:}
The MILP can have multiple optimal solutions, some of which may not be realizable for the test agent. If the GR(1) formula is unrealizable, we exclude the solution and re-solve the MILP until we find a realizable GR(1) formula. In particular, every new set of optimal cuts \(C\) that is unrealizable is added to the set \(\mathtt{C}_{\text{ex}}\). Then, the MILP is resolved with an additional set of affine constraints as follows,
\begin{equation}
\label{eq:counterexample_constraint}
\tag{c15}
\sum_{e \in C} d^e \leq \vert C \vert - 1, \: \forall C \in \mathtt{C}_{\text{ex}}.
\end{equation}
This corresponds to preventing all edges in an excluded solution $C$ from being cut at the same time. The adapted MILP is then defined as follows:

\noindent
\rule{\linewidth}{0.6pt}
\textbf{\textproc{MILP-agent}:}
\begin{equation}
\label{milp:mixed_adapted}
\begin{aligned}
    \max_{\substack{\fvec, \dvec, \dvec_{\text{state}}, \bm{\mu}, \\\fvec_{\sys}^{(q,\mathtt{s})} \,\forall q \in \mc{B}_{\pi}.Q,\, \\ \forall \mathtt{s} \in \src_{\mc{G}_{\sys}}(q).}} &F - \frac{1}{\vert E \vert} \sum_{v \in V} d^v_{\text{state}} - \frac{1}{\vert E \vert^2} \sum_{e \in E} d^e \\
   &\text{s.t.} \quad \text{\eqref{eq:binary_cuts}-\eqref{eq:static_map_cuts_in_G}}, \text{\eqref{eq:d_state_constraint}}, \text{\eqref{eq:counterexample_constraint}}.
   \end{aligned}
\end{equation}
\rule{\linewidth}{0.6pt}

This process is repeated until a strategy is synthesized or the \textbf{\textproc{MILP-agent}} becomes infeasible. 

\begin{lemma}
\label{lem:sameness_of_strategies}
Let $\pi_{\tester}$ be the test agent strategy and let $\mathtt{Obs}$ be the set of static obstacles synthesized from the optimal solution $C^*$ of \textbf{\textproc{MILP-agent}} according to the GR(1) formula with assumptions~\eqref{eq:sys_init}--\eqref{eq:no_collision_asm} and guarantees~\eqref{eq:testagent_init}--\eqref{eq:only_transiently_block}. Let \(\pi_{\test}\) be the reactive test strategy corresponding to the optimal cuts \(C^*\). Then $\pi_{\tester}$ and $\mathtt{Obs}$ realize $\pi_\test$.
\end{lemma}
\begin{proof}
By construction in Eqs.~\eqref{eq:c_map},~\eqref{eq:static_obs},~\eqref{eq:reactive_obs_map}, we have that \(\mc{C}(q) = \mc{R}(q) \cup \mathtt{Obs}\) for all history variables \(q \in \mc{B}_{\pi}.Q\). Due to guarantee~\eqref{eq:realize_cuts}, the synthesized test agent strategy restricts the transitions in \(\mc{R}(q)\). The test agent is also prohibited from restricting any other transitions by the guarantee~\eqref{eq:do_not_overconstrain}. Therefore, at each step of the test execution, the system actions restricted as a result of \(\pi_{\tester}\) and static obstacles \(\mathtt{Obs}\) exactly correspond to those restricted by the test strategy $\pi_\test$. 
\end{proof}

\begin{theorem}
\label{thm:alg2_solves_p2}
Algorithm~\ref{alg:synth_test_agent_strategy} solves Problem~\ref{prob:match_agent}.
\end{theorem}
\begin{proof}
The test agent strategy is synthesized to satisfy guarantees~\eqref{eq:testagent_init}-\eqref{eq:only_transiently_block}.
The guarantees~\eqref{eq:testagent_init}-\eqref{eq:turn_change_asm} specify the dynamics of the test agent, which satisfies \textbf{A1}. The safety guarantee~\eqref{eq:no_collision_guar} satisfies \textbf{A2}. Guarantees~\eqref{eq:realize_cuts} and
~\eqref{eq:do_not_overconstrain} realize the optimal cuts from \textbf{\textproc{MILP-agent}}. Due to constraint~\eqref{eq:feasibility_flow_on_Gsys} the optimal cuts ensure that there always exists a path on $G_{\sys}$. Together with guarantee~\eqref{eq:only_transiently_block}, this results in $\pi_\tester$ satisfying assumptions \textbf{A3} and \textbf{A4}. By Lemma~\ref{lem:sameness_of_strategies}, $\pi_\tester$ is a realization of a least-restrictive feasible $\pi_\test$.
\end{proof}

The test agent strategy and obstacles, $\pi_\tester$ and $\mathtt{Obs}$ correspond to the least-restrictive reactive test strategy $\pi_\test$ possible for that test environment. Other test environments might result in different least-restrictive reactive test strategies. 

\section{Complexity}
Our framework comprises three parts: i) graph construction, ii) routing optimization, and iii) reactive synthesis. For graph construction, we first need to construct B\"uchi automata from specifications. In the worst case, this construction has doubly-exponential complexity, \(2^{2^{\vert \phi\vert}}\), in the length of the formula $\phi$~\cite{baier2008principles}. Then, graph construction involves computing a Cartesian product of two graphs \(T_{\sys}\) and \(\mc{B}_{\pi}\), and has a worst-case time complexity of \(O(\vert T_{\sys}.S \vert^2\cdot \vert \mc{B}_{\pi}.Q\vert^2)\). For a more efficient implementation, we construct this product by expanding into states that are reachable from the source \(\src\). In the reactive synthesis part of the framework, we use GR(1) synthesis which is known to have a complexity of \(O(\vert N\vert)^3\), where \(N\) is the number of states required to define the formula. In this section, we will establish the computational complexity of the routing optimization and show that the associated decision problem is NP-hard.

\begin{figure*}
\centering
 \vspace{2mm}
\begin{minipage}{.2\textwidth}
    \includegraphics[width=\linewidth,trim={0.0cm 0.0cm -0.0cm 0.0cm}]{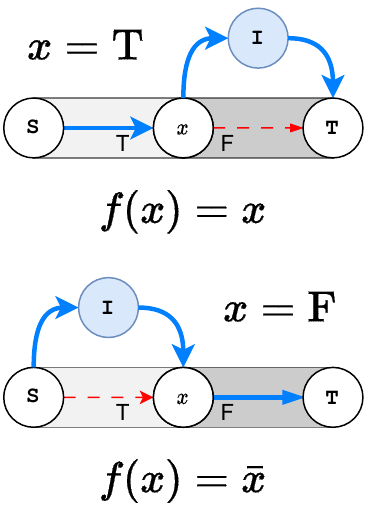}
    \subcaption{\label{fig:reduction_legend} Graphs matching formulae with a single variable $x$.}
  \end{minipage}
  \hspace{3mm}
  \begin{minipage}{.7\textwidth}
    \centering
\includegraphics[width=\linewidth,trim={0.0cm 0.4cm 0cm 0.0cm}]{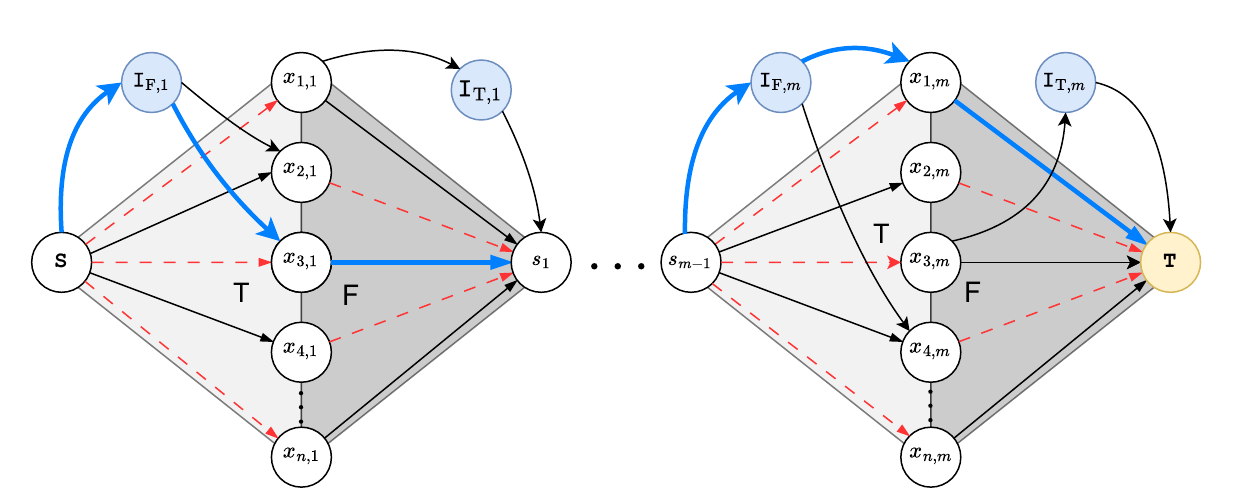}
    \subcaption{\label{fig:reduction} Graph resulting from a reduction of the 3-SAT formula $F(x_1, \dots, x_5)$, where the resulting edge cuts correspond to the truth assignment of the variables $x_1, \dots, x_5$.}
  \end{minipage}
\caption{Graphs constructed from a 3-SAT formula, where a truth assignment for the variables can be found using the network flow approach for static obstacles.}
\label{fig:complexity}
\end{figure*}

\begin{figure*}
\centering
\begin{minipage}{.48\textwidth}
    \includegraphics[width=\linewidth,trim={0.0cm 0.0cm -0.0cm 0.0cm}]{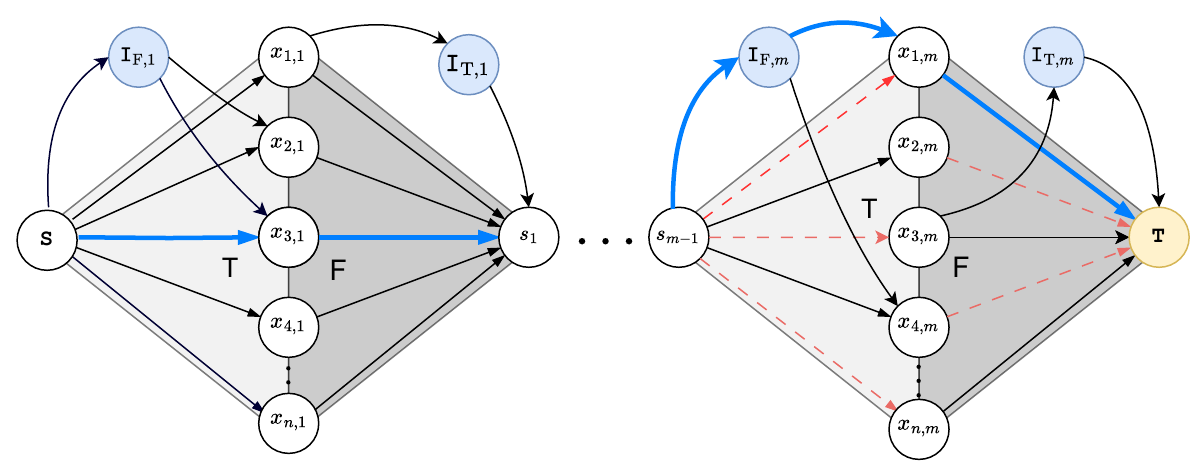}
    \subcaption{\label{fig:Greactive} Graph $G$ according to Construction~\ref{constr:formula_2_graphs} for the reactive case.}
  \end{minipage}
  \begin{minipage}{.48\textwidth}
    \centering
\includegraphics[width=\linewidth,trim={0.0cm 0.4cm 0cm 0.0cm}]{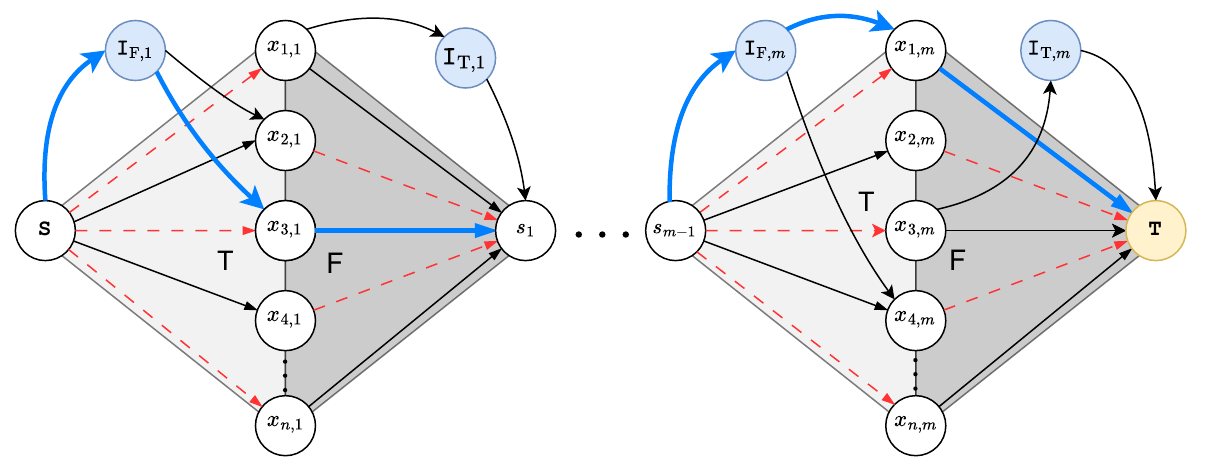}
    \subcaption{\label{fig:Gsys_reactive} Graph $G_{\sys}$ according to Construction~\ref{constr:formula_2_graph}.}
  \end{minipage}
\caption{Graphs $G$ and $G_{\sys}$ constructed from a 3-SAT formula, where a truth assignment for the variables can be found using the flow approach for reactive obstacles.}
\label{fig:reactive_complexity}
\end{figure*}

To prove the computational complexity of finding the cuts on the graph, we first prove the computational complexity in the special case of static obstacles. As defined in sections~\ref{sec:graph_construction} and~\ref{sec:optimization}, the problem data is a graph \(G = (V,E)\) with node groups \(\src\), \(\inter\), \(\sink\), and the corresponding flow network $\mc{G}$. For some edge $e \in E\setminus E(\inter)$, the binary variable $d^e$ indicates whether the edge is cut: $d^e = 1$. The set \(C\subset E\) represents the set of edges with \(d^e = 1\). For static obstacles, the edges are grouped by the corresponding transition in $T_{\sys}$. The grouping $\mathtt{Gr}_{\text{static}}: T_{\sys}.E \rightarrow G.E$, and defined as follows,
\begin{equation}
\label{eq:group_static}
\mathtt{Gr}_{\text{static}}((s,s')) \coloneqq \{ (u,v) \in G.E \: \vert \:\: u.s = s, v.s=s'\}.
\end{equation}
For some \((s,s') \in T_{\sys}.E\), all edges \(e \in \mathtt{Gr}_{\text{static}}((s,s'))\) have the same \(d^e\) value, i.e., if \(d^e = 1\) for some edge \(e\) in the group, then all edges in this group will have \(d^e\) set to 1.
A bypass path on \(G\) is some \(Path(\src,\sink)\) which does not visit the intermediate $\inter$. The flow value \(F\) on \(\mc{G}\) is defined from the source \(\src\) to target \(\sink\), with each edge having unit capacity. A valid set of edge cuts $C$ is such that i) there does not exist a bypass path, ii) there exists a path from $\src$ to $\sink$, and iii) edges respect the grouping $\mathtt{Gr}_{\text{static}}$.

\begin{problem}[Static Obstacles Optimization Problem]
     Given a graph \(G\), find a valid set of edge cuts $C$ such that the resulting maximum flow $F$ is maximized over all possible sets of edge cuts, and such that $\vert C \vert$ is minimized for the flow $F$.
    \label{prob:opt_static_obstacles}
\end{problem}

This corresponds to finding the valid set of edge cuts $C$ that as first priority, maximizes the flow $F$, and subsequently chooses the set of edge cuts $C$ with the smallest cardinality $\vert C \vert$ (i.e. breaking ties between all valid edge cuts that realize $F$).
For static obstacles, Problem~\ref{prob:opt_static_obstacles} corresponds to the following decision problem.

\begin{problem}[Static Obstacles Decision Problem]
    Given a graph \(G\) and an integer \(M\geq 0\), does there exist a valid set of edge cuts \(C\) such that $\vert C \vert\, \leq M$?
    \label{prob:decision_static_obstacles}
\end{problem}

\begin{lemma}
\label{lem:search_2_decision_static}
Any solution to Problem~\ref{prob:opt_static_obstacles} can be used to construct a solution for Problem~\ref{prob:decision_static_obstacles} in polynomial time.
\end{lemma}

Lemma \ref{lem:search_2_decision_static} implies that if there exists a polynomial-time algorithm to compute a solution to Problem \ref{prob:opt_static_obstacles}, then there also exists a polynomial-time algorithm to solve Problem \ref{prob:decision_static_obstacles}. Thus, if we can show that Problem~\ref{prob:decision_static_obstacles} belongs to the class of NP-hard problems (i.e., there exists a polynomial-time reduction from any arbitrary problem in NP to Problem~\ref{prob:decision_static_obstacles}), that would imply that there exists a polynomial-time algorithm to solve Problem~\ref{prob:decision_static_obstacles} only if \(P=NP\). This in turn would support the MILP approach we provide to solve Problem~\ref{prob:opt_static_obstacles}. To show that Problem~\ref{prob:decision_static_obstacles} is NP-hard, we construct a polynomial-time reduction from 3-SAT to Problem~\ref{prob:decision_static_obstacles}. This polynomial-time reduction maps any instance of 3-SAT to Problem \ref{prob:decision_static_obstacles} such that the solution of the constructed instance of Problem~\ref{prob:decision_static_obstacles} corresponds to a solution of the instance of the 3-SAT problem. 

\begin{definition}[3-SAT \cite{cook2023complexity}]
    Let \(f(x_1,\ldots, x_n) \coloneqq \bigwedge_{j=1}^{m} c_j\) be a propositional logic formula over Boolean propositions \(x_1, \ldots, x_n\) in conjunctive normal form (CNF) in which each clause $c_j$ is a disjunction of three Boolean propositions or their negations. A solution to the 3-SAT problem is an algorithm that returns \emph{True} if there exists a satisfying Boolean assignment to \(f(x_1, \ldots, x_n)\) and \emph{False} otherwise.
\end{definition}
We first introduce a construction which maps any clause in a propositional logic formula to some sub-graph of a graph. We will then connect these sub-graphs to obtain the graph which will allows a reduction of any instance of 3-SAT to Problem \ref{prob:decision_static_obstacles}. In turn, we will show that we can use any algorithm that solves Problem~\ref{prob:decision_static_obstacles} to solve the 3-SAT problem, showing that Problem~\ref{prob:decision_static_obstacles} is polynomial-time only if there exists a polynomial-time algorithm to solve 3-SAT, implying P=NP.

\begin{construction}[Clause to Sub-graph]
\label{constr:clause_2_subgraph}
Given a 3-SAT clause $c_j$, we can construct a sub-graph representing this clause as follows. For each clause $c_j$, we introduce nodes $s_{j-1}$ and $s_j$. Then, we add the nodes $x_{1,j}, \dots x_{n,j}$ corresponding to variables $x_{1}, \ldots, x_{n}$ in the 3-SAT formula. We add the following directed edges for each $x_{i,j}$ node --- an incoming edge from node $s_{j-1}$ to $x_{i,j}$, and an outgoing edge from $x_{i,j}$ node to node $s_{j}$.
Then we add two nodes, denoted by $\inter_{\text{T},j}$ and $\inter_{\text{F},j}$, to this sub-graph. If $x_i$ appears in the clause $c_j$, then we connect the $\inter_{\text{T},j}$ node by bypassing the edge from $x_{i,j}$ to $x_j$, and if $\bar{x}_i$ appears in $c_j$, then we connect $\inter_{\text{F},j}$ to bypass the edge from $s_{j-1}$ to $x_{i,j}$ (as shown in Fig.~\ref{fig:reduction_legend}).
\end{construction}

Constructing a sub-graph for a clause $c_j$ via Construction~\ref{constr:clause_2_subgraph} allows us to relate the edge cuts to the Boolean assignment for the variables $x_0, \dots, x_n$. If the incoming edge into $x_{i,j}$ is cut, then the corresponding Boolean assignment to $x_i$ is \emph{False}, and if the outgoing edge from $x_{i,j}$ is cut, then the corresponding Boolean assignment to $x_i$ is \emph{True}. This ensures that a satisfying assignment for the clause corresponds to edge cuts such that all \(Paths(s_{j-1}, s_j)\) are routed through intermediate nodes \(\{\inter_{\text{T},j}, \inter_{\text{F},j}\}\). An assignment that evaluates the clause \(c_j\) to \emph{False} corresponds to edge cuts in the sub-graph such that there is no path from \(s_{j-1}\) to \(s_j\).

\begin{construction}[Reduction of 3-SAT to Problem \ref{prob:decision_static_obstacles}]
\label{constr:formula_2_graph}
Suppose we have an instance of the 3-SAT problem with $n$ variables $x_1, \dots, x_n$ and $m$ clauses $c_1, \dots c_m$. First, we construct the sub-graphs for each clause according to Construction~\ref{constr:clause_2_subgraph}. Let \(M\coloneqq m\times n\). We denote the node $s_0$ as the source $\src$, and $s_{m}$ as the sink $\sink$. The resulting graph is a series of sub-graphs representing each clause $c_j$ of the 3-SAT formula.
For every variable $x_i$ in the formula, we maintain two groups of edges: i) incoming edges $\{(s_{j-1},x_{i,j})\:\vert \: 1 \leq j \leq m\}$, and ii) outgoing edges $\{(x_{i,j}, s_j) \: \vert \:  1 \leq j \leq m\}$. All edges in a group share the same edge cut value, corresponding to \(\mathtt{Gr}_{\text{static}}\). This ensures that every variable has the same Boolean assignment across clauses.
\end{construction}
This allows us to construct a graph corresponding to a 3-SAT formula in polynomial time via the procedure outlined in Construction~\ref{constr:formula_2_graph}, also illustrated in Fig.~\ref{fig:complexity}.

\begin{theorem}
\label{thm:static_NP_hard}
Problem~\ref{prob:decision_static_obstacles} is NP-complete.
\end{theorem}
\begin{proof}
We will show that Problem~\ref{prob:decision_static_obstacles} is NP-hard by showing that Construction \ref{constr:formula_2_graph} is a correct polynomial-time reduction of the 3-SAT problem to Problem~\ref{prob:decision_static_obstacles} i.e., any polynomial-time algorithm to solve Problem \ref{prob:decision_static_obstacles} can be used to solve 3-SAT in polynomial-time. Consider the graph constructed by Construction~\ref{constr:formula_2_graph} for any propositional logic formula.
The valid set of edge cuts $C$ on this graph with cardinality $\vert C \vert \leq M$ is a witness for  Problem~\ref{prob:decision_static_obstacles}. A witness for the 3-SAT formula is an assignment of the variables $x_1,\dots, x_n$. A witness to a problem is \emph{satisfying} if the problem evaluates to \emph{True} under that witness. Next, we show that a valid set of edge cuts \(C\) is a satisfying witness for Problem~\ref{prob:decision_static_obstacles} iff the corresponding assignment to variables $x_1, \dots, x_n$ is a satisfying witness for the 3-SAT formula.

First, consider a satisfying witness for Problem~\ref{prob:decision_static_obstacles}. 
By Construction~\ref{constr:formula_2_graph}, the cardinality of the witness, \(|C|=m\times n\) will be exactly \(M\), which is the minimum number of edge cuts required to ensure no bypass paths on the constructed graph. This implies that each variable $x_i$ has a Boolean assignment. By Construction~\ref{constr:clause_2_subgraph}, a strictly positive flow on the sub-graph of clause \(c_j\) implies that \(c_j\) is satisfied. By Construction~\ref{constr:formula_2_graph}, a strictly positive flow through the entire graph implies that all clauses in the 3-SAT formula are satisfied.
Therefore, a satisfying witness to the 3-SAT formula can be constructed in polynomial-time from a satisfying witness for an instance of Problem~\ref{prob:decision_static_obstacles}.

Next, we consider a satisfying witness for the 3-SAT formula. 
The Boolean assignment for each variable $x_i$ corresponds to edge cuts on the graph (see Fig.~\ref{fig:reduction}). Any Boolean assignment ensures that there is no bypass path on the graph since either all incoming edges or all outgoing edges for each variable \(x_{i}\) are cut. This also corresponds to the minimum number of edge cuts required to cut all bypass paths, corresponding to $\vert C \vert = m \times n$. By Construction~\ref{constr:clause_2_subgraph}, a satisfying witness corresponds to a Path(\(s_{j-1}, s_j\)) on the sub-graph for each clause $c_j$. By Construction~\ref{constr:formula_2_graph}, observe that there exists a strictly positive flow on the graph. Thus, we can construct a satisfying witness to an instance of Problem~\ref{prob:decision_static_obstacles} in polynomial time from a satisfying witness to the 3-SAT formula.
Therefore, any 3-SAT problem reduces to an instance of Problem~\ref{prob:decision_static_obstacles}, and thus, Problem~\ref{prob:decision_static_obstacles} is NP-hard. Additionally, Problem~\ref{prob:decision_static_obstacles} is NP-complete since we can check the cardinality of \(C\), and whether \(C\) is a valid set of edge cuts in polynomial time. 
\end{proof}

\begin{corollary}
\label{cor:opt_static_obstacles_FNP}
Problem~\ref{prob:opt_static_obstacles} is NP-hard~\cite{papadimitriou2003computational}.
\end{corollary}
\begin{proof}
By Theorem~\ref{thm:static_NP_hard}, Problem~\ref{prob:decision_static_obstacles} is NP-complete, and therefore by Lemma~\ref{lem:search_2_decision_static}, Problem~\ref{prob:opt_static_obstacles} is NP-hard.
\end{proof}
Additionally, we can identify the computational complexity for the reactive setting. For the reactive setting, a valid set of edge cuts is similar to the static setting, except in how edges are grouped, which is discussed in Remark~\ref{remark:reactive_grouping_on_G}. Fig.~\ref{fig:reactive_complexity} illustrates the graphs used for establishing the computational complexity in this setting. The optimization problem and its corresponding decision problem can be stated as follows.


\begin{problem}[Reactive Obstacles Optimization Problem]
\label{prob:opt_reactive_obstacles}
Given a graph \(G\), identify a valid set of edge cuts $C$ such that the resulting flow $F$ is maximized over all possible sets of edge cuts, and such that $\vert C \vert$ is minimized for the flow F.
\end{problem}
Note that a \emph{valid} set of edge cuts for the reactive problem is different from a valid set of edge cuts for the static problem.

\begin{problem}[Reactive Obstacles Decision Problem]
\label{prob:decision_reactive_obstacles}
Given a graph \(G\), and an integer $M \geq 0$, does there exist a valid set of cuts \(C\) such that $|C| \leq M$?
\end{problem}

Once again, we prove a reduction from 3-SAT, but to an instance of Problem~\ref{prob:decision_reactive_obstacles} with a single history variable $q$. Given a 3-SAT formula, the construction of the graph follows from the static setting, but with a few key differences.
\begin{construction}[Reduction from 3-SAT to Problem \ref{prob:decision_reactive_obstacles} with single history variable $q$]
\label{constr:formula_2_graphs}
Suppose we have an instance of the 3-SAT problem with $n$ variables $x_1, \dots, x_n$ and $m$ clauses $c_1, \dots c_m$. Let \(M \coloneqq n\). Using Construction~\ref{constr:formula_2_graph}, setup two graphs: \(G\) and a copy \(G^{(q,\src)}\) for source \(\src\) and the single history variable \(q\). The key difference is that \(G^{(q,\src)}\) follows Construction~\ref{constr:formula_2_graph} exactly, while in \(G\), edges in a group need not have the same cut value. Furthermore, for each group in \(G^{(q,\src)}\), the cut value is set to the maximum edge-cut value in the corresponding group in \(G\). 
\end{construction}

\begin{theorem}
Problem~\ref{prob:decision_reactive_obstacles} is NP-complete and Problem~\ref{prob:opt_reactive_obstacles} is NP-hard.
\end{theorem}
\begin{proof}
The proof follows similarly from Theorem~\ref{thm:static_NP_hard}. In this setting, a witness for Problem~\ref{prob:decision_reactive_obstacles} comprises the maximum edge cut value of each group in \(G\). Construction~\ref{constr:formula_2_graphs} relates edge cuts on \(G\) and \(G^{(q,\src)}\). This implies that edge cuts on \(G\) are found under the condition that there is a strictly positive flow on \(G^{(q,\src)}\) under a static mapping of edges. The minimum set of edge cuts which ensures no bypass paths on \(G\) has cardinality \(n\), corresponding to only one of the sub-graphs having edge cuts. Furthermore, for each \(x_i\), there will be one edge-cut in one of the two groups (incoming or outgoing edges). Therefore, for each \(x_i\), only the incoming or the outgoing edge group will have a maximum edge cut value of 1, corresponding to the Boolean assignment for $x_i$. A minimum cut on \(G\) found under the conditions of no bypass paths on $G$ and a positive flow on \(G^{(q,\src)}\) results in a Boolean assignment that is a satisfying witness to the 3-SAT formula. Thus, we have polynomial-time construction of a satisfying witness to the 3-SAT formula from a satisfying witness to Problem~\ref{prob:decision_reactive_obstacles}. This follows similarly to Theorem~\ref{thm:static_NP_hard}. 

Likewise, a satisfying witness to the 3-SAT formula can be mapped to edge cuts on one of the sub-graphs of \(G\). These edge cuts will be such that there is no bypass path on \(G\), and will be the minimum set of edge cuts to accomplish this task, corresponding to $\vert C \vert = n$. Additionally, by construction of the graphs, this will correspond to a strictly positive flow on \(G^{(q,\src)}\). Thus, we can construct a satisfying witness to Problem~\ref{prob:decision_reactive_obstacles} in polynomial time from a satisfying witness of the 3-SAT formula. Therefore, any 3-SAT problem reduces to an instance of Problem~\ref{prob:decision_reactive_obstacles}. As a result, Problem~\ref{prob:decision_reactive_obstacles} is NP-complete and following similarly to Corollary~\eqref{cor:opt_static_obstacles_FNP}, Problem~\ref{prob:opt_reactive_obstacles} is NP-hard. 
\end{proof}

\section{Experiments}
In this section, we demonstrate our framework on simulated and hardware experiments, and include runtime analysis.
In the following experiments, examples with static test environments solve the routing optimization \textbf{\textproc{MILP-static}} to find the test strategy. Similarly, examples with reactive test environments solve \textbf{\textproc{MILP-reactive}}, and those with reactive dynamic agents solve \textbf{\textproc{MILP-agent}}, unless otherwise stated. These optimizations are solved using Gurobipy~\cite{gurobi}. The reactive test agent strategies are synthesized using the temporal logic planning toolbox TuLiP~\cite{wongpiromsarn2011tulip}.  

In simulations and hardware experiments, we utilize Unitree A1 quadrupeds as both the system and test agents. The low-level control of the quadruped is managed through a motion primitive layer, which abstracts the underlying dynamics and facilitates transitions between primitives as described in \cite{ubellacker2023icra}. This includes behaviors such as lying down, standing, walking, jumping, and reduced-order model-based waypoint tracking using a unicycle or single integrator model. These behaviors can be directly commanded by the autonomy layer provided by TuLiP. Individual motion primitives are implemented within our C++ motion primitive framework, with control laws, sensing, and estimation executed at 1kHz.


\begin{figure*}
\centering
  \begin{minipage}{.22\textwidth}
    \centering
\includegraphics[width=\linewidth,trim={0.0cm 0.0cm 0cm 0.0cm}]{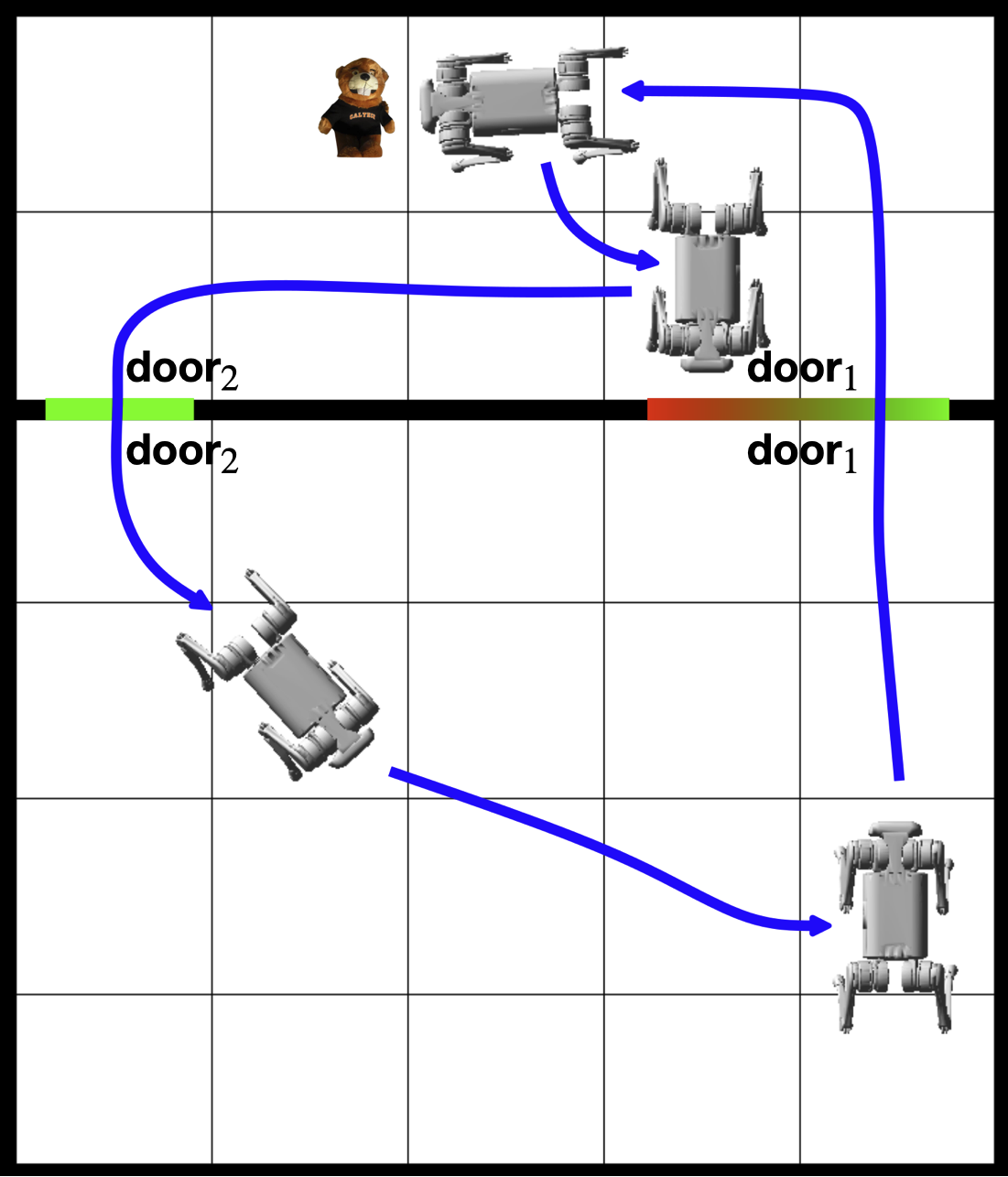}
    \subcaption{\label{fig:beaver_sim} Beaver rescue.}
  \end{minipage}
   \begin{minipage}{.22\textwidth}
    \centering
    \includegraphics[width=\linewidth,trim={0.0cm 0.0cm 0cm 0.0cm}]{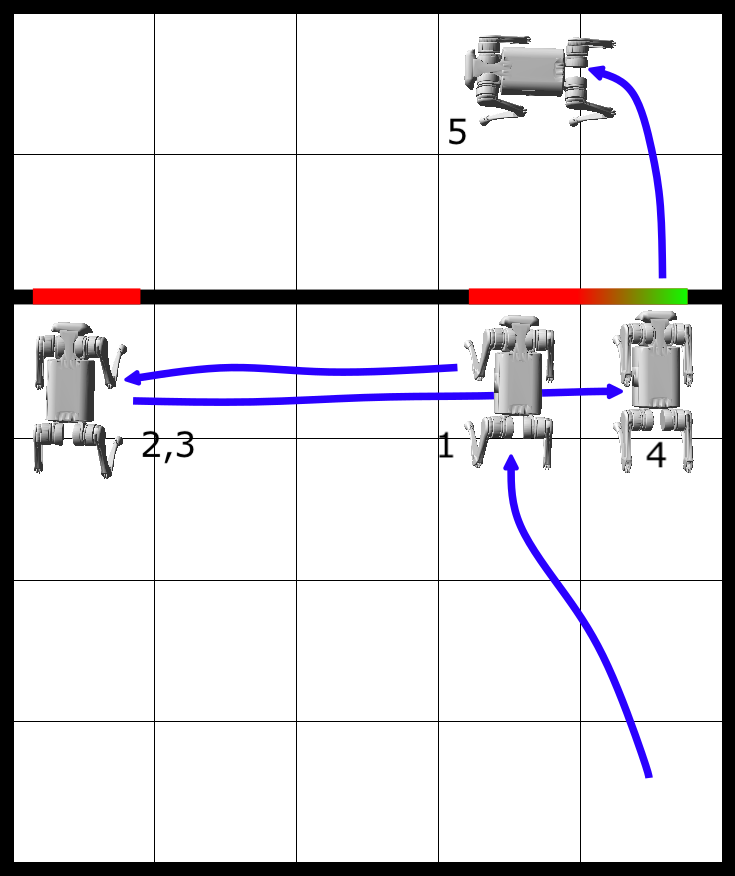}
    \subcaption{\label{fig:motionprim_sim} Motion primitive example. }
  \end{minipage}
  \begin{minipage}{.22\textwidth}
    \centering
\includegraphics[width=\linewidth,trim={0.0cm 0.0cm 0cm 0.0cm}]{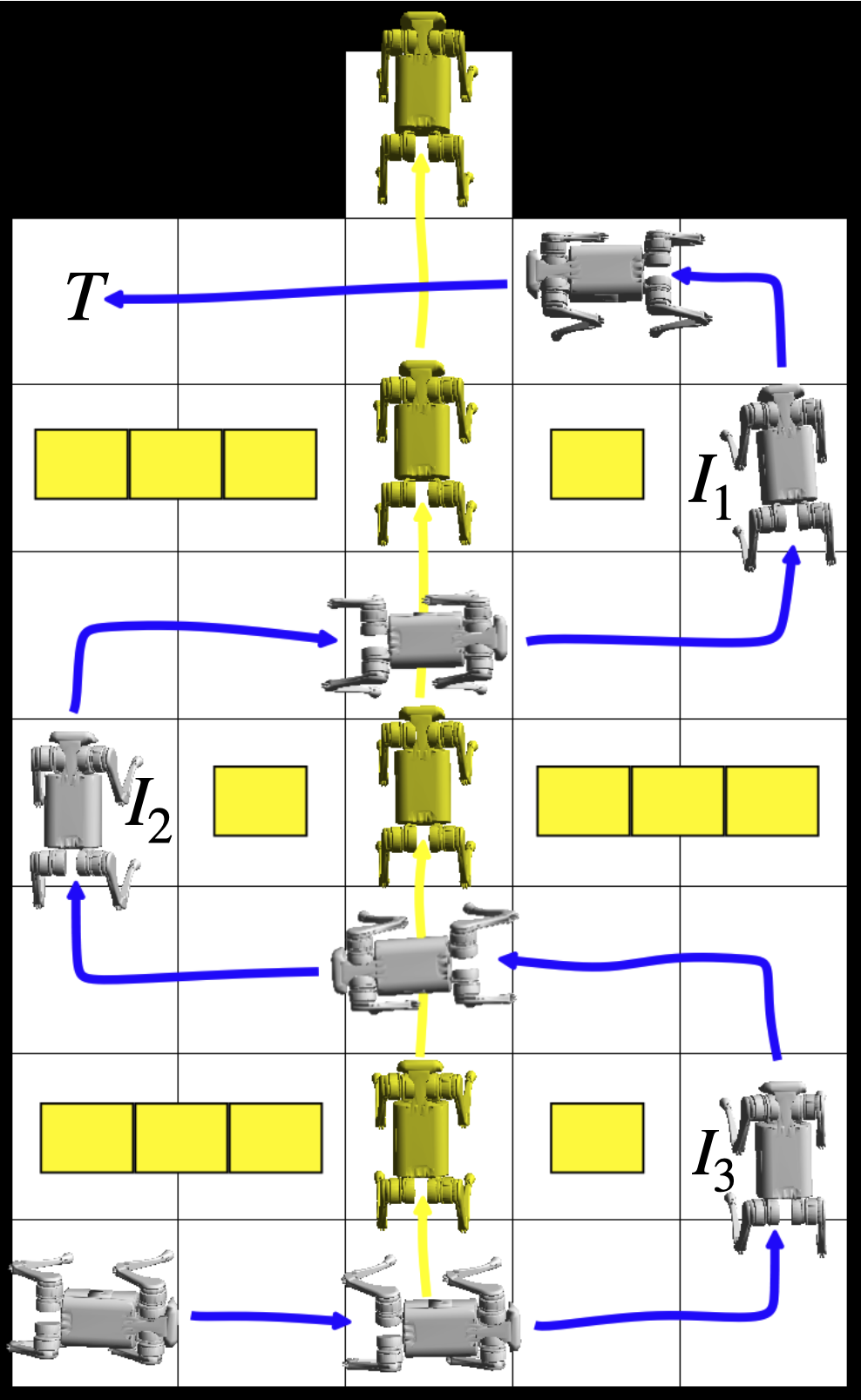}
    \subcaption{\label{fig:two_quads_sim} Maze 1.}
  \end{minipage}
  \begin{minipage}{.3\textwidth}
    \includegraphics[width=\linewidth,trim={0.0cm 0.0cm -0.0cm 0.0cm}]{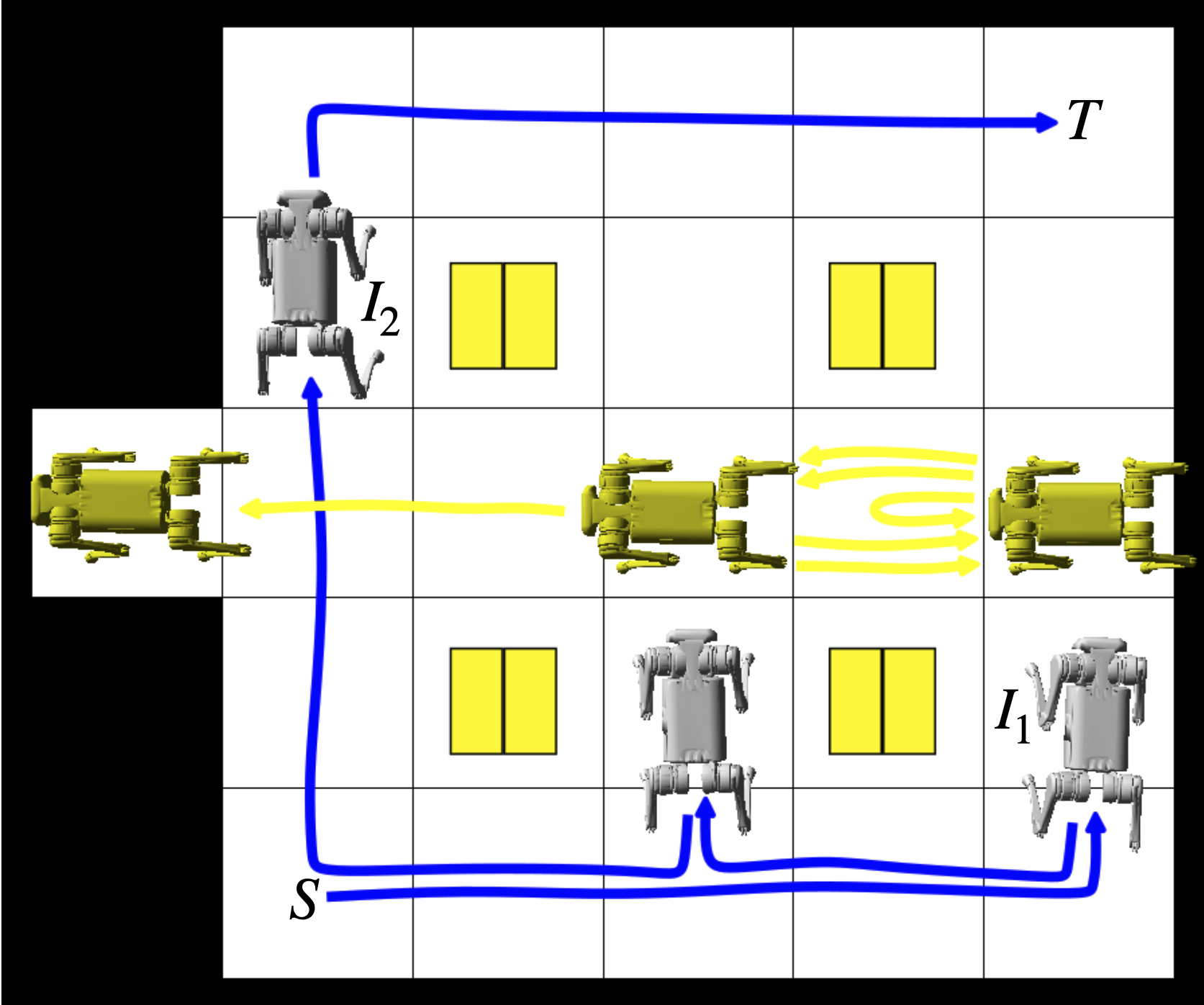}
\subcaption{\label{fig:quad_plus_alternative} Simulated alternative trace,  Maze 2.}
  \end{minipage}
\caption{Simulated experiment results. Yellow boxes are obstacles to indicate states that are not navigable in \(T_{\sys}\). Gray quadruped is the system, and yellow quadruped in (c) and (d) is the test agent. In (b), system demonstrates primitives in the order: stand (1), stand (2), jump (3), and lie (4), before advancing to goal (5). In (c) and (d), the test agent chooses to navigate off-grid after the test objective is realized.}
\label{fig:simulations}
\end{figure*}

\subsection{Simulation}
\smallbreak
\noindent
\textbf{Reactive Test Environment:}
The following two reactive examples were demonstrated on hardware in previous work in~\cite{badithela2023synthesizing}. The updated framework in this paper resulted in simulated test traces (see Figs.~\ref{fig:beaver_sim},~\ref{fig:motionprim_sim}) that are qualitatively similar to the hardware demo in~\cite{badithela2023synthesizing}. Additionally, using the updated framework reduced the time to solve the optimization by three orders of magnitude. 
\subsubsection{Beaver Rescue}
The quadruped's task is to rescue the beaver from the hallway and return it to the lab. The system objective is given as $\varphi_{\text{sys}} = \Feventually (\text{beaver} \wedge \Feventually \text{goal})$, where `beaver' corresponds to the quadruped reaching the beaver, and `goal' corresponds to the quadruped and the beaver reaching the safe location in the lab.
The test objective is given as $\varphi_{\text{test}} = \Feventually \text{door}_1 \land \Feventually \text{door}_2$, ensuring that the quadruped will use different doors on the way to the beaver and back into the lab. The resulting test execution first shows the quadruped using $\text{door}_2$ to exit the lab into the hallway, then after it reaches the beaver, $\text{door}_2$ is shut and the quadruped walks to $\text{door}_1$ to finally return to the lab. The reactive aspect here can be observed as follows --- if the quadruped chose to enter the hallway through $\text{door}_1$, then the resulting test execution would constrain access to $\text{door}_1$ when the quadruped is attempting to re-enter the lab with the beaver. The simulated test trace is shown in Fig.~\ref{fig:beaver_sim}.

\subsubsection{Motion Primitive Example}
In this example, we test the motion primitives of the quadruped given as `lie', `jump', and `stand'. The goal for the quadruped is to reach the beaver in the hallway.
The test objective is given as $\varphi_{\text{test}} = \Feventually \text{jump} \wedge \Feventually \text{lie} \wedge \Feventually \text{stand}$, which ensures that each motion primitive is tested at least once; and the system objective is $\varphi_\sys =\Feventually \text{goal}$, where `goal' corresponds to the beaver location. The test setup includes doors that might be unlocked by the system demonstrating specific motion primitives. Our framework will decide whether the doors will be locked or unlocked according to which motion primitives have already been observed during the test. This is where the reactivity of this framework becomes apparent, if the quadruped chose a different set of doors and motion primitives, the resulting test execution would have been different. The simulated test trace is shown in Fig.~\ref{fig:motionprim_sim}.
\smallbreak

\noindent
\textbf{Test Environment with Dynamic Agent}
\subsubsection{Maze 1}
The system (gray quadruped) wants to reach its goal location in the top left corner of the grid, and the test agent wants to route it through a series of states, labeled $I_1, I_2$, and $I_3$, shown in Fig.~\ref{fig:two_quads_sim}.
The system specification and test objective are given as
$\varphi_{\text{sys}}=\Feventually \text{goal}$ and $\varphi_{\text{test}}=\Feventually I_1 \land \Feventually I_2 \land \Feventually I_3.$ 
The test agent (yellow quadruped) can move up on the center column of the grid, and its strategy is found using the flow-based synthesis framework. Observe that it blocks specific cells such that the quadruped cannot directly navigate to its goal through the center of the grid. Instead, the system quadruped is forced to visit the labeled cells, and only then, the test agent moves into the parking state off the grid to not excessively constrain the system. The resulting test execution is shown in Fig.~\ref{fig:two_quads_sim}.
\subsection{Hardware Experiments}
\begin{figure}[h!]
    \centering
\includegraphics[width=0.7\columnwidth]{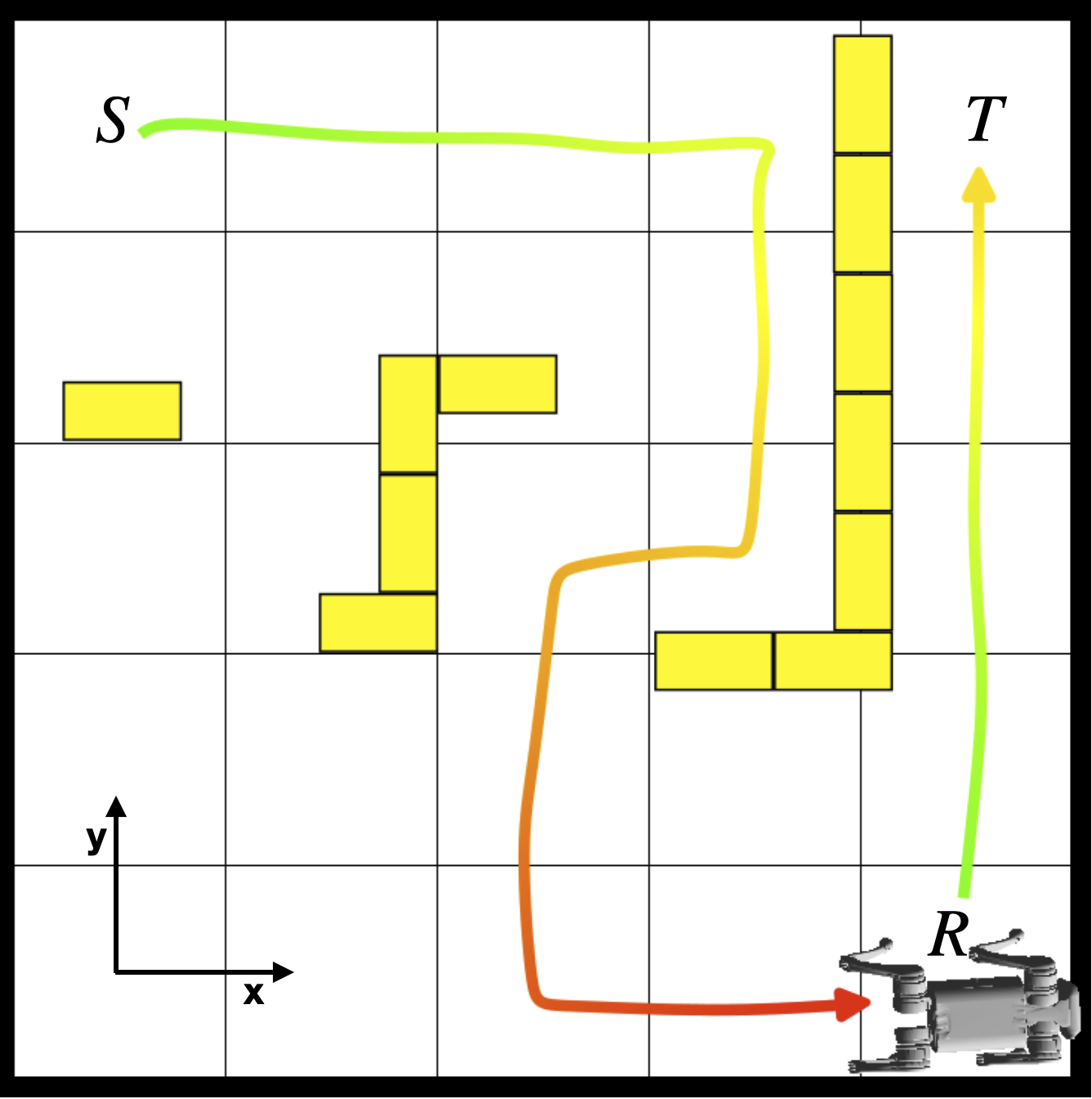}
    \caption{Refueling example experiment trace with yellow boxes representing static obstacles \(\mathtt{Obs}\).}
    \label{fig:refuel_trace}
\end{figure}

\begin{figure*}
\centering
  \begin{minipage}{.33\textwidth}
    \centering
\includegraphics[width=\linewidth,trim={0.0cm 0.0cm 0cm 0.0cm}]{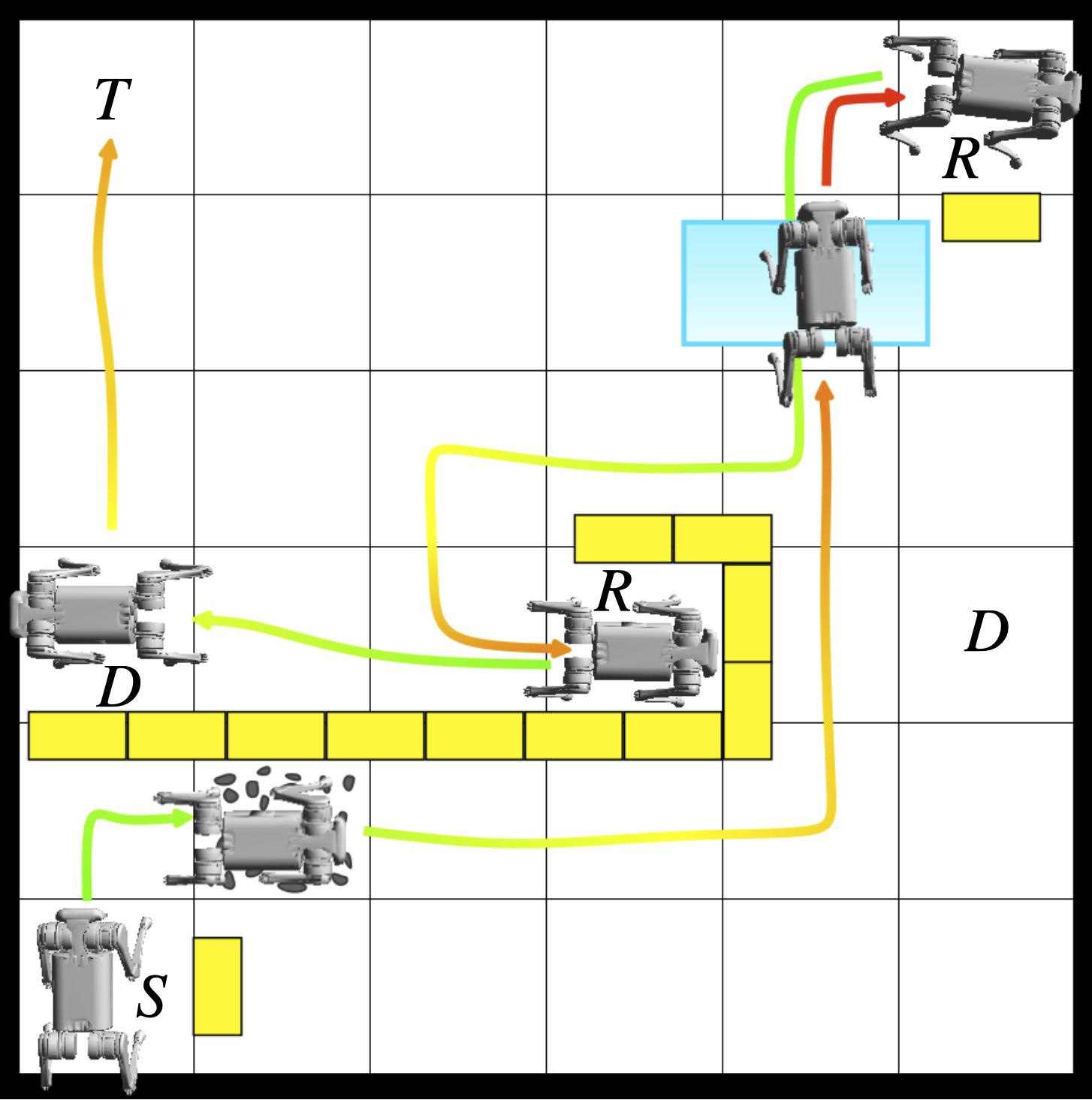}
    \subcaption{\label{fig:mars_expl_trace} Mars exploration experiment trace.}
  \end{minipage}
  \begin{minipage}{.65\textwidth}
    \centering
\includegraphics[width=\linewidth,trim={0.0cm 0.0cm 0cm 0.0cm}]{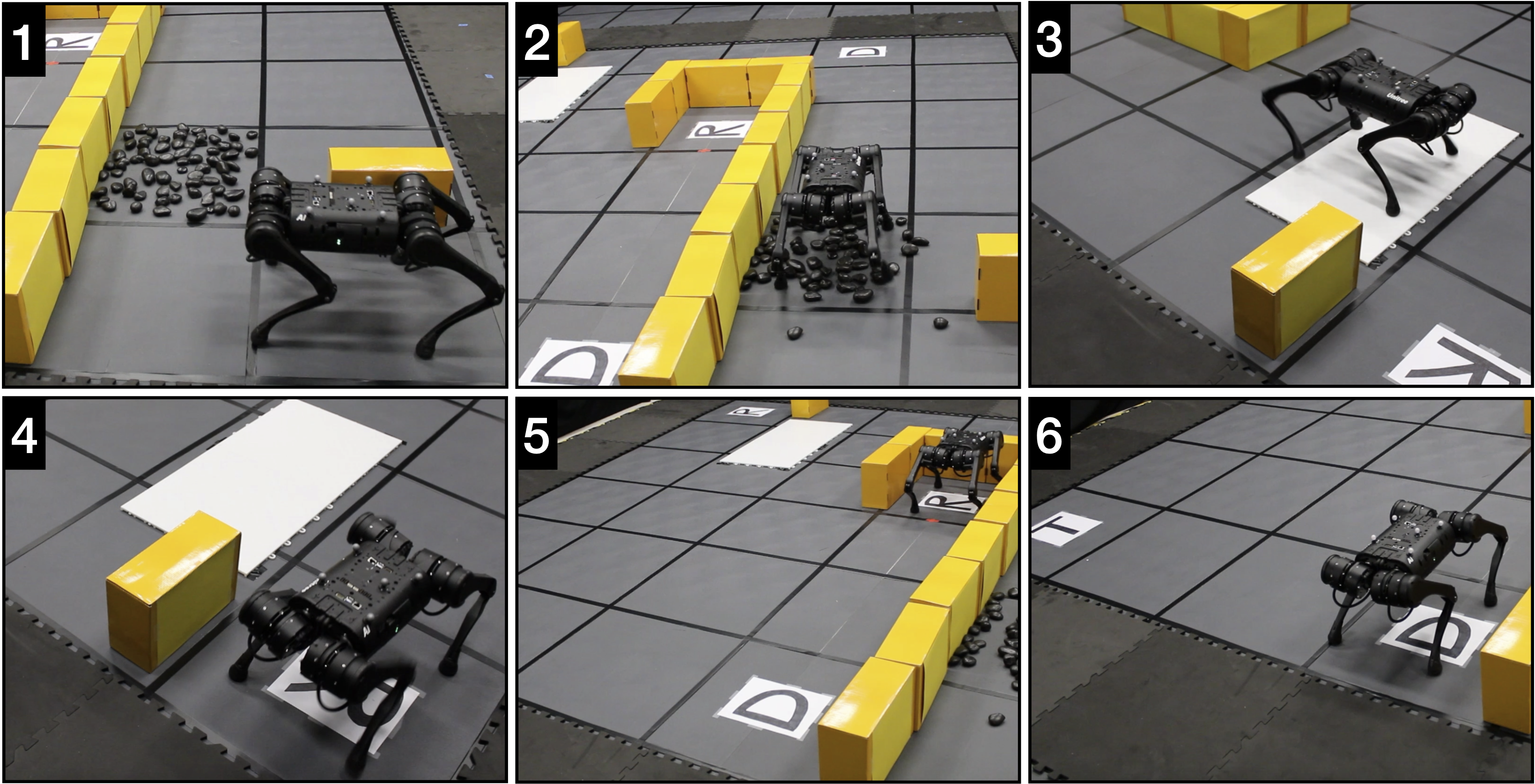}
    \subcaption{\label{fig:mars_expl_snapshopt} Mars exploration experiment snapshots.}
  \end{minipage}
\caption{Resulting test execution on the Unitree A1 quadruped for static test environments.}
\vspace{-4mm}
\label{fig:mars_results}
\end{figure*}

\smallbreak
\noindent
\textbf{Static Test Environment:}
\subsubsection{Running Example}
\begin{figure}
    \centering
    \includegraphics[width=0.75\columnwidth]{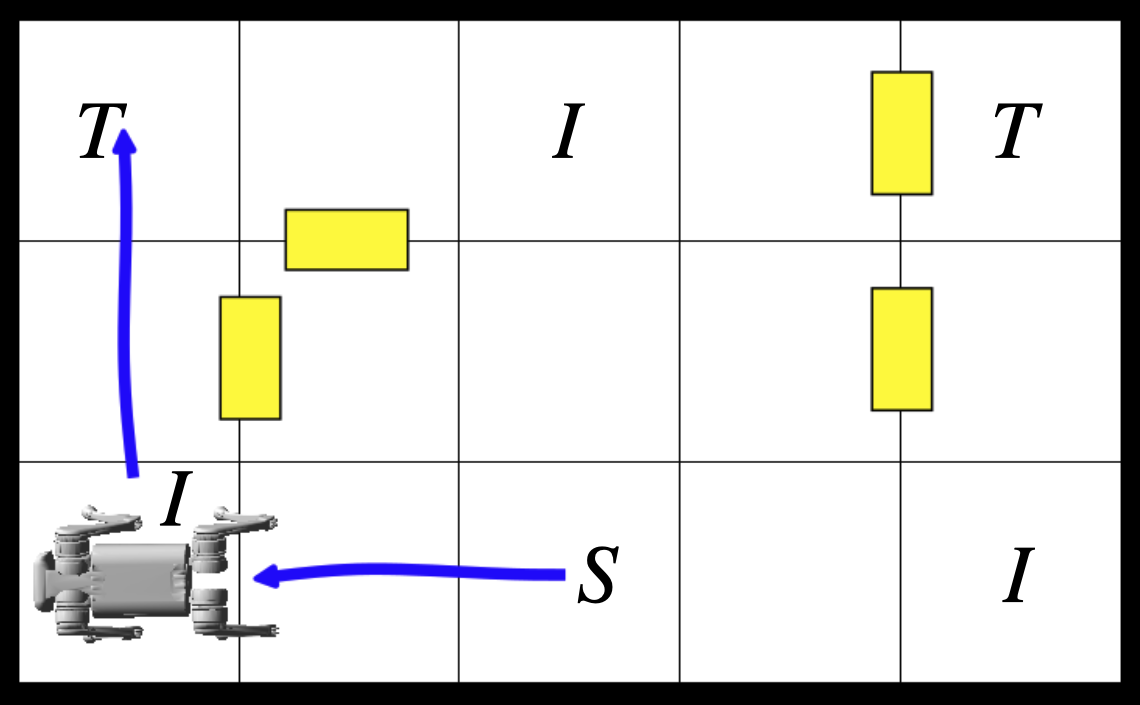}
    \caption{Experiment trace for Example~\ref{ex:med_ex}.}
    \label{fig:med_ex_trace}
\end{figure}

For this experiment we implemented Example~\ref{ex:med_ex} on the quadruped. The resulting test trace is shown in Fig.~\ref{fig:med_ex_trace}.

For the following two examples, the system state also contains the fuel level. Thus, the auxiliary bidirectional constraints in \textbf{\textproc{MILP-static}} are such that the fuel level is abstracted away, meaning if a transition is cut for a specific fuel level, it is cut for all fuel levels.
\subsubsection{Refueling}
This example highlights that intermediate nodes need not always represent poses of the system. In addition to the coordinates \(\mathbf{x} = (x,y)\), the quadruped state also tracks the fuel level \(f\). A full fuel tank consists of 10 units of fuel. Every move on the grid reduces the fuel level by 1 and reaching the refueling station (in the bottom right corner of the grid) resets the fuel tank to full. The desired test behavior is to have the system visit a state that is too far away to reach the goal state with its available fuel, specifically we want to see the system be in the lower three rows of the grid with a fuel level of lower than $2$.
The system objective is given as $\varphi_\sys = \Feventually \text{goal} \land \square \neg (f=0)$ and the test objective is $\varphi_\test = \Feventually (y < 4 \land f < 2)$. 
Note that this test objective also includes states where the fuel tank is empty, $f=0$, but the MILP will not route the test execution through these \emph{unsafe} states, but will automatically only route it through the states where $f=1$ instead.
Snapshots and the trace of the test execution are shown in Fig.~\ref{fig:refuel_trace}. The color of the trace corresponds to the fuel level, and we observe that the obstacle configuration is such that for the quadruped to successfully reach its goal location it is required to visit the refueling station.

\subsubsection{Mars Exploration} In this example the system is tested for a combination of reachability, reaction and avoidance sub-tasks. This example is inspired by a planetary rover's exploration of the Martian surface. Consequently, the grid world has states designated as `rock', `ice', and `drop-off', denoting sample locations and the drop-off position, respectively. In addition to the coordinates \(\mathbf{x}\), the quadruped state also contains the fuel level \(f\) that decreases by 1 for every transition on the grid. The maximum fuel capacity is 10 units and is reset to full at the refueling locations labeled `R'.
The system objective states that the quadruped must reach its goal location, labeled `T', and if it picks up a sample, it shall drop it off at the drop-off location, while not running out of fuel. This is captured in the system objective $$\varphi_\sys = \Feventually T \land \square \neg (f = 0) \land \square(\text{ice} \lor \text{rock} \rightarrow \Feventually \text{drop-off}).$$

The test objective corresponds to the triggers of the reaction sub-task. Specifically, the quadruped is required to collect a `rock' sample and an `ice' sample, and is routed such that a successful run requires the quadruped to refuel: $$\varphi_\test = \Feventually \text{rock} \land \Feventually \text{ice} \land \Feventually (d > f),$$
where $d = \vert \mathbf{x} - \mathbf{x}_{\text{goal}}\vert$ is the distance to the goal and $f$ is the fuel level.
The experiment trace and snapshots of the hardware test execution are shown in Figs.~\ref{fig:mars_expl_trace} and \ref{fig:mars_expl_snapshopt}. From the experiment trace, the static obstacles are placed such that the quadruped has to pick up rock and ice samples, refuel twice, and then drop off samples before reaching its goal. 
The test environment for the hardware run in Fig.~\ref{fig:mars_results} corresponds to a sub-optimal solution of \textbf{\textproc{MILP-static}} with a flow of 1. This sub-optimal solution still ensures that the system is routed in a manner that the test objective is still satisfied. In Table~\ref{tab:exp_rt_no_agent}, we list the runtimes for getting the optimal solution for this example.

\begin{figure*}
    \centering
    \begin{minipage}{0.23\linewidth}
    \includegraphics[width=\linewidth]{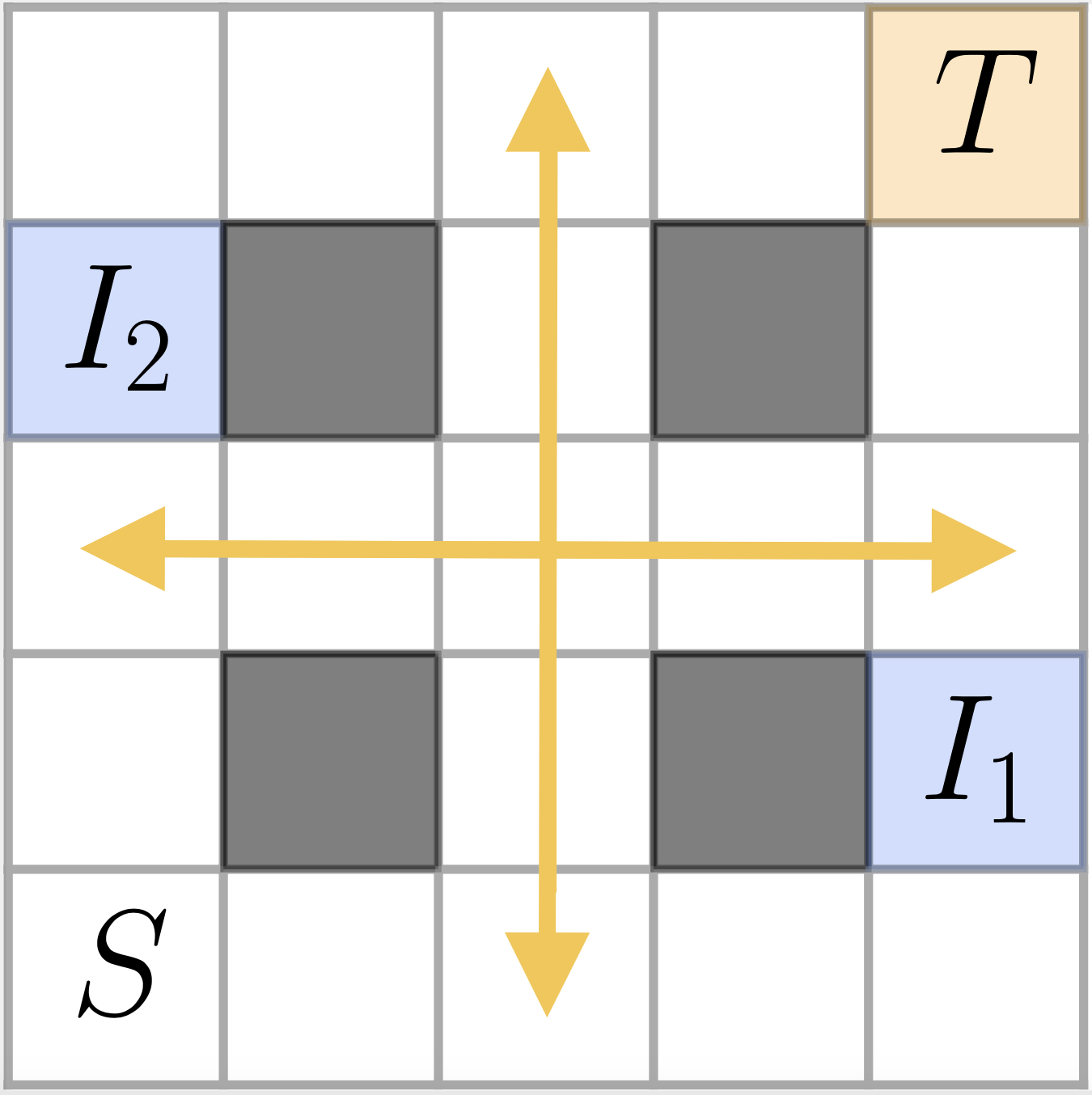}
    \subcaption{Grid world layout.}
    \label{fig:quad_plus_empty}
    \end{minipage}
    \hspace{1mm}
    \begin{minipage}{0.23\linewidth}
    \includegraphics[width=\linewidth]{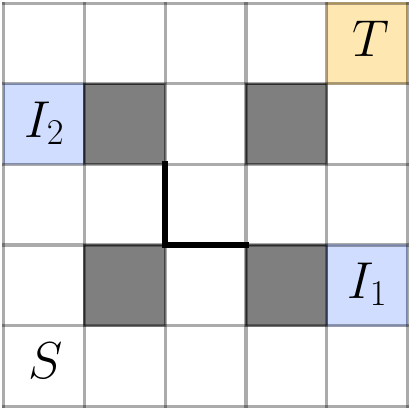}
    \subcaption{Reactive cuts in q$0$.}
    \label{fig:quad_plus_cuts_q0}
    \end{minipage}
    \hspace{1mm}
    \begin{minipage}{0.23\linewidth}
    \includegraphics[width=\linewidth]{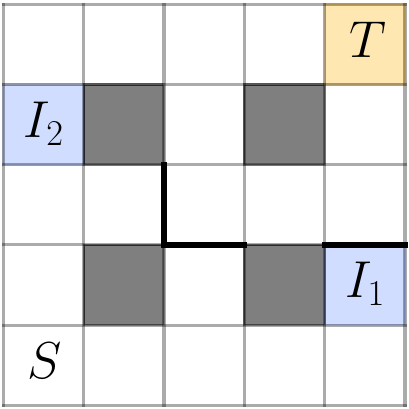}
    \subcaption{Reactive cuts in q$6$.} 
    \label{fig:quad_plus_cuts_q6}
    \end{minipage}
    \hspace{1mm}
    \begin{minipage}{0.23\linewidth}
    \includegraphics[width=\linewidth]{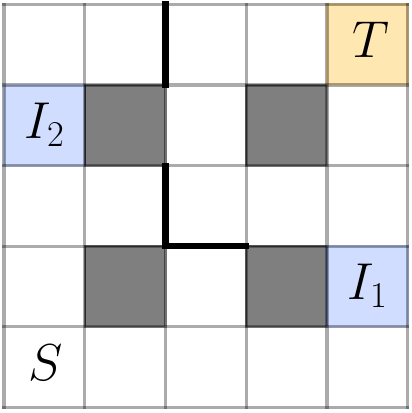}
    \subcaption{Reactive cuts in q$7$.}
    \label{fig:quad_plus_cuts_q7}
    \end{minipage}
    \caption{(a) Grid world layout with cells traversible by the test agent marked. Dark gray cells are not traversible by either agent. (b) Black edges indicate reactive cuts corresponding to the history variables for the Maze 2 experiment. Note that the cuts are not bidirectional. The history variable states q$0$, q$6$, and q$7$ can be inferred from \(\mc{B}_{\pi}\) illustrated in Fig.~\ref{fig:small_reactive_Bpi}, and correspond to initial state, visiting $I_1$ first, and visiting $I_2$ first.}
    \label{fig:quad_plus_cuts}
\end{figure*}

\begin{figure*}
\centering
  \begin{minipage}{.33\textwidth}
    \includegraphics[width=\linewidth,trim={0.0cm 0.0cm -0.0cm 0.0cm}]{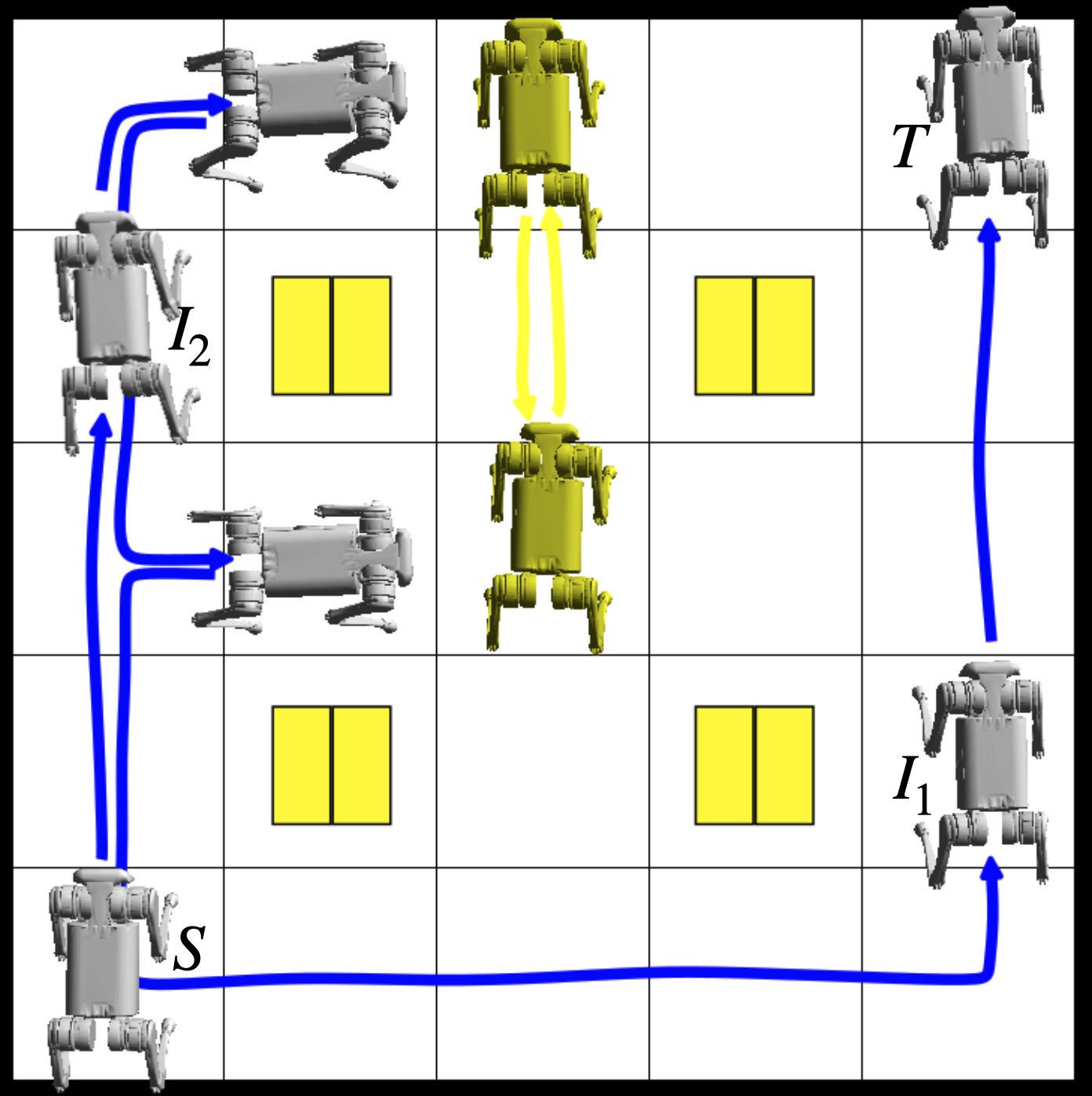}
    \subcaption{\label{fig:maze2_sim} Maze 2 trace.}
  \end{minipage}
  \begin{minipage}{.65\textwidth}
    \centering
\includegraphics[width=\linewidth,trim={0.0cm 0.0cm 0cm 0.0cm}]{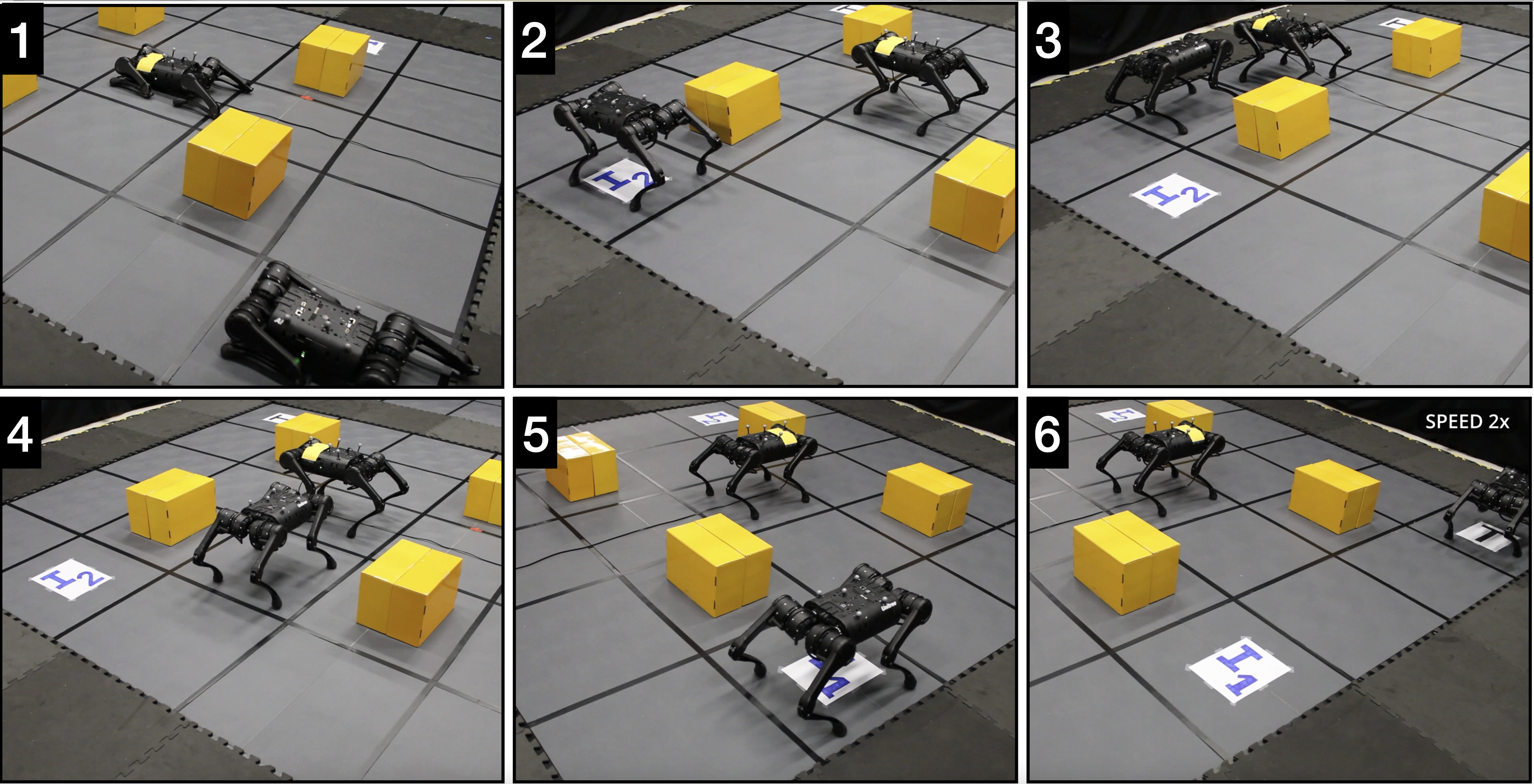}
    \subcaption{\label{fig:maze2_snaps} Maze 2  experiment snapshots.}
  \end{minipage}
\caption{Resulting test execution for the Maze 2 experiment with a dynamic test agent.}
\vspace{-4mm}
\label{fig:results}
\end{figure*}
\smallbreak
\noindent
\textbf{Reactive Dynamic Agent:}
\subsubsection{Patrolling}
This example is similar to the static refueling example, except that the test environment now consists of a test agent and static obstacles (see Fig.~\ref{fig:overview}). The system (gray quadruped) starts in the lower right corner and must reach its goal in the lower left corner of the grid without running out of fuel, which is encoded in the system objective:
$\varphi_\sys = \Feventually T \land \square \neg (f = 0).$
The refueling station is denoted `R' in Fig.~\ref{fig:overview}.
Once again, the test objective routes the system through a state from which a successful test execution requires it to refuel,
$$\varphi_\test = \Feventually (d > f),$$
where $d = \vert \mathbf{x} - \mathbf{x}_{\text{goal}}\vert$ is the distance to the goal.
The test agent can move up and down the third column of the grid, and can leave the grid from the first and last rows to a parking state. 
As shown in the trace and hardware snapshots in Fig.~\ref{fig:overview}, our framework chooses to place a static obstacle near the start state, and the test agent blocks the system from directly navigating to the goal (see panels \(2\), \(3\) and \(4\) in Fig.~\ref{fig:overview}) until its fuel level is low enough, thus requiring it to refuel. For this experiment, we solve \textbf{\textproc{MILP-agent}} with the objective~\eqref{eq:objective} for numerical stability in Gurobi.

\subsubsection{Maze 2} In this example, the system quadruped starts in the bottom left corner of the grid, and must reach its goal location in the top right corner. The grid world is a $5\times 5$ grid, with a symmetric obstacle configuration shown in yellow in Fig.~\ref{fig:quad_plus_empty}. In this example, the test environment consists of a test agent that can traverse along the center row and center column of the grid. While the test environment can also place static obstacles, it realizes the test strategy entirely via the test agent. 
The system objective is given as follows
$\varphi_\sys = \Feventually T$.
The test objective consists of two visit tasks in arbitrary order, encoded as
$$\varphi_\test = \Feventually I_1 \land \Feventually I_2,$$ where $I_1$ and $I_2$ correspond to the designated locations on the grid. The specification product is the same as shown in Fig.~\ref{fig:small_reactive_Bpi}, we can see that to route the test execution through the test objective acceptance states, we need to find cuts for the history variables q$0$, q$6$, and q$7$. The reactive cuts found by the flow-based synthesis procedure are shown in Figs.~\ref{fig:quad_plus_cuts_q0}-\ref{fig:quad_plus_cuts_q7}. The trace and snapshots of the resulting test execution is shown in Figs.~\ref{fig:maze2_sim} and \ref{fig:maze2_snaps}. We observe that the system quadruped decides to take the top path first, visits $I_2$ (see panel $2$ in Fig.~\ref{fig:maze2_snaps}), and is blocked by the test agent (see panel $3$). It then decides to try navigating through the center of the grid, and is again blocked by the test agent (see panel $4$). Subsequently, it decides to try the bottom path, visits $I_1$ (see panel $5$), and successfully reaches the goal without any further test agent intervention. 
If the system decided to visit $I_1$ first, the adaptive test agent strategy would have blocked the system from reaching the goal directly from $I_1$ until it visits $I_2$. This is an example with a maximum flow of \(F=2\), corresponding to the two unique ways for the system to reach the goal. For an alternative system controller in which the system chooses to approach the goal through \(I_1\), the simulated trace resulting from the test agent strategy is shown in Fig.~\ref{fig:quad_plus_alternative}. 


\subsection{Runtimes}
Table~\ref{tab:exp_rt_no_agent} showcases runtimes for simulated and hardware experiments involving static or reactive obstacles. Table~\ref{tab:exp_rt_w_agent} shows runtimes for simulated and hardware experiments with a dynamic test agent. The size of the automata and graphs reported in these tables corresponds to the tuple \((|V|, |E|)\), where \(V\) is the number of nodes, and \(E\) the number of edges. To evaluate the scalability of this framework, we include runtimes on randomized grid worlds for specification sub-tasks in Table~\ref{tab:rand_rt_static} for static obstacles, and in Table~\ref{tab:rand_rt_reactive} for reactive obstacles. These experiments were conducted on an Apple M2 Pro with 16 GB of RAM. The Mars exploration example corresponds to solving an MILP with over $13,000$ binary variables, for which the solver takes $46.6$s to find the optimal solution. For examples involving a dynamic agent such as Maze 1 and Maze 2, our framework iterates through counterexamples that are not dynamically feasible for the test agent until it finds a solution. 
\begin{table*}
\centering
\setlength{\tabcolsep}{1.5pt}
\caption{Graph Construction Runtimes (with mean and standard deviation) for Random Grid World Experiments}
\label{tab:random_graph_rt}
\begin{tabular}{|lc|c|c|c|c|}
\hline
\multicolumn{2}{|c|}{Experiment} & \multicolumn{1}{c|}{$5 \times 5$} & \multicolumn{1}{c|}{$10 \times 10$} & \multicolumn{1}{c|}{$15 \times 15$} & \multicolumn{1}{c|}{$20 \times 20$} \\
\hline
\multicolumn{1}{|c}{$\vert AP \vert$} & {$\vert \mc{B}_{\pi} \vert$}  & \multicolumn{4}{|c|}{Graph Construction [s]}\\
\hline
\hline
\multicolumn{6}{|l|}{\textbf{Reachability}:}\\ 
\hline
2 & (4, 9) & 0.046$\,\pm\,$ 0.001 & 0.224$\,\pm\,$ 0.0056 & 0.554$\,\pm\,$ 0.009 & 1.078$\,\pm\,$ 0.011 \\

3 & (8, 27) & 0.344$\,\pm\,$ 0.007  & 1.661$\,\pm\,$ 0.022 & 4.004$\,\pm\,$ 0.048 & 7.376$\,\pm\,$ 0.061  \\

4 & (16, 81) & 1.997$\,\pm\,$ 0.077 & 9.895$\,\pm\,$ 0.109 & 23.512$\,\pm\,$ 0.179 & 43.188$\,\pm\,$ 0.454 \\

\hline
\hline
\multicolumn{6}{|l|}{\textbf{Reachability \& Reaction}: }\\
\hline
3 & (6, 21) & 0.090$\,\pm\,$ 0.001 & 0.424$\,\pm\,$ 0.016 & 1.037$\,\pm\,$ 0.004 & 2.044$\,\pm\,$ 0.013\\
5 & (20, 155)  & 1.628$\,\pm\,$ 0.087 & 7.560$\,\pm\,$ 0.023 & 18.019$\,\pm\,$ 0.129  & 33.539$\,\pm\,$ 0.144   \\
7 & (68, 1065) & 44.809$\,\pm\,$ 0.996 & 209.612$\,\pm\,$ 1.732 & 488.611$\,\pm\,$ 6.308 & 869.060$\,\pm\,$ 16.870 \\
\hline
\hline

\multicolumn{6}{|l|}{\textbf{Reachability \& Safety}:}\\ 
\hline
3 & (6, 18) & 0.102$\,\pm\,$ 0.002 & 0.508$\,\pm\,$ 0.010 & 1.278$\,\pm\,$ 0.022  & 2.557$\,\pm\,$ 0.023 \\ 

4 & (6, 18) & 0.116$\,\pm\,$ 0.002  & 0.590$\,\pm\,$ 0.009  & 1.485$\,\pm\,$ 0.024 & 2.918$\,\pm\,$ 0.046 \\

5 & (6,18) & 0.179$\,\pm\,$ 0.027 & 0.960$\,\pm\,$ 0.037 & 2.329$\,\pm\,$ 0.072 & 4.482$\,\pm\,$ 0.116 \\
\hline
\end{tabular}
\end{table*}
\begin{table*}
\centering
\setlength{\tabcolsep}{2.5pt}
\caption{Run Times (with mean and standard deviation) for Random Grid World Experiments solving \textbf{\textproc{MILP-reactive}}}
\label{tab:rand_rt_reactive}
\begin{tabular}{|lc|cc|cc|cc|cc|}
\hline
\multicolumn{2}{|c|}{Experiment} & \multicolumn{2}{c|}{$5 \times 5$} & \multicolumn{2}{c|}{$10 \times 10$} & \multicolumn{2}{c|}{$15 \times 15$} & \multicolumn{2}{c|}{$20 \times 20$} \\
\hline
\multicolumn{1}{|c}{$\vert AP \vert$} & {$\vert \mc{B}_{\pi} \vert$}  & \multicolumn{8}{|c|}{Optimization[s], Success Rate (\%)}\\
\hline
\hline
\multicolumn{10}{|l|}{\textbf{Reachability}:}\\ 
\hline
2 & (4, 9) & 5.63$\,\pm\,$13.43  & 100 & 64.62$\,\pm\,$38.75 & 100 & 67.38$\,\pm\,$25.47 & 100  & 68.63$\,\pm\,$31.12 & 100 \\

3 & (8, 27) & 23.36$\,\pm\,$38.15 & 100 & 61.68$\,\pm\,$35.12 &  100 & 91.54$\,\pm\,$31.41 & 100 & 117.82$\,\pm\,$34.89 & 100 \\

4 & (16, 81) & 22.49$\,\pm\,$36.33 & 100 & 83.52$\,\pm\,$29.25 & 100 & 171.49$\,\pm\,$50.72 & 100 & 317.62$\,\pm\,$89.08 & 100 \\
\hline
\hline
\multicolumn{10}{|l|}{\textbf{Reachability \& Reaction}:}\\
\hline
3 & (6, 21) & 5.97$\,\pm\,$13.21 & 100 & 61.06$\,\pm\,$34.67 & 100 & 71.64$\,\pm\,$41.03 & 100 & 85.20$\,\pm\,$19.49 & 100\\
5 & (20, 155) & 17.19$\,\pm\,$25.51 & 100 & 78.44$\,\pm\,$34.71 & 100 & 159.91$\,\pm\,$76.63 & 100 & 279.86$\,\pm\,$148.23 & 90 \\
7 & (68, 1065) & 52.71$\,\pm\,$41.23 & 100 & 331.32$\,\pm\,$187.28 & 90 & 585.21$\,\pm\,$67.58 & 15 & \textbf{600.00$\,\pm\,$0.00} & \textbf{0}\\
\hline
\hline
\multicolumn{10}{|l|}{\textbf{Reachability \& Safety}: }\\
\hline
3 & (6, 18)  & 0.76$\,\pm\,$1.52 & 100  & 70.82$\,\pm\,$89.70 & 100 & 63.68$\,\pm\,$27.54 & 100 &  80.58$\,\pm\,$20.79 & 100 \\ 
4 & (6, 18) & 0.15$\,\pm\,$0.29 & 100 & 71.47$\,\pm\,$80.61 & 100 & 59.59$\,\pm\,$38.92 & 100  & 76.02$\,\pm\,$27.11 & 100 \\
5 & (6, 18) & 0.12$\,\pm\,$0.18 & 100 & 94.68$\,\pm\,$88.04 & 100 & 71.34$\,\pm\,$30.89 & 100 & 82.54$\,\pm\,$22.69 & 100 \\
\hline
\end{tabular}
\end{table*}
\begin{table*}
\centering
\setlength{\tabcolsep}{2.5pt}
\caption{Run Times (with mean and standard deviation) for Random Grid World Experiments solving \textbf{\textproc{MILP-static}}. }
\label{tab:rand_rt_static}
\begin{tabular}{|lc|cc|cc|cc|cc|}
\hline
\multicolumn{2}{|c|}{Experiment} & \multicolumn{2}{c|}{$5 \times 5$} & \multicolumn{2}{c|}{$10 \times 10$} & \multicolumn{2}{c|}{$15 \times 15$} & \multicolumn{2}{c|}{$20 \times 20$}\\
\hline
\multicolumn{1}{|c}{$\vert AP \vert$} & {$\vert \mc{B}_{\pi} \vert$}  & \multicolumn{8}{|c|}{Optimization [s], Success Rate (\%)}\\
\hline
\hline
\multicolumn{10}{|l|}{\textbf{Reachability}:}\\ 
\hline
2 & (4, 9) & 8.17$\,\pm\,$13.14 & 100 & 54.07$\,\pm\,$17.98 & 100 & 60.17$\,\pm\,$0.12 & 100 & 60.17$\,\pm\,$0.10 & 100 \\
3 & (8, 27) &  27.78$\,\pm\,$21.71 & 100 & 60.17$\,\pm\,$0.10 & 100 & 60.48$\,\pm\,$0.86 & 100 & 74.02$\,\pm\,$38.70 & 100 \\
4 & (16, 81) & 52.60$\,\pm\,$14.05 & 100 & 60.42$\,\pm\,$0.34 & 100 & 82.02$\,\pm\,$41.26 & 100 & 265.41$\,\pm\,$203.51 & 80 \\
\hline
\hline
\multicolumn{10}{|l|}{\textbf{Reachability \& Reaction}:}\\
\hline
3 & (6, 21) & 10.62$\,\pm\,$14.85 & 100 & 60.09$\,\pm\,$0.06 & 100 & 60.23$\,\pm\,$0.24 & 100 & 60.34$\,\pm\,$0.46 & 100 \\
5 & (20, 155) & 20.41$\,\pm\,$19.21 & 100 & 67.77$\,\pm\,$31.90 & 100 & 95.31$\,\pm\,$116.65 & 95 & 268.50$\,\pm\,$222.14 & 75\\
7 & (68, 1065) & 36.64$\,\pm\,$23.34 & 100 & 110.63$\,\pm\,$92.81 & 100 & 419.77$\,\pm\,$214.30 & 55 & \textbf{556.38$\,\pm\,$131.06} & \textbf{10}\\
\hline
\hline
\multicolumn{10}{|l|}{\textbf{Reachability \& Safety}:}\\
\hline
3 & (6, 18) & 1.27$\,\pm\,$1.47 & 100 & 60.08$\,\pm\,$0.06 & 100 &  57.27$\,\pm\,$12.61 & 100 &  60.32$\,\pm\,$0.24 & 100 \\
4 & (6, 18) & 0.17$\,\pm\,$0.23 & 100 & 60.06$\,\pm\,$0.05 & 100 & 60.14$\,\pm\,$0.10 & 100 &  60.30$\,\pm\,$0.19 & 100 \\
5 & (6, 18) & 0.11$\,\pm\,$0.16 & 100 & 54.15$\,\pm\,$17.80 & 100 & 60.17$\,\pm\,$0.09 & 100 & 60.29$\,\pm\,$0.26 & 100 \\
\hline
\end{tabular}
\end{table*}
\begin{table*}
\centering
\caption{Runtimes for Simulated and Hardware Experiments showing sizes of the automata and graphs}
\label{tab:exp_rt_no_agent}
\begin{tabular}{|l||*{9}{c}c|}
\hline
{Experiment} & {$\vert \mc{B}_{\pi} \vert$} & $\vert T_{\sys} \vert$ & {$\vert G \vert$} &  {$G$}[s]  & {$\vert$BinVars$\vert$} & {$\vert$ContVars$\vert$} & {$\vert$Constraints$\vert$} & {Opt}[s] & Flow & $\vert C \vert$\\
\hline
\hline
Example~\ref{ex:med_ex} & (4, 9) & (15, 53) & (27, 96) & 0.0270 & 73 & 87 & 540 & 0.0003 & 3.0 & 14 \\
Refueling&(6, 18) & (265, 1047) & (332, 1346) & 0.6655 & 1014 & 1261 & 19819 & 0.8682 & 2.0 & 199\\
Mars Exploration & (36, 354) & (376, 1522) & (4073, 17251) & 75.8313 & 13178 & 16604 & 1646480 & 46.6209 & 2.0 & 1641 \\
Example~\ref{ex:small_reactive} &(8, 27) & (6, 17) & (20, 56) & 0.0452 & 25 & 115 & 409 & 0.0003 & 2.0 & 4 \\
Beaver Rescue & (12, 54) & (7, 19) & (15, 39) & 0.0470 & 8 & 154 & 441 & 0.0001 & 2.0 & 2 \\
Motion Primitives & (16, 81) & (15, 42) & (72, 207) & 0.4286 & 106 & 761 & 2606 & 0.0005 & 3.0 & 15 \\
\hline
\end{tabular}
\end{table*}

\begin{table*}
\centering
\caption{Runtimes for Simulated and Hardware Experiments with Dynamic Agents}
\label{tab:exp_rt_w_agent}
\begin{tabular}{|l||*{9}{c}c|}
\hline
{Experiment} & {$\vert \mc{B}_{\pi} \vert$} & $\vert T_{\sys} \vert$ &  {$\vert G \vert$} & {$G$}[s]  & $\vert$ BinVars $\vert$ & {Opt}[s] & {Controller}[s] & $\vert \mathtt{C}_{\text{ex}} \vert$ & Flow & $\vert C \vert$\\
\hline
\hline
Maze 1 & (16, 81) & (26, 80) &  (196, 604) & 1.6226 & 355 &  0.0007 & 68.7052 & 3 & 1.0 & 3  \\
Patrolling & (6, 18) & (386, 1539) & (210, 831) & 0.4573 & 621 &  6.0535 & 16.1191 & 0 & 1.0 & 13 \\
Maze 2 & (8, 27) & (21, 66) & (80, 252) & 0.2195 & 176 & 0.0160 & 5.0072 & 5 & 2.0 & 8 \\
\hline
\end{tabular}
\end{table*}
For randomized experiments, we time out if the Gurobi fails to find a feasible solution to the MILP within 10 min. If it finds a feasible solution within 10 minutes, we allocate an additional minute for the optimizer reach the optimal, otherwise terminating the optimization with a feasible solution. For these experiments, we increase the length of the system and test objective for the following three classes of specification patterns: i) reachability, ii) reachability and reaction, and iii) reachability and safety. For reachability patterns, the set AP comprises of atomic propositions needed to describe the system and test objectives as follows, $\varphi_\sys = \Feventually p_0$ and $\varphi_\test = \bigwedge_{i=1}^n \Feventually p_i$, and the total number of atomic propositions are $\vert AP \vert = \vert \{p_0, \ldots, p_n\}\vert = n+1$. Similarly, for reachability and reaction patterns (case ii), we have $\varphi_\sys = \Feventually p_1 \wedge \bigwedge_{i=2}^n \square(p_i \rightarrow \Feventually q_i)$ and $\varphi_\test = \Feventually p_0 \wedge \bigwedge_{i=2}^n \Feventually p_i$, with $\vert AP \vert = \vert \{p_0, \ldots, p_n, q_2,\ldots, q_n\} \vert = 2n$. In the reachability and safety case (iii), only the length of the system objective changes: $\varphi_\sys = \Feventually p_1 \wedge \bigwedge_{i=2}^n \square \neg p_i $ and $\varphi_\test = \Feventually p_0$, with $\vert AP \vert = \vert\{p_0, \ldots, p_n\}\vert=n+1$. 
Improving the runtimes for graph construction and controller synthesis subroutines is orthogonal to the focus of this paper. Since the test synthesis framework is carried out offline, we observe reasonable runtimes for medium-sized problems with hundreds to thousands of integer variables. An interesting direction for future research involves identifying good convex relaxations of the MILPs to further improve scalability.

\section{Comparison to Reactive Synthesis}
We presented an approach to solve Problems~\ref{prob:reactive_test_strategy} and~\ref{prob:match_agent} leveraging tools from automata theory and network flow optimization. In particular, for Problem~\ref{prob:match_agent}, we rely on the optimization solution to construct a GR(1) specification to reactively synthesize a test agent strategy. One indication of the optimization step being necessary is the computational complexity of the problem. If the problem data are consistent, there exists a GR(1) specification for the test agent that would solve the problem, but directly expressing this specification is impractical. Essentially, the challenge is in finding the restrictions on system actions, which are then captured in the sub-formulae of the GR(1) specification. In this section, we argue that we cannot solve Problems~\ref{prob:reactive_test_strategy} and~\ref{prob:match_agent} solely via synthesis from an LTL specification. 

To the authors' knowledge, directly capturing the different perspectives of the system and the test agent in this neither fully adversarial nor fully cooperative setting is not possible with current state-of-the-art approaches in GR(1) synthesis. Particularly in the reactive setting, the test strategy must ensure that from the system's perspective, there always exists a path to the system goal. To capture this constraint, we reason over a second product graph that represents the system perspective. It is not obvious how this semi-cooperative setting can be directly encoded as a synthesis problem in common temporal logics.

\begin{figure}
    \centering
    \includegraphics[width=\columnwidth]{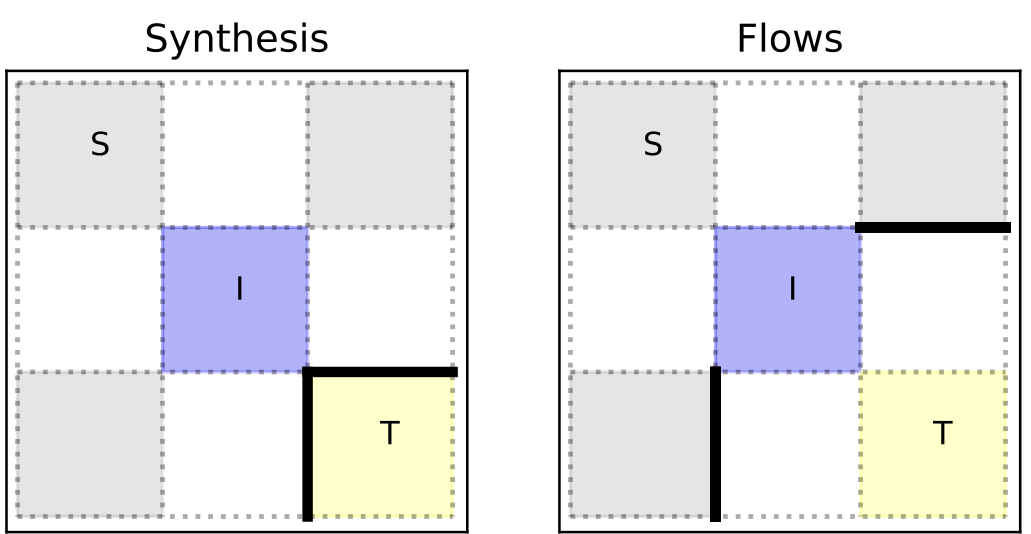}
    \caption{Solution returned by GR(1) synthesis and the network flow optimization in the case of static constraints}
    \label{fig:comparison_gr1_flows}
\end{figure}

In the static setting, the problem can be posed on a single graph. However, it is difficult to find the set of static obstacles directly from GR(1) synthesis. Every state in the winning set describes an edge-cut combination, but qualitative GR(1) synthesis cannot maximize the flow or minimize the cuts. Furthermore, the winning set can include states that vacuously satisfy the formula, i.e., not allowing the system any path to the goal. Finally, the combinatorial complexity of the problem would manifest as follows. Although the time complexity of GR(1) synthesis is \(O(N^3)\) in the number of states \(N\), we require an exponential number of states to characterize the GR(1) formula. For example, in Figure~\ref{fig:comparison_gr1_flows}, this is illustrated for the GR(1) formula:
\begin{equation*}
    \square \varphi^{\text{dyn}}_{\sys} \wedge \square \Feventually \mathtt{T} \rightarrow \square \varphi^{\text{dyn}}_{\test} \wedge \square \varphi^{\text{aux}\_\text{dyn}}_{\test} \wedge \square \Feventually I_{\text{aux}},
\end{equation*}
where \(\varphi^{\text{dyn}}_{\sys}\) captures the system transitions on the grid world,  \(\varphi^{\text{dyn}}_{\test}\) are the dynamics of the test environment, and $\varphi^{\text{aux}\_\text{dyn}}_{\test}$ and $I_{\text{aux}}$ capture the $\Feventually I$ condition in GR(1) form. In this example, each edge in the system transition system \(T_{\sys}\) can take 0/1 values, and once an edge is cut, it remains cut and the system cannot take a transition that corresponds to a cut edge. Due to this, the number of states \(N\) to describe the GR(1) formula includes the \(2^{\vert T_{\sys}.E \vert}\) states that characterize the edge cuts. As seen in Figure~\ref{fig:comparison_gr1_flows}, the direct GR(1) synthesis approach returns a trivial solution corresponding to an impossible setting for the system. Finally, even when an acceptable solution is returned, the problem being at least NP-hard will result in the combinatorial complexity manifesting in the synthesis approach.

One key advantage of the network flow optimization is reasoning over flows as opposed to paths, which allows for tractable implementations. These insights from network flow optimization in this work can help in driving further research along these directions.

\section{Conclusion and Future Work}
We presented a framework to synthesize least-restrictive strategies for test environments according to specified system and test objectives. To do this, we formulate a network flow-based MILP corresponding to the types of agents available in the test environment. In the case of a dynamic test agent, we parse the solution of the MILP to synthesize a test agent strategy via reactive synthesis. Furthermore, we use a counterexample-guided approach to find a realizable test agent strategy. Our problem is shown to be NP-hard, yet the MILP can handle medium-sized problem instances. Our test strategies are such that the system is minimally restricted while routing the test execution through the test objective without creating a livelock. Therefore, a test execution in which the system fails to meet the system objective is solely the fault of the system, and not due to the test environment. 

There are several exciting future directions. First, we aim to extend this framework to automatically select dynamic test agents from a library. This selection can optimized to meet user-defined metrics such as testing effort or cost. 
Secondly, we wish to improve the runtime of our algorithm by using symbolic methods to speed up graph construction and exploring convex relaxations to the MILP. More broadly, we want to investigate how to incorporate test metrics such as coverage and difficulty into our framework. 
\section*{Acknowledgment}
The authors acknowledge Emily Fourney, Chris Umans, Scott Livingston, Joel Burdick, Ioannis Filippidis, Mani Chandy, and Lijun Chen for useful discussions.

\bibliographystyle{ieeetr}
\balance
\bibliography{references}
\newpage
\begin{IEEEbiography}[{\includegraphics[width=1in,height=1.25in,clip,keepaspectratio]{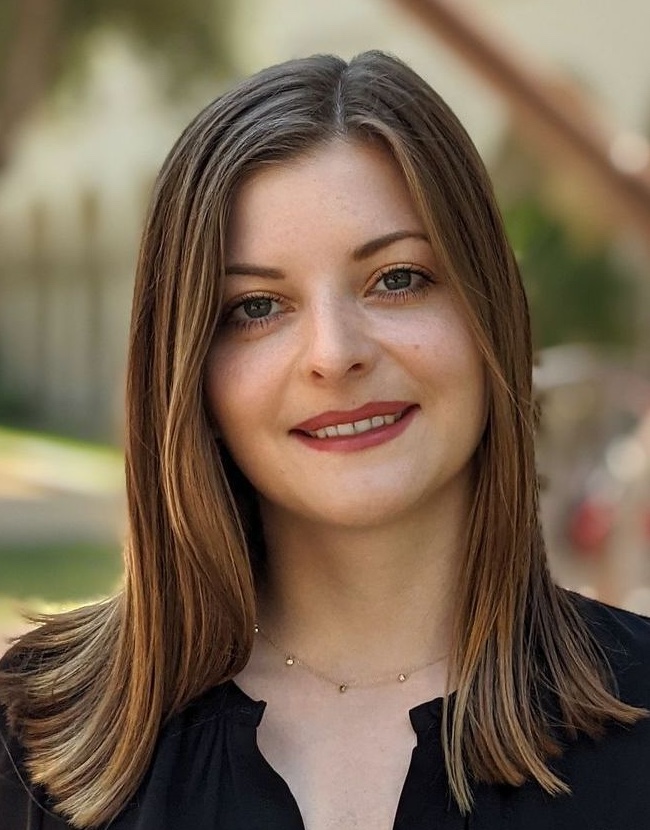}}]{Josefine B. Graebener} (Student Member, IEEE) received a B.Eng. in Aerospace Engineering in 2017 from the Aachen University of Applied Sciences (FH Aachen) in Aachen, Germany, and a M.S. in Space Engineering from California Institute of Technology in 2019. Currently, she is a Ph.D. candidate in Space Engineering with a minor in Computer Science at the California Institute of Technology. Her research interest lies in using formal methods for test and evaluation of autonomous systems, and system diagnostics.
\end{IEEEbiography}
\begin{IEEEbiography}[{\includegraphics[width=1in,height=1.25in,clip,keepaspectratio]{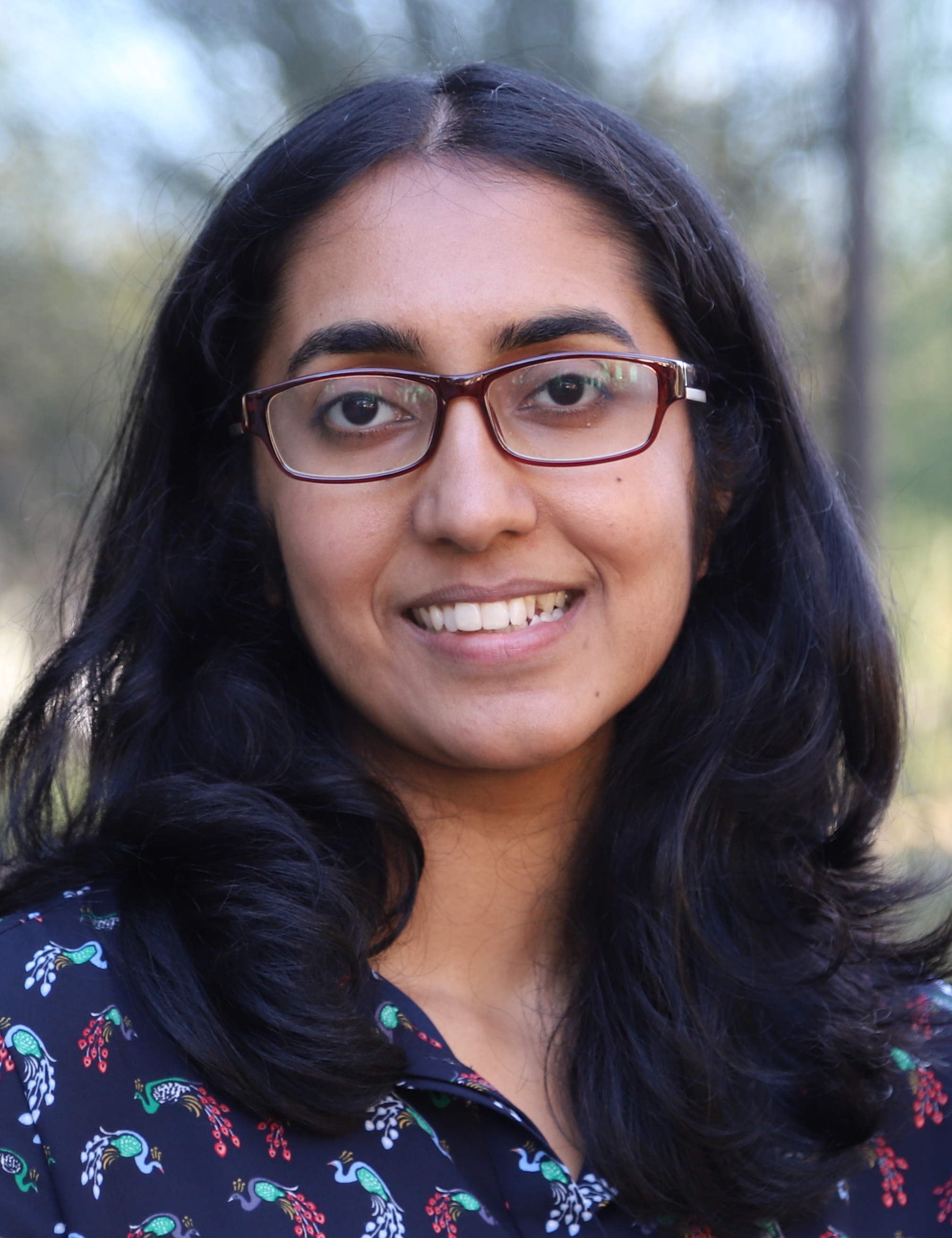}}]{Apurva Badithela}(Student Member, IEEE) received a Bachelors degree in Aerospace Engineering and Mechanics in 2018 from the University of Minnesota, Twin-Cities. Currently, she is a Ph.D. candidate in Control and Dynamical Systems at the California Institute of Technology. Her dissertation work focuses on Formal Test Synthesis and System-level Evaluation for Safety-Critical Autonomous Systems.  
\end{IEEEbiography}
\begin{IEEEbiography}[{\includegraphics[width=1in,height=1.25in,clip,keepaspectratio]{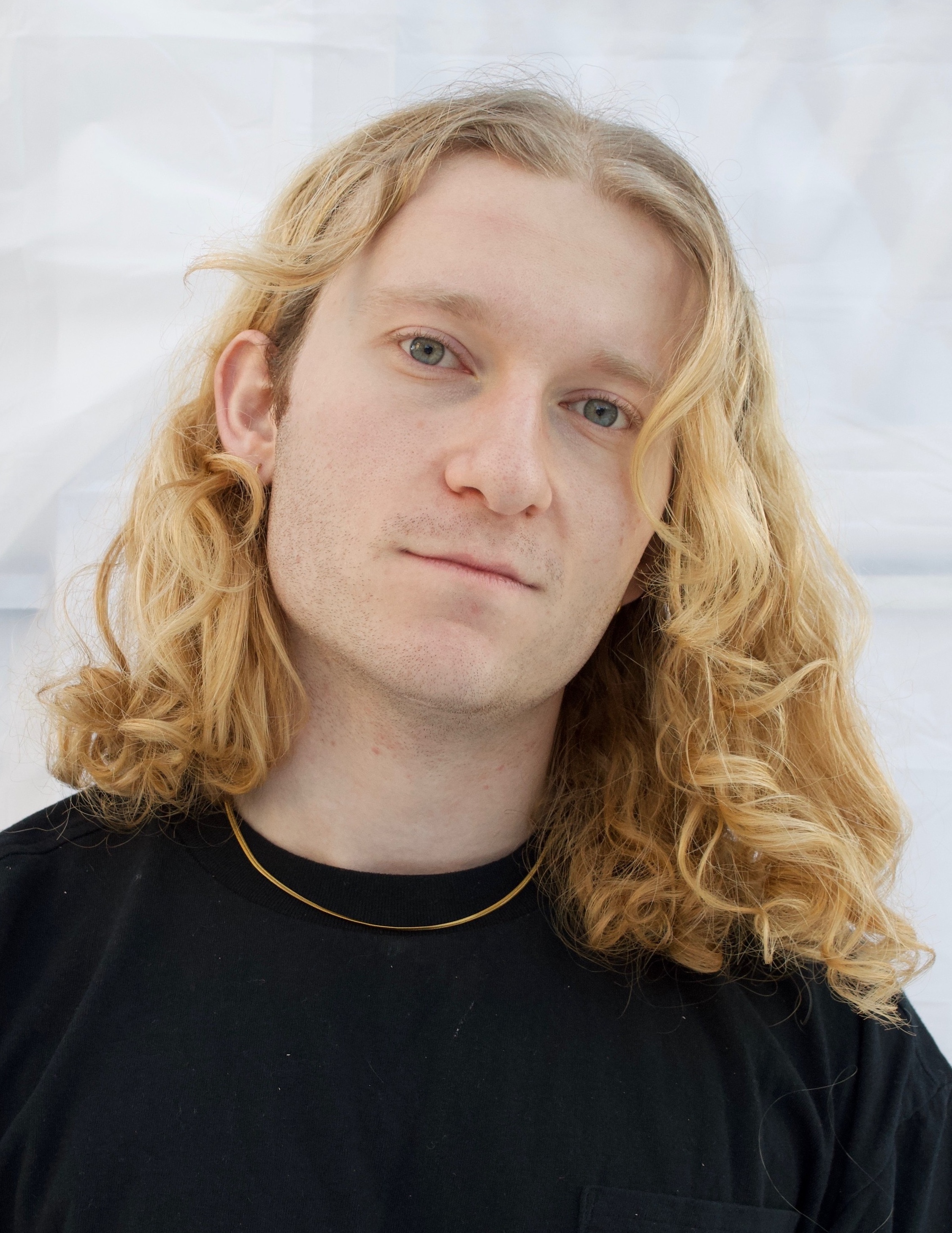}}]{Denizalp Goktas} (Student Member, IEEE), 
Denizalp Goktas is a Ph.D. Candidate in Computer Science at Brown University. His research focuses on artificial intelligence, particularly how it intersects with economics and computer science. His research seeks to create algorithms for games and markets, aiming to use these to tackle problems practical problem such as in economics and robotics. He is supported by a JP Morgan AI fellowship.
Previously, Denizalp earned his BA in Computer Science and Statistics from Columbia University and another BA in Political Science and Economics from Sciences Po. His past research experience includes internships at Google DeepMind and JP Morgan, and a visiting scholar position at UC Berkeley’s Simons Institute.
\end{IEEEbiography}
\begin{IEEEbiography}[{\includegraphics[width=1in,height=1.25in,clip,keepaspectratio]{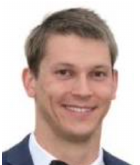}}]
{Wyatt Ubellacker} (Student Member, IEEE), earned his B.S. and M.S. degrees in Mechanical Engineering from the Massachusetts Institute of Technology in 2013 and 2016, respectively. Prior to joining Caltech in 2019, he was a Robotics Technologist at the Jet Propulsion Laboratory, where he wrote autonomy and control algorithms for the Mars Perseverance Rover.

Currently, he is a Ph.D. candidate in Control and Dynamical Systems at Caltech. His research interests focus on control and autonomy for dynamic platforms, with a special emphasis on robotic morphologies that are capable of exhibiting a wide variety of behaviors.
\end{IEEEbiography}
\begin{IEEEbiography}[{\includegraphics[width=1in,height=1.25in,clip,keepaspectratio]{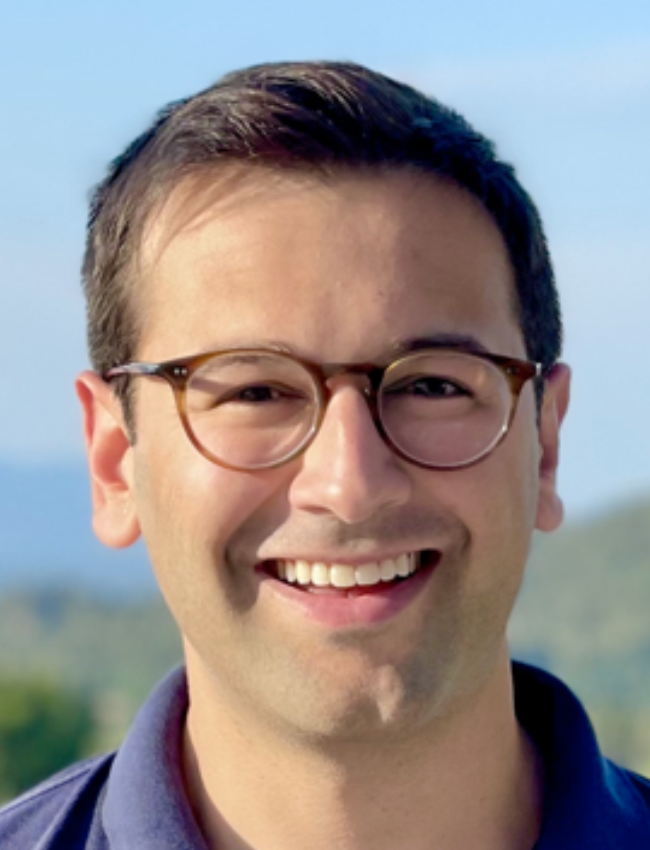}}]
{Eric V. Mazumdar} (Member, IEEE) received a B.S. degree in Computer Science from the Massachusetts Institute of Technology (MIT) in 2015, and a Ph.D in Electrical Engineering and Computer Science from UC Berkeley in 2021.

Currently he is an Assistant Professor in Computing and Mathematical Sciences and Economics at Caltech. His research interests lie at the intersection of machine learning and economics, focusing on developing theoretical foundations and tools to confidently deploy machine learning algorithms into societal systems, particularly in settings with uncertain, dynamic environments in which learning algorithms interact with strategic agents.

Dr. Mazumdar received the NSF CAREER Award in 2023 as well as a Research Fellowship for Learning in Games from the Simons Institute for Theoretical Computer Science.
\end{IEEEbiography}

\begin{IEEEbiography}[{\includegraphics[width=1in,height=1.25in,clip,keepaspectratio]{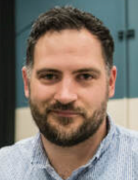}}]
{Aaron D. Ames} (Fellow, IEEE) received a B.S. degree in Mechanical Engineering and a B.A. degree in Mathematics from the University of St. Thomas in 2001, and a M.A. degree in Mathematics and a Ph.D. in Electrical Engineering and Computer Sciences from UC Berkeley in 2006.

Currently he is the Bren Professor of Mechanical and Civil Engineering and Control and Dynamical Systems at the California Institute of Technology. Prior to joining Caltech, he was an Associate Professor in Mechanical Engineering and Electrical \& Computer Engineering at the Georgia Institute of Technology. He was as a Postdoctoral Scholar in Control and Dynamical Systems at Caltech from 2006 to 2008, and began is faculty career at Texas A\&M University in 2008. His research interests span the areas of robotics, nonlinear control and hybrid systems, with a special focus on applications to bipedal robotic walking—both formally and through experimental validation.

Dr. Ames was the recipient of the 2005 Leon O. Chua Award for achievement in nonlinear science at UC Berkeley, and the 2006 Bernard Friedman Memorial Prize in Applied Mathematics. Dr. Ames received the NSF CAREER award in 2010, and the Donald P. Eckman Award in 2015, and the 2019 Antonio Ruberti Young Researcher Prize.\end{IEEEbiography}

\begin{IEEEbiography}[{\includegraphics[width=1in,height=1.25in,clip,keepaspectratio]{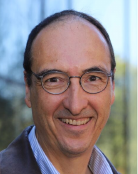}}]
{Richard M. Murray} (Fellow, IEEE) received the B.S. degree in Electrical Engineering from California Institute of Technology in 1985 and the M.S. and Ph.D. degrees in Electrical Engineering and Computer Sciences from the University of California, Berkeley, in 1988 and 1991, respectively.

He is currently the Thomas E. and Doris Everhart Professor of Control \& Dynamical Systems and Bioengineering at Caltech. Murray's research is in the application of feedback and control to networked systems, with applications in synthetic biology and autonomy. Current projects include design and implementation of synthetic cells and design, verification, and test synthesis for discrete decision-making protocols for safety-critical, reactive control systems.

Dr. Murray's professional awards include the Richard P. Feynman-Hughes Faculty Fellowship in 1993, awarded annually to an outstanding young faculty member in Engineering and Applied Science at Caltech, the National Science Foundation Early Faculty Career Development (CAREER) Award in 1995, the Office of Naval Research Young Investigator Award in 1995 and the Donald P. Eckman Award in 1997. He is a Fellow of the Institute for Electrical and Electronics Engineers (IEEE) and holds an honorary doctorate from Lund University in Sweden. He is an elected member of the National Academy of Engineering (2013). Dr. Murray received the IEEE Bode Lecture Prize in 2016, the IEEE Control Systems Award in 2017, and the AACC John R. Ragazzini Education Award in 2019.
\end{IEEEbiography}

\end{document}